\documentstyle[psfig,aps]{revtex}
\tightenlines
\def\Journal#1#2#3#4{{#1} {\bf #2}, #3 (#4)}

\def\AP{{\em Ann. Phys.}}
\def\APPB{{\em Acta Phys. POlonica} B}
\def\CMP{{\em Comm. Math. Phys.}}
\def\IJMPA{{\em Int. J. of Mod. Phys.}}

\def\JHEP{{\em J. High of Energy Phys.}}
\def\JMP{{\em J. Math. Phys.}}

\def\JPC{{\em J. Phys.} C}

\def\JSP{{\em J. Stat. Physics.}}
\def\MPLA{{\em Mod. Phys. Lett.} A}
\def\NCA{{\em Nuovo Cimento} A}
\def\NPB{{\em Nucl. Phys.} B}

\def\PHA{{\em Physica} A}

\def\PLA{{\em Phys. Lett.} A}
\def\PLB{{\em Phys. Lett.} B}
\def\PRL{\em Phys. Rev. Lett.}
\def\PR{{\em Phys. Rev.}}
\def\PRA{{\em Phys. Rev.} A}
\def\PRB{{\em Phys. Rev.} B}
\def\PRD{{\em Phys. Rev.} D}

\def\PREP{{\em Phys. Rep.}}
\def\PREPC{{\em Phys. Rep.} C}

\def\PTP{{\em Progr. of Theor. Phys.}}

\def\RMP{{\em Rev. Mod. Phys.}}
\def\RNC{{\em Riv. Nuovo Cimento}}

\def\ZPC{{\em Z. Phys.} C}
\def\YF{{\em Yad. Fiz.}}

\newcommand{\be}{\begin{equation}}
\newcommand{\ee}{\end{equation}}
\newcommand{\bea}{\begin{eqnarray}}
\newcommand{\eea}{\end{eqnarray}}

\newcommand{\hf} {{1\over2}}
\newcommand{\nonu}{\nonumber\\}
\newcommand{\br}  {\hskip -0.25cm /}
\def\dk{\Delta k}
\def\eq#1{(\ref{#1})}
\def\tr{{\mathrm tr}}
\def\mr#1{{\mathrm{#1}}}
\def\ra{\rangle}
\def\la{\langle}
\def\ord#1{{\cal O}\left(#1\right)}
\def\ih{{1\over\hbar}}
\def\psib{\bar\psi}
\def\jb{\bar j}
\def\br{\hskip -0.2cm /}
\def\tGa{\tilde\Gamma}

\begin{document}
\title{Lectures on the functional renormalization\\ group method}
\author{Janos Polonyi\thanks{E-mail: polonyi@fresnel.u-strasbg.fr}}
\address{Laboratory of Theoretical Physics,\\ Louis Pasteur
University, Strasbourg, France \and Department of Atomic Physics, \\
L. E\"otv\"os University, Budapest, Hungary}

\date{\today}
\maketitle
\begin{abstract}
These introductory notes are about functional
renormalization group equations and some of their applications. It
is emphasised that the applicability of this method extends well
beyond critical systems, it actually provides us a general purpose
algorithm to solve strongly coupled quantum field theories. The
renormalization group equation of F. Wegner and A. Houghton is
shown to resum the loop-expansion. Another version, due to J.
Polchinski, is obtained by the method of collective coordinates
and can be used for the resummation of the perturbation series.
The genuinely non-perturbative evolution equation is obtained by a
manner reminiscent of the Schwinger-Dyson equations. Two variants
of this scheme are presented where the scale which determines the
order of the successive elimination of the modes is extracted from
external and internal spaces. The renormalization of composite
operators is discussed briefly as an alternative way to arrive at
the renormalization group equation. The scaling laws and fixed
points are considered from local and global points of view.
Instability induced renormalization and new scaling laws are shown
to occur in the symmetry broken phase of the scalar theory. The
flattening of the effective potential of a compact variable is
demonstrated in case of the sine-Gordon model. Finally, a
manifestly gauge invariant evolution equation is given for QED.
\end{abstract}

\tableofcontents
\newpage

\section{Introduction}
The origin of renormalization goes back to hydrodynamics, the
definition of the mass and other dynamical characteristics of
bodies immersed into fluids. The more systematic elaboration of
this concept is based on the (semi) group property of changing the
observational scale in a theory. The renormalization group (RG)
method has already been used with a number of different goals.
Some of the more important directions are (i) to remove U.V.
divergences \cite{rgqft}, (ii) to describe the scale dependence of
physical parameters and to classify the parameters of a theory
around a critical point according to their impact on the dynamics
\cite{rgcrs}, (iii) to express the highly singular product of
local field variables \cite{opprod}, (iv) to resum the
perturbation expansion in Quantum Field Theory \cite{resum} and in
the case of differential equations \cite{prde} and finally (v) to
solve strongly coupled theories. Furthermore, the RG method offers
a systematic way of organizing our understanding of a complicated
dynamics by successively eliminating degrees of freedom. The
distinctive feature of the method is that it retains the influence
of the eliminated subsystem on the rest.

The main subject of these lectures is application (v)
in the framework of the functional formalism,
the search of a general purpose algorithm to handle
non-perturbative and not necessarily critical theories.

The strategy of the RG scheme for the resummation of the perturbation
expansion is to introduce and to evolve effective vertices
instead of dealing with higher order corrections. The result is an iterative
procedure where the contributions to the effective vertices
computed at a given step in the leading order of the perturbation expansion
and inserted in a graph at the next step reproduce the higher order
corrections. In the traditional implementation
of the RG procedure one follows the evolution of few coupling
constants only. Our concern will be an improvement on this method in
order to follow the evolution of a large number of coupling constants.
This is realized by means of the functional formalism where the
evolution is constructed for the generator function for all coupling
constants.

Before starting let us review the conventional multiplicative RG schemes
as used in Quantum Field Theory. This scheme is based on the relation
\be\label{mult}
G_R^{(n)}(p_1,\cdots,p_n;g_R,\mu)=
Z^{-n/2}G_B^{(n)}(p_1,\cdots,p_n;g_B,\Lambda)+\ord{{\mu^2\over\Lambda^2}}
+\ord{{p_j^2\over\Lambda^2}},
\ee
for the Green functions, where $\Lambda$ and $\mu$ denote the U.V. cut-off
and the substraction (observational) scales, respectively. $Z$ depends
on either $g_R$ or $g_B$ and the ratio $\Lambda/\mu$ (the mass is treated
as one of the parameters $g$). The ignored terms stand for non-universal,
cut-off dependent interactions.

\underline{Bare RG equation}:
The renormalized theory is independent of the choice of the \hbox{cut-off},
\be\label{trba}
\Lambda{d\over d\Lambda}G_R^{(n)}(p_1,\cdots,p_n;g_B,\Lambda)
=\Lambda{d\over d\Lambda}\left[Z^{n/2}(g_B,\Lambda)
G_B^{(n)}(p_1,\cdots,p_n;g_B,\Lambda)\right]=0,
\ee
where the renormalized coupling constants and the substraction scale
are kept fixed in computing the derivatives. The equation involving
the renormalized quantities is not too useful. It is the multiplicative
renormalization scheme which leads to the second equation,
expressing the possibility of compensating the change of
the substraction scale by the modification of
the coupling constants $g_B(\Lambda)$ arising from the solution of the
differential equation by the method of characteristics. Notice that
the compensation is possible only if the list of the coupling constants
$\{g_B\}$ includes all relevant coupling constants of the appropriate
scaling regime, around the U.V. fixed point.

\underline{Renormalized RG equation}:
The bare theory is independent of the choice of the substraction scale,
\be
\mu{d\over d\mu}G_B^{(n)}(p_1,\cdots,p_n;g_R,\mu)
=\mu{d\over d\mu}\left[Z^{-n/2}(g_R,\mu)
G_R^{(n)}(p_1,\cdots,p_n;g_R,\mu)\right]=0,
\ee
where the bare coupling constants and the cut-off are kept fixed.

\underline{Callan-Symanzik equation}:
The change of the mass in the propagator is governed by the expression
\be\label{propmd}
{d\over dm^2}{1\over p^2-m^2}={1\over p^2-m^2}\cdot{1\over p^2-m^2}
\ee
which can be used to find out the dependence on the renormalized mass
when the cut-off and all bare parameters except the bare mass
are kept fixed. The resulting evolution equation \cite{casy}
is similar to the two previous RG equations except that the derivative
is with respect to the bare mass instead of the cut-off
or the substraction scale and the right hand side is
non-vanishing. This latter feature indicates that
contrary to the first two schemes the evolution $g_R(m^2)$ in the
Callan-Symanzik equation is not a renormalized trajectory, it
connects theories with different mass.

The serious limitation of these equations is that they are
asymptotic, i.e., are applicable in the regime
$\mu^2,p^2<<\Lambda^2$ only. In fact, the omission of the
non-universal terms in the multiplicative renormalization scheme
\eq{mult} requires that we stay away from the cut-off. In models
with IR instability (spontaneous symmetry breaking, dynamical mass
generation,...) another limitation arises, $m^2_{dyn}<<\mu^2,p^2$,
to ensure that we stay in the U.V. scaling regime. This is because
the IR scaling regime may have relevant operators which are
irrelevant on the U.V. side \cite{glrg}. In more realistic models
with several scaling regimes which have different relevant
operator sets we need non-asymptotic methods which can interpolate
between the different scaling regimes. This is achieved by the
functional extensions of the RG equation based on the
infinitesimal blocking steps.

\section{Functional RG equations}
There are different avenues to arrive at a functional RG equation.
The simplest is to follow Wilson-Kadanoff blocking in continuous
space-time. It leads to a functional differential equation
describing the cut-off dependence of the bare action.
Better approximation schemes can be worked out when the modes
are suppressed smoothly as the cut-off changes. The parameters
of the bare action which are followed by these RG equations are
related to observables qualitatively only. It is more advantageous
to transform the RG equation for the effective action whose
parameters have direct relation with observables.

For the sake of simplicity we consider an Euclidean theory in $d$
dimensions for a scalar field $\phi_x$ governed by the action
$S_B[\phi]$. An $O(d)$ invariant U.V. cut-off $\Lambda$ is
introduced in the momentum space by requiring $\phi_p=0$ for
$|p|>\Lambda$ to render the generator functional \be\label{cggf}
e^{\ih W[j]}=\int D[\phi]e^{-\ih S_B[\phi]+\ih j\cdot\phi} \ee
finite. We shall use the following notation conventions, stated
below: In order to render the functional manipulations well
defined we always assume discrete spectrum, i.e., the presence of
U.V. and IR regulators, say a lattice spacing $a$ and system volume
$V=a^dN^d$. The space-time integrals are $\int
d^dx=a^d\sum_x=\int_x$, $\int_xf_xg_x=f\cdot g$, $\int
d^dp/(2\pi)^d=V^{-1}\sum_p=\int_p$, $f_p=\int_xe^{-ipx}f_x$, and
$f_x=\int_pe^{ipx}f_p$. The Dirac-deltas
$\delta_{x,y}=a^{-d}\delta_{x,y}^K$ and
$\delta_{p,q}=V\delta_{p,q}^K$ are expressed in terms of the
Kronecker-deltas $\delta_{x,y}^K$ and $\delta_{p,q}^K$.

\subsection{Resumming the loop expansion}
We start with the simplest form of infinitesimal blocking-step RG
equations which is the functional extension of the bare RG scheme
mentioned above \cite{wh}.

\subsubsection{Blocking in continuous space-time}\label{blcspts}
We shall denote the moving U.V. cut-off by $k$. Its lowering $k\to k-\dk$
leads to the blocking transformation of the action which preserves the
generator functional \eq{cggf}.
Due to the presence of the source this blocking introduces an explicit
source dependence in the action,
\be\label{sourcebl}
\int D_k[\phi]e^{-\ih S_k[\phi;j]+\ih j\cdot\phi}
=\int D_{k-\dk}[\phi]e^{-\ih S_{k-dk}[\phi;j]+\ih j\cdot\phi},
\ee
where $D_k[\phi]$ stands for the integration measure over the functional
space ${\cal F}_k$ consisting of functions whose Fourier transform
is non-vanishing for $|p|\le k$. In order to avoid this complication
one usually assumes $j\in{\cal F}_{k-\dk}$. In this case the blocking
becomes a mapping $S_k[\phi]\to S_{k-\dk}[\phi]$ and it is enough
impose the invariance of the partition function
\be\label{obl}
e^{-\ih S_{k-\dk}[\phi]}=\int D[\tilde\phi]e^{-\ih S_k[\phi+\tilde\phi]},
\ee
where $\phi\in{\cal F}_{k-\dk}$ and
$\tilde\phi\in{\cal F}_k\backslash{\cal F}_{k-\dk}$.

The evaluation of the path integral by means of the loop-expansion
gives immediately the functional RG equation,
\be\label{elsowh}
S_{k-\dk}[\phi]=S_k[\phi+\tilde\phi_{cl}]
+{\hbar\over2}\tr\ln{\delta^2S_k[\phi+\tilde\phi_{cl}]\over
\delta\tilde\phi\delta\tilde\phi}+\ord{\hbar^2},
\ee
where $\tilde\phi_{cl}$ denotes the saddle point. The
trace is over the functional space
$\tilde\phi\in{\cal F}_k\backslash{\cal F}_{k-\dk}$ and $\ord{\hbar^2}$
represents the higher loop contributions. We write this equation as
\be\label{pbl}
S_k[\phi]-S_{k-\dk}[\phi]
=\underbrace{S_k[\phi]-S_k[\phi+\tilde\phi_{cl}]}_\mr{tree\ level}
-\underbrace{{\hbar\over2}\tr\ln{\delta^2S_k[\phi+\tilde\phi_{cl}]\over
\delta\tilde\phi\delta\tilde\phi}+\ord{\hbar^2}}_\mr{loop\ contributions}.
\ee

When can one safely assume that the saddle point is trivial,
$\tilde\phi_{cl}=0$? Let us suppose that
$\phi_x$ is weakly and slowly varying, ie $\phi_x=\Phi+\eta_x$
where $\eta_x\approx0$ and the characteristic momentum of $\eta$
is small with respect to the cut-off $k$. Then the
fluctuating component $\eta$ appears as an external source breaking
translation invariance. The saddle point,
$\tilde\phi\in{\cal F}_k\backslash{\cal F}_{k-\dk}$,
being inhomogeneous exists as $\eta\to0$ only if the external
space-time symmetry is broken dynamically.

But a $\dk$-dependent small saddle point, $\tilde\phi_\mr{cl}=\ord{\dk^n}$,
$n>0$ may occur without breaking external symmetries. In this case we
expand in $\tilde\phi_\mr{cl}$ and find the Wegner-Houghton (WH) equation
\cite{wh}
\bea\label{morgwh}
S_k[\phi]-S_{k-\dk}[\phi]&
= & \hf{\delta S_k[\phi]\over\delta\tilde\phi}\cdot
\left({\delta^2S_k[\phi]\over\delta\tilde\phi\delta\tilde\phi}\right)^{-1}
\cdot{\delta S_k[\phi]\over\delta\tilde\phi} \\
& & -{\hbar\over2}\tr\ln{\delta^2S_k[\phi]\over
\delta\tilde\phi\delta\tilde\phi}+\ord{\hbar^2}+\ord{\dk^2}, \nonumber
\eea
where all functional derivatives are taken at $\tilde\phi=0$.
The saddle point can be omitted in the argument of the logarithmic
function because the trace brings a factor of $\dk$ as we shall see below.
The discussion of the tree-level renormalization when $\tilde\phi_\mr{cl}$
is non-vanishing and finite as $\dk\to0$ is deferred to section \ref{tree}.

\begin{figure}
\centerline{\psfig{file=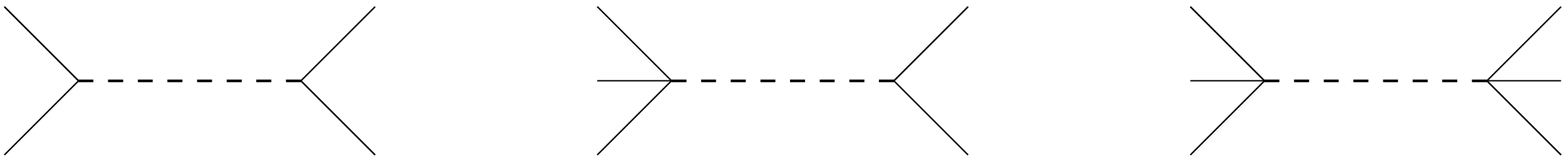,height=1.4cm,width=10.cm,angle=0}}
\centerline{(a)}

\centerline{\psfig{file=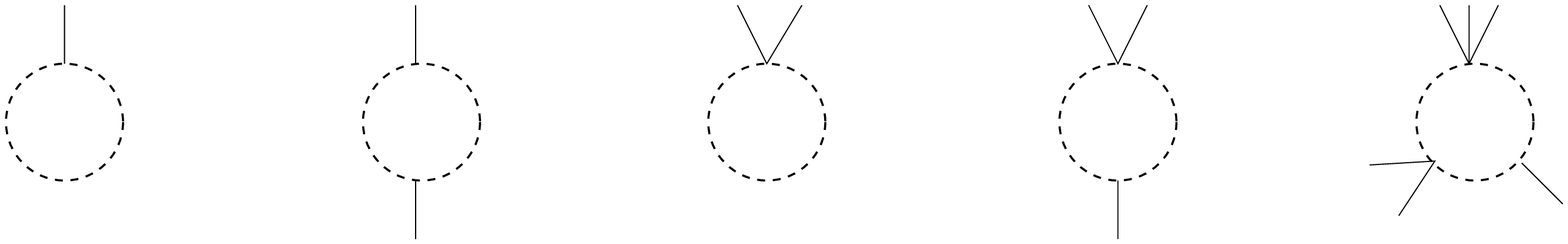,height=2.cm,width=14.cm,angle=0}}
\centerline{(b)} \caption{Graphs contributing to the blocking,
(a): tree-level, (b): one-loop corrections. The dashed line stands
for a particle of momentum $|p|=k$ and the solid lines represent
$\phi$.\label{whgrf}}
\end{figure}

One can understand the loop contribution better by
splitting the action into the sum of the quadratic part and the rest,
$S=\hf\phi\cdot G^{-1}_0\cdot\phi+S_i$ and expanding in $S_i$,
\be\label{morgwhe}
\tr\ln{\delta^2S_k[\phi]\over\delta\tilde\phi\delta\tilde\phi}=
\tr\ln G^{-1}_0-\sum_{n=1}^\infty{(-1)^n\over n}\tr\left(
{\delta^2S_k[\phi]\over\delta\tilde\phi\delta\tilde\phi}\cdot G_0\right)^n.
\ee
We recovered the sum of one-loop graphs.
The loop corrections close on all possible pair of $\tilde\phi$ legs with
the propagator $G_0$. The tree-level piece describes the feedback
of the change of the cut-off on the dynamics of the classical background field
$\phi$. This is classical contribution because the cut-off controls
the configuration space, the number of degrees of freedom.
Some of the graphs contributing to the right hand side
are shown in Fig. \ref{whgrf}.

The point of central importance is that $\dk$ serves here as a new
small parameter to suppress the higher-loop contributions. As an
example consider the simplest case, $\phi_x=\Phi$, $\tilde\phi_{cl}=0$
with the ansatz
\be\label{grexpa}
S_k[\phi]=\int_x\left[\hf Z_k(\phi_x)(\partial_\mu\phi_x)^2+U_k(\phi_x)\right],
\ee
for the action with $Z(\phi)=1$,
\be\label{spbl}
U_{k-\dk}(\Phi)=U_k(\Phi)+{\hbar\over2}\int_{k-\dk<|p|<k}
\ln[p^2+U''_k(\Phi)]+\ord{(\hbar\dk)^2}.
\ee
Each loop-integral is over the shell $k-\dk<|p|<k$
in momentum space. So long the propagator is non-singular
in the integration domain the n-loop integrals will
be proportional to the $n$-th power of the integration volume, giving
a dimensionless small suppression parameter $\approx(\dk/k)^n$,the proportion
of the modes eliminated with those left. The question of the singularity
will be considered in section \ref{singli}. The higher loop contributions
are suppressed in the infinitesimal blocking step limit and the
one-loop evolution equation turns out to be an exact functional
equation. The limit $\dk\to0$ is safe for the loop contributions
but more care is needed when the saddle point is non-trivial because the
tree-level contribution might be singular in this limit.

Notice that the convergence of the loop expansion was assumed in the
argument above. Another word of caution is in order about
Eq. \eq{pbl}. Not only the trace but the logarithm function itself
is considered within the subspace
$\tilde\phi\in{\cal F}_k\backslash{\cal F}_{k-\dk}$, a rather formal
observation which becomes essential later.

It is sometimes  easy to arrive at exact, but unmanageable
equations. The problem is rather to find an approximation, an
acceptable compromise between precision and simplicity. By
assuming the absence of non-local interactions and the homogeneity
of the vacuum the blocked action is usually expanded in the
gradient of the field, leading to the ansatz \eq{grexpa} for the
action where contributions $\ord{\partial^4}$ are neglected. This
step is the key to the success of the RG method. The point is that
the blocked action can be identified by evaluating the integral
\eq{obl} for different background configurations. The
determination of the blocked action is an over-constrained problem
because there are 'more' background field configurations than
coupling constants in a given ansatz. {\em The tacit assumption,
that the effective action is well defined, i.e., independent of
the set of configurations used to read off its coupling constants,
gives the real power of the RG method: it enables us to make
predictions}.

\subsubsection{Local potential approximation}
In order to relate the WH equation to conventional perturbation
expansion we shall use the ansatz \eq{grexpa} with $Z=1$, in the local
potential approximation. In order to determine the only non-trivial piece
of this action the potential $U_k(\phi)$
it is sufficient to consider a homogeneous field $\phi(x)=\Phi$
in \eq{pbl}. The saddle point is vanishing as mentioned above,
$\tilde\phi_{cl}=0$, so long as the external space-time symmetries
are unbroken. The blocking equation simplified in this manner yields Eq.
\eq{spbl}, the projection of the WH equation into the restricted space
of actions,
\be\label{wh}
\dot U_k(\Phi)=-{\hbar\Omega_dk^d\over2(2\pi)^d}\ln[k^2+U^{(2)}_k(\Phi)]
\ee
where the dot stands for $k\partial_k$ and $\Omega_d=2\pi^{d/2}/\Gamma(d/2)$
denotes the solid angle.

\begin{figure}
\vspace{14bp}
\centerline{\psfig{file=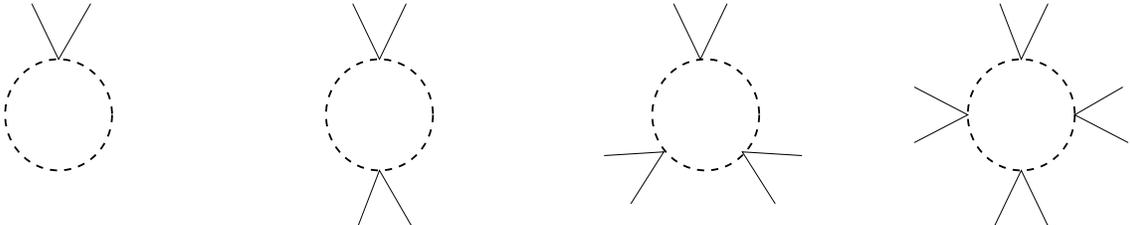,height=3cm,width=15cm,angle=0}}
\caption{The first four one-loop graphs contributing to the WH
equation in the local potential approximation when the potential
$U(\phi)$ is truncated to the terms $\phi^2$ and $\phi^4$. The
dashed line corresponds a particle of momentum
$|p|=k$.\label{effpotgr}}
\end{figure}

The expansion of the logarithmic function leads to the series
\be
\dot U_k(\Phi)=-{\hbar\Omega_dk^d\over2(2\pi)^d}\left\{
\ln[k^2+m^2]-\sum_{n=1}^\infty{(-1)^n\over n}\left(
{U^{(2)}_k(\Phi)-m^2\over k^2+m^2}\right)^n\right\},
\ee
where $m^2=U^{(2)}(0)$, the sum of the one-loop graphs whose
external legs have vanishing momentum and the internal lines
carry momentum $|p|=k$, cf Fig. \ref{effpotgr}, the leading order
contributions to the renormalization of the coupling constants $g_n$
in the loop expansion.

The partial resummation of the perturbation series performed by the WH
equation can easily be seen by comparing the solution of Eq. \eq{wh}
with those obtained in the independent mode approximation, where the
$k$-dependence is ignored in the right hand side of the equation,
\be\label{ukvef}
U_k(\Phi)=U_\lambda(\Phi)-{\hbar\over2}\int_{k<|p|<\Lambda}
\ln[p^2+U^{(2)}_\Lambda(\Phi)].
\ee
One recovers the one-loop effective potential at the IR fixed point,
$V_\mr{eff}(\Phi)=U_{k=0}(\Phi)$. This is not by accident and remains valid
for the complete solution, as well, because $e^{-VV_\mr{eff}(\Phi)}$
is the distribution of the homogeneous mode and the blocking \eq{obl}
at $k=0$ retains the homogeneous mode only, $S_{k=0}[\Phi]=VU_{k=0}(\Phi)$.
We shall see in section \ref{tree} that this simple argument is
valid in the absence of large amplitude inhomogeneous instabilities only.

Let us consider as an example the differential equation
\be
\dot x_k=f(x_k), \ \ x_\Lambda=x_B,
\ee
where $f(x)$ is weakly varying analytic function given in its expanded
form around the base point $x=0$,
\be
\dot x_k=\sum_n{f^{(n)}(0)\over n!}x_k^n.
\ee
The RG method allows us to perform the expansion at the current point,
\be
x_{k-\dk}=x_k-\dk f(x_k)+\ord{\dk^2},
\ee
cf Eq. \eq{elsowh}. The $k$-dependence of the right hand side
represents the accumulation of the information obtained during the
integration. In a similar manner, the virtue of the functional RG scheme
\eq{pbl} is the expansion of the path integral around around the
current action, instead of the Gaussian fixed point.

\begin{figure}
\centerline{\psfig{file=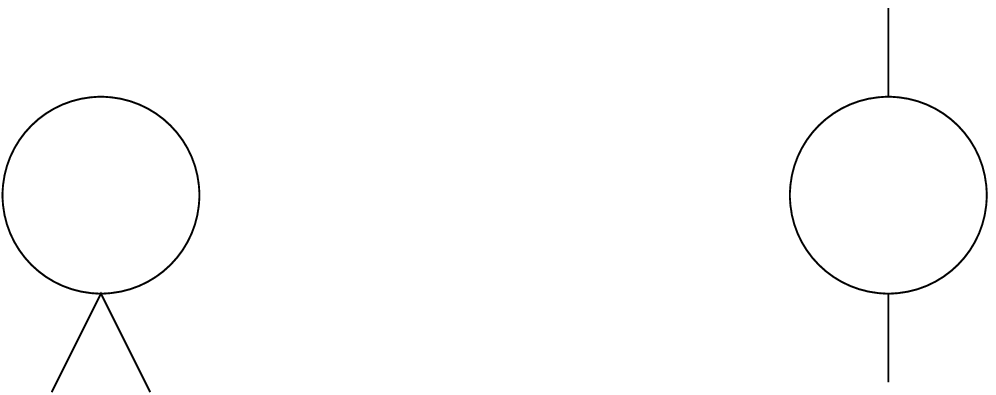,height=2cm,width=6.33cm,angle=0}}
\centerline{(a)}
\centerline{\psfig{file=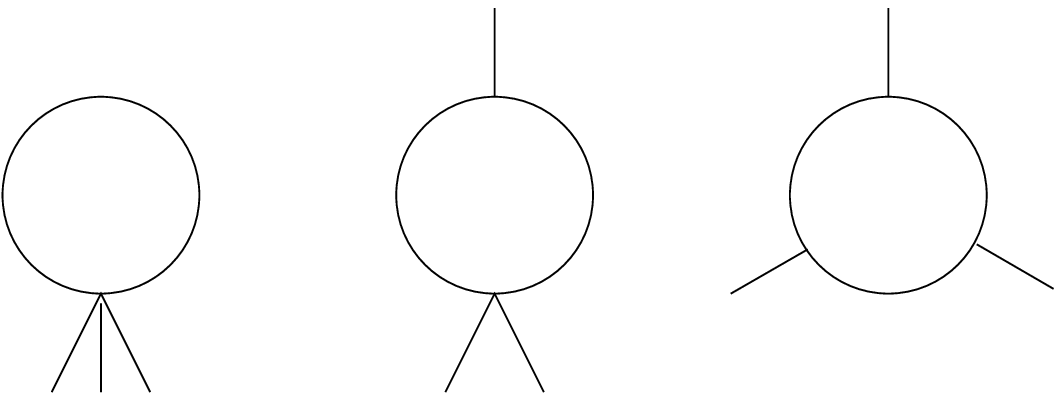,height=2cm,width=6.33cm,angle=0}}
\centerline{(b)}
\centerline{\psfig{file=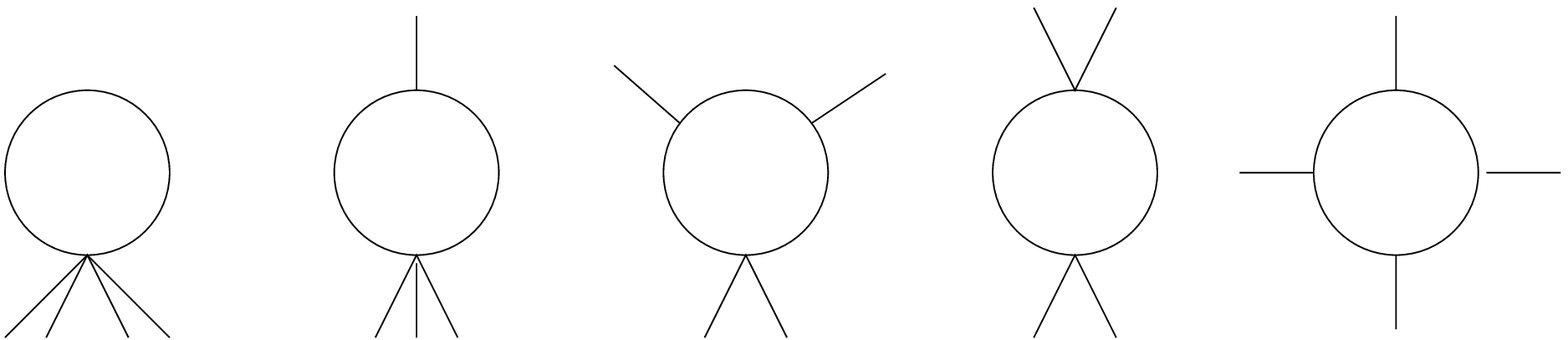,height=2cm,width=10cm,angle=0}}
\centerline{(c)}
\caption{Graphs contributing to $\beta_n$,
(a): $n=2$, (b): $n=3$ and (c): $n=4$.\label{betf}}
\end{figure}

A more detailed picture of the RG flow can be obtained by inspecting
the beta functions. For this end we parametrize the potential as
\be\label{locpot}
U_k(\phi)=\sum_n{g_n(k)\over n!}(\phi-\phi_0)^n,
\ee
and write the beta functions corresponding to $\phi_0=\la\phi\ra$ as
\be\label{befunct}
\beta_n=k\partial_k g_n(k)=k\partial_k\partial_\phi^nU_k(\phi_0)
=\partial_\phi^nk\partial_kU_k(\phi_0)
=-{\hbar\Omega_dk^d\over2(2\pi)^d}
\partial_\phi^n\ln[k^2+\partial_\phi^2U_k(\phi)]
\ee
It is easy to see that $\beta_n$ is an $n$-th order polynomial
\be
\beta_n=-{\hbar\Omega_dk^d\over2(2\pi)^d}{\cal P}_n
[G_3,\cdots,G_{n+2}],
\ee
of the expression $G_n=g_n/(k^2+g_2)$, eg
\bea\label{betapol}
{\cal P}_2&=&G_4-G_3^2,\nonu
{\cal P}_3&=&G_5-3G_3G_4+2G_3^3,\nonu
{\cal P}_4&=&G_6-4G_3G_5+12G_3^2G_4-3G_4^2-6G_3^4,\\
{\cal P}_5&=&G_7-5G_3G_6+20G_3^2G_5-10G_4G_5-60G_3^3G_4
+30G_3G_4^2+24G_3^5,\nonu
{\cal P}_6&=&G_8-6G_3G_7+30G_3^2G_6-5G_5G_6-120G_3^3G_5+120G_3G_4G_5\nonu
&&-270G_3^2G_4^2-10G_5^2-10G_4G_6+360G_3^4G_4+30G_4^3
-120G_3^6,\nonumber
\eea
etc. The correspondence between these expressions
and Feynman graphs contributing to the beta functions
in the traditional renormalization scheme is indicated in Fig. \ref{betf}.

We shall need later the beta functions corresponding to coupling constants
made dimensionless by the running cut-off. By means of the parametrization
$\phi=k^{(d-2)/2}\tilde\phi$, $g_n=k^{[g_n]}\tilde g_n$,
\be\label{gdim}
[g_n]=d+n\left(1-{d\over2}\right),
\ee
$U(\phi)=k^d\tilde U(\tilde\phi)$ one finds
\be\label{dlbefunct}
\tilde\beta_n(\tilde\phi)=\dot{\tilde g}_n(k)=
k^{-[g_n]}\beta_n(\phi)-[g_n]\tilde g_n
=-{\hbar\Omega_d\over2(2\pi)^d}{\cal P}_n
[\tilde G_3,\cdots,\tilde G_{n+2}]-[g_n]\tilde g_n,
\ee
where $\tilde G_n=\tilde g_n/(1+\tilde g_2)$.

An instructive way to read Eqs. \eq{locpot} and \eq{wh} is that the
potential $U(\phi)$ is the generator function for the coupling constants.
The functional RG method is an economical book-keeping method
for the computation of graphs and their symmetry factors
by keeping track of their generator function(al).

\begin{figure}
\centerline{\psfig{file=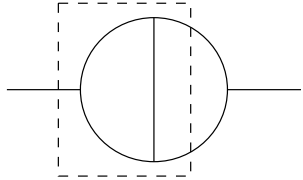,height=2.4cm,width=4.cm,angle=0}}
\caption{A two-loop contribution to $g_2$.\label{twlf}}
\end{figure}

The higher loop contributions to the running coupling constants are generated
by the integration of the differential equations $k\partial_kg_n=\beta_n$.
The us consider the two-loop graph depicted in Fig. \ref{twlf} and
expand the subgraph in the square in the momenta of its external
legs. The zeroth order contributions is the last graph of Fig. \ref{betf}b.
By going higher order in the gradient expansion one can, in principle,
generate the full momentum dependence of the subgraph. In general,
the $n$-th loop contributions appear in the integration after $n$
step $k\to k+\dk$ and all loops are resummed in the limit $\dk\to0$.

\subsubsection{Gradient expansion}\label{whap}
The natural expansion for the long distance properties
of systems with homogeneous ground state is the expansion in the
inverse characteristic length of the fluctuations, in the gradient
operator acting on the field variable.

\underline{Functional derivatives of local functionals:}
We consider the following generalization of the $\ord{\partial^2}$ ansatz
\eq{grexpa},
\be\label{gradact}
S[\phi]=\int_x\left[\hf Z(\phi_x)\partial_\mu\phi_xK^{-1}(-\Box)
\partial_\mu\phi_x+U(\phi_x)\right].
\ee We shall use the notation $Z_x=Z(\phi_x)$, $U_x=U(\phi_x)$ for
the coefficient functions and
$Z^{(n)}_x=\partial_\phi^nZ(\phi_x)$,
$U^{(n)}_x=\partial_\phi^nU(\phi_x)$ for their derivatives. The
first two functional derivatives of the action are
\bea\label{elso}
V{\delta
S\over\delta\phi_p}&=&\sum_ye^{ipy}{\delta
S\over\delta\phi_y}\nonu
&=&\int_{x,y}e^{ipy}\Biggl[\delta_{x,y}U^{(1)}_x
+\hf\delta_{x,y}Z^{(1)}_x\partial_\mu\phi_xK^{-1}\partial_\mu\phi_x \\
& &
+\hf Z_x\left(\partial_\mu\delta_{x,y}K^{-1}\partial_\mu\phi_x
+\partial_\mu\phi_xK^{-1}\partial_\mu\delta_{x,y}\right)\Biggr], \nonumber
\eea and
\bea V^2{\delta^2S\over\delta\phi_p\delta\phi_q}
&=&\int_{x,y,z}e^{ipy+iqz}\Biggl\{\hf\delta_{x,y}\delta_{x,z}Z^{(2)}_x
\partial_\mu\phi_xK^{-1}\partial_\mu\phi_x \nonu
&& +\hf Z_x\left(
 \partial_\mu\delta_{x,y}K^{-1}\partial_\mu\delta_{x,z}
+\partial_\mu\delta_{x,z}K^{-1}\partial_\mu\delta_{x,y}\right)\nonu
&&+\hf Z^{(1)}_x\biggl[\delta_{x,y}
\left(\partial_\mu\delta_{x,z}K^{-1}\partial_\mu\phi_x
+\partial_\mu\phi_xK^{-1}\partial_\mu\delta_{x,z}\right) \nonu
&& \hbox{\hfill{}} +\delta_{x,z}\left(\partial_\mu\delta_{x,y}K^{-1}
\partial_\mu\phi_x+\partial_\mu\phi_xK^{-1}\partial_\mu\delta_{x,y}\right)
\biggr]\nonu
&&+\delta_{x,y}\delta_{x,z}U^{(2)}_x\Biggr\}.
\eea
We split the field into a homogeneous background and fluctuations,
$\phi_x=\Phi+\eta_x$ and find up to $\ord{\eta^2}$,
\bea\label{elsoi}
V{\delta S\over\delta\phi_p}&=&\int_{x,y}e^{ipy}\Biggl\{\delta_{x,y}
\left(U^{(1)}+U^{(2)}\eta_x+\hf U^{(3)}\eta_x^2\right)
+\hf\Biggl[\delta_{x,y}Z^{(1)}\partial_\mu\eta_xK^{-1}\partial_\mu\eta_x\nonu
&&+(Z +Z^{(1)}\eta_x)\left(\partial_\mu\delta_{x,y}K^{-1}\partial_\mu\eta_x
+\partial_\mu\eta_xK^{-1}\partial_\mu\delta_{x,y}\right)\Biggr]\Biggr\}
\nonu
&=&\int_{x,y}e^{ipy}\delta_{x,y}\Biggl\{
\left(U^{(1)}+U^{(2)}\int_re^{ir\cdot x}\eta_r+\hf U^{(3)}\int_{r,s}
e^{i(r+s)\cdot x}\eta_r\eta_s\right)\nonu
&&-\hf\Biggl[Z^{(1)}\int_{r,s}e^{i(r+s)\cdot x}
r\cdot s\eta_rK^{-1}_s\eta_s \nonu
&& +\left(Z+Z^{(1)}\int_re^{ir\cdot x}\eta_r\right)
\int_sp\cdot se^{is\cdot x}\eta_s\left(K^{-1}_s+K^{-1}_p\right)\Biggr]\Biggr\}
\nonu
&=&\delta_{p,0}U^{(1)}+U^{(2)}\eta_{-p}
+\hf p^2Z(K^{-1}_{-p}+K^{-1}_p)\eta_{-p}
\nonu
&&+\hf\int_{r,s}\delta_{p+r+s,0}\eta_r\eta_s\left[U^{(3)}
-Z^{(1)}\left(r\cdot sK^{-1}_s+p\cdot rK^{-1}_r+p\cdot sK^{-1}_p\right)\right]
\eea
and
\bea\label{masodik}
V^2{\delta^2S\over\delta\phi_p\delta\phi_q}
&=&\int_{x,y,z}e^{ipy+iqz}\Biggl\{\delta_{x,y}\delta_{x,z}
\left(U^{(2)}+U^{(3)}\eta_x+\hf U^{(4)}\eta^2_x\right) \nonu
&&+\hf\Biggl[\delta_{x,y}\delta_{x,z}Z^{(2)}
\partial_\mu\eta_xK^{-1}\partial_\mu\eta_x\nonu
&&+\delta_{x,y}(Z^{(1)}+Z^{(2)}\eta_x)
\left(\partial_\mu\delta_{x,z}K^{-1}\partial_\mu\eta_x
+\partial_\mu\eta_xK^{-1}\partial_\mu\delta_{x,z}\right)\nonu
&&+\delta_{x,z}(Z^{(1)}+Z^{(2)}\eta_x)\left(
\partial_\mu\delta_{x,y}K^{-1}\partial_\mu\eta_x
+\partial_\mu\eta_xK^{-1}\partial_\mu\delta_{x,y}\right)\nonu
&&+\left(Z+Z^{(1)}\eta_x+\hf Z^{(2)}\eta^2_x\right)\left(
 \partial_\mu\delta_{x,z}K^{-1}\partial_\mu\delta_{x,y}
+\partial_\mu\delta_{x,y}K^{-1}\partial_\mu\delta_{x,z}\right)
\Biggr]\Biggr\}\nonu
&=&\int_{x,y,z}e^{ipy+iqz}\delta_{x,y}\delta_{x,z}\Biggl\{-\hf\Biggl[Z^{(2)}
\int_{r,s}r\cdot se^{irx}\eta_rK^{-1}_se^{isx}\eta_s\nonu
&&+\left(Z^{(1)}+Z^{(2)}\int_re^{irx}\eta_r\right)\int_se^{isx}\eta_s\left(
q\cdot s(K^{-1}_s+K^{-1}_q)+p\cdot s(K^{-1}_s+K^{-1}_p)\right)\nonu
&&+p\cdot q\left(Z+Z^{(1)}\int_re^{irx}\eta_r
+\hf Z^{(2)}\int_{r,s}e^{i(r+s)x}\eta_r\eta_s\right)
(K^{-1}_p+K^{-1}_q)\Biggr]\nonu
&&+\left(U^{(2)}+U^{(3)}\int_re^{irx}\eta_r
+\hf U^{(4)}\int_{r,s}e^{i(r+s)x}\eta_r\eta_s\right)\Biggr\}\nonu
&=&\delta_{p+q,0}\left[U^{(2)}-\hf Zp\cdot q(K^{-1}_p+K^{-1}_q)\right]\nonu
&&+\int_r\delta_{p+q+r,0}\eta_r\biggl[U^{(3)}
-\hf Z^{(1)}\biggl(p\cdot q(K^{-1}_p+K^{-1}_q) \nonu
&& +q\cdot r(K^{-1}_r+K^{-1}_q)+p\cdot r(K^{-1}_r+K^{-1}_p)\biggr)\biggr]\nonu
&&+\hf\int_{r,s}\delta_{p+q+r+s,0}\eta_r\eta_s\biggl[U^{(4)}
-Z^{(2)}\biggl(\hf p\cdot q(K^{-1}_p+K^{-1}_q)+r\cdot sK^{-1}_s \nonu
&&+q\cdot s(K^{-1}_s+K^{-1}_q)+p\cdot r(K^{-1}_p+K^{-1}_r)\biggr)\biggr],
\eea
where $Z^{(n)}=Z^{(n)}(\Phi)$, $U^{(n)}=U^{(n)}(\Phi)$ and $K_s=K(s^2)$.

\underline{WH equation in $\ord{\eta^2}:$}
We set $K=1$ in \eq{gradact}, use sharp cut-off, i.e. all momentum
integral is for $k-\dk<|p|<k$,
\bea\label{masodikwh}
V^2{\delta^2S\over\delta\phi_p\delta\phi_q}
&=&\delta_{p+q,0}\left(U^{(2)}-Zp\cdot q\right)
+\int_r\delta_{p+q+r,0}\eta_r\Biggl[U^{(3)}
-Z^{(1)}(p\cdot q+q\cdot r+p\cdot r)\Biggr]\nonu
&&+\hf\int_{r,s}\delta_{p+q+r+s,0}\eta_r\eta_s\Biggl[
U^{(4)}
 -Z^{(2)}(p\cdot q+r\cdot s+q\cdot s+p\cdot s+q\cdot r+p\cdot r)
\Biggr]\nonu
&=&G^{-1}_{p,q}+A_{p,q}+B_{p,q},
\eea
with
\bea
G^{-1}_{p,q}&=&\delta_{p+q,0}\left[U^{(2)}+\hf Z(p^2+q^2)\right],\nonu
A_{p,q}&=&\int_r\delta_{p+q+r,0}\eta_r\Biggl[U^{(3)}
+\hf Z^{(1)}(p^2+q^2+r^2)\Biggr],\nonu
B_{p,q}&=&\hf\int_{r,s}\delta_{p+q+r+s,0}\eta_r\eta_s\Biggl[
U^{(4)}+\hf Z^{(2)}(p^2+q^2+r^2+s^2)\Biggr].
\eea

As emphasised above, the logarithmic function
on the right hand side of the WH equation is computed within the
space of field configurations whose Fourier transform is
non-vanishing for $k-\dk<|p|<k$ only. In order to keep this
constrain in an explicit manner we introduce the projection operator
${\cal P}$ corresponding to this space and write
the rest of the $\ord{\eta^2}$ evolution equation as
\bea
\partial_kS[\phi]&=&-{\hbar\over2\dk}\tr\log[G^{-1}+A+B]\nonu
&=&-{\hbar\over2\dk}\tr{\cal P}\log G^{-1}
-{\hbar\over2\dk}\tr[{\cal P}G(A+B)]
+{\hbar\over4\dk}\tr[{\cal P}GA{\cal P}GA].
\eea
Since the propagator does not lead out from the function space,
$[{\cal P},G]=0$, it is sufficient to make the replacement
$G\to{\cal P}G$ on the right hand side of the evolution equation.
The term $\ord{\eta}$ drops owing to the vanishing of the
homogeneous component of the fluctuation, $\eta_{p=0}=0$ and we find
\bea
\hf\int_r\eta_{-r}\eta_r\left[\dot Zr^2+\dot U^{(2)}\right]+V\dot U
&=&-{\hbar k^d\over2(2\pi)^d}\int de\Biggl\{V\log[Zk^2+U^{(2)}] \nonu
&&-\int_r\eta_{-r}\eta_r{\cal P}_{r-ek,ek-r}
{\biggl[U^{(3)}+\hf Z^{(1)}(k^2+(ek-r)^2+r^2)\biggr]^2
\over2(Zk^2+U^{(2)})(Z(ek-r)^2+U^{(2)})}\nonu
&&+\int_r\eta_{-r}\eta_r{U^{(4)}+Z^{(2)}(k^2+r^2)\over2(Zk^2+U^{(2)})}\Biggr\},
\eea
including an integration over the direction of $k$, the unit vector $e$.

The identification of the terms $\ord{\eta^0r^0}$ results the equations
\be\label{whuz}
\dot U=-{\hbar k^d\Omega_d\over2(2\pi)^d}\log[Zk^2+U^{(2)}],
\ee
c.f. Eq. \eq{wh}. The terms $\ord{\eta^2r^2}$ give
\bea\label{whzz}
\dot Z\int_r\eta_{-r}\eta_rr^2&=&-{\hbar k^d\over2(2\pi)^d}
\int_r{\eta_{-r}\eta_r\over Zk^2+U^{(2)}}\int de
\Biggl\{Z^{(2)}r^2 \nonu
&&-{\cal P}_{r-ek,ek-r}\biggl[2r^2Z^{(1)}{U^{(3)}+Z^{(1)}k^2\over Zk^2+U^{(2)}}
-Zr^2{(U^{(3)}+Z^{(1)}k^2)^2\over(Zk^2+U^{(2)})^2}\nonu
&&+k^2(er)^2\left({Z^{(1)2}\over Zk^2+U^{(2)}}
-ZZ^{(1)}{U^{(3)}+Z^{(1)}k^2\over(Zk^2+U^{(2)})^2}\right)\biggr]\Biggr\}.
\eea
Finally, in $\ord{\eta^2r^0}$ we find
\bea\label{whudd}
\dot U^{(2)}\int_r\eta_{-r}\eta_r
&=&-{\hbar k^d\over2(2\pi)^d}\int_r{\eta_{-r}\eta_r\over Zk^2+U^{(2)}}\int de
\biggl[Z^{(2)}k^2 \nonu
&&+U^{(4)}-{\cal P}_{r-ek,ek-r}
{(Z^{(1)}k^2+U^{(3)})^2\over Zk^2+U^{(2)}}\biggr].
\eea

Let us make the replacement ${\cal P}_{r-k,k-r}\to1+[{\cal P}_{r-k,k-r}-1]$
in Eqs. \eq{whuz}-\eq{whudd} and call the contributions corresponding
to $1$ and ${\cal P}_{k-r,k-r}-1$ regular and irregular, respectively.
The regular contribution can be obtained from the one-loop graphs. This
has already been seen for Eq. \eq{whuz} and the regular part of
Eq. \eq{whudd} is just the second derivative of Eq. \eq{whuz}.

But there are problems with the irregular contributions which represent
the cut-off in the multi-loop integrals. One obvious
problem is the inconsistency between Eqs. \eq{whuz} and \eq{whudd}.
Another problem is that the irregular contributions include a truncated
$r$-integration and as a result the left and right hand sides of the
equations have different $\eta$-dependence. This could be avoided
by imposing $r<<\dk$ for the momentum of the fluctuations. But the
price is unacceptable high since it would indicate that the radius of
convergence for the gradient expansion is smaller then the infinitesimal $\dk$.

Another problem the projection operator leads to is non-locality.
It is known that the bare action is non-local at the scale
of the U.V. cut-off, i.e. it contains higher order derivatives up
to a finite order. In fact, the gradient $\partial^n/\partial x^{\mu n}$
represents couplings up to $n$ lattice spacing distances when
lattice regularization is used. But terms in the action which are
non-polynomial in the gradient induce correlations at finite, i.e.
cut-off independent distances and are not acceptable.
Returning to the gradient expansion we note
that the operator ${\cal P}_{r-k,k-r}$ restricts the integration
domain for $r$ in such a non-isotropical manner that not only
$r^2$ but $|r|$ appears, too. This latter corresponds
to the non-local operator $\sqrt{\Box}$ in the gradient expansion.

This problem, inherent in the sharp cut-off
procedure in the momentum space is rather general. The one-loop
correction for the two-particle scattering amplitude in the $\phi^4$
model, represented by the fourth graph in Fig. \ref{betf}c
has a non-local contribution when the momenta in
both internal lines are properly restricted by the U.V. cut-off.
Such non-local effects, disregarded in the usual textbooks,
might well be negligible in a renormalized theory because they
are represented by non-renormalizable, irrelevant operators
such as $\phi^3\sqrt{\Box}\phi$. Therefore one hopes that their
effects become weak when the cut-off is removed.
The more careful analysis is rather involved
since the details of the cut-off procedure within the momentum
regime $(1-\epsilon)k<p<(1+\epsilon)k$ influences the dynamics
on the distance scale $\hbar/(2\epsilon k)$ which is a quantity of the
form $0\cdot\infty$.

But it is not so easy to dispel the doubts.
First, the irrelevance and unimportance of a coupling constant
are quite different concepts. An irrelevant coupling constant
approaches its I.R. fixed point value as we move in the I.R.
direction. But its value at the fixed point may be strong,
indicating that the coupling constant in question is important.
It is the scale dependence only what becomes unimportant for an
irrelevant coupling constant and not its presence. The higher
order vertices induced at low energies in a strongly coupled self
interacting scalar field theory in $d<4$ may serve as an example of this
difference. Second, the global view of the
renormalization group, outlined in Section \ref{grgs}
suggest that a coupling constant which is irrelevant at the
U.V. fixed point does not necessarily remain so at around
the I.R. fixed point. Finally, this problem is obviously
present in effective theories where the cut-off reaches
physical scales.

\subsection{Resumming the perturbation expansion}\label{resperts}
The basic idea leading to the WH equation is the successive elimination
of the degrees of freedom. This procedure produces the blocked
action, the integrand of the functional integral of the blocked theory.
This integrand must be well defined, ie each mode must either be left intact
or be completely eliminated during the blocking. This is the positive side
of the sharp cut-off in momentum space. The negative side is that it
generates non-local interactions and spoils the gradient expansion.
The simplest cure, to smear out the regulator and to use smooth cut-off
procedure, is not a valid alternative. This is because the blocking based on
smooth cut-off suppresses the modes partially and the integrand
for the functional integral of the blocked theory is ill defined.
Despite this general sounding argument one can proceed and generalize
the successive elimination process for smooth cut-off
in a rather surprising manner.

\subsubsection{Polchinski equation}
Let us start with the partition function
\be\label{zor}
Z_k=\int D[\Phi]e^{-{1\over2\hbar}\Phi G^{-1}_k\Phi-\ih S^I_k[\Phi]},
\ee
where the interaction functional $S^I_k[\Phi]$ corresponds to the U.V. cut-off
$k$. We split the propagator and the field variable into IR and U.V. components,
\be\label{split}
G_k=G_{k-\dk}+\tilde G_k,\ \ \Phi=\phi+\tilde\phi,
\ee
with the intention that the fields $\phi$ and $\tilde\phi$
should propagate with $G_{k-\dk}$ and $\tilde G_k$, respectively.
In other words, the kinetic energy contribution to the action
is supposed to be of the form
$(\phi G^{-1}_{k-\dk}\phi+\hf\tilde\phi\tilde G^{-1}_k\tilde\phi)/2$
when written in terms of $\phi$ and $\tilde\phi$.

The introduction of the fields $\phi$ and $\tilde\phi$ separated by a
smooth cut-off makes both $\phi_p$ and $\tilde\phi_p$ non-vanishing and
appears as a double counting of the degrees of freedom.
In order to ensure the proper integration measure we introduce a
dummy field, $\tilde\Phi$, in such a manner that a suitable, momentum
dependent linear combination of $\phi_p$ and $\tilde\phi_p$ reproduces
$\Phi_p$ and $\tilde\Phi_p$,
\be\label{distr}
\pmatrix{\Phi\cr\tilde\Phi}=\pmatrix{A_{1,1}(\Box)&A_{1,2}(\Box)\cr
A_{2,1}(\Box)&A_{2,2}(\Box)}
\pmatrix{\phi\cr\tilde\phi}
\ee
and decouple $\tilde\Phi$,
\be\label{decoupl}
\hf\phi G^{-1}_{k-\dk}\phi+\hf\tilde\phi\tilde G^{-1}_k\tilde\phi=
\hf\Phi G^{-1}_k\Phi+\hf\tilde\Phi\tilde G^{-1}_D\tilde\Phi,
\ee
with a freely chosen dummy propagator $\tilde G_D$.
Owing to the second equation in \eq{split} $A_{1,1}=A_{1,2}=1$.
The condition \eq{decoupl} gives
\be\label{matr}
\pmatrix{G^{-1}_{k-\dk}&0\cr0&\tilde G^{-1}_k}
=\pmatrix{1&A_{2,1}(\Box)\cr1&A_{2,2}(\Box)}\pmatrix{G^{-1}_k&0\cr0&\tilde G^{-1}_D}
\pmatrix{1&1\cr A_{2,1}(\Box)&A_{2,2}(\Box)},
\ee
whose solution is
$A_{2,1}(\Box)=\sqrt{\tilde G_D\tilde G_k/G_{k-\dk}G_k}$,
$A_{2,2}(\Box)=\sqrt{\tilde G_DG_{k-\dk}/\tilde  G_kG_k}$.
The transformation \eq{distr} is non-singular so long as
$\tilde G_k\not=G_{k-\Delta k}$, the propagations of $\phi$ and $\tilde\phi$
are distinguishable.

We are finally in the position to describe the blocking procedure
which consists of the following steps:
\begin{enumerate}
\item Redouble the number of degrees of freedom,
$\Phi\to(\Phi,\tilde\Phi)$.
\item Rotate the new, dummy degrees of freedom
into the dynamics in such manner that the rotated fields split the original
field variable and follow the prescribed propagation.
\item Integrate out $\tilde\phi$.
\end{enumerate}

In order to complete step 3. we insert a field independent
constant into the partition function,
\be\label{ccint}
Z_k=\int D[\Phi]D[\tilde\Phi]e^{-{1\over2\hbar}\tilde\Phi\tilde G^{-1}_D
\tilde\Phi-{1\over2\hbar}\Phi G^{-1}_k\Phi-\ih S^I_k[\Phi]}.
\ee
The redistribution of the degrees of freedom by the inverse of
the transformation \eq{distr} yields
\be
Z_k=\int D[\phi]D[\tilde\phi]
e^{-{1\over2\hbar}\phi G^{-1}_{k-\dk}\phi
-{1\over2\hbar}\tilde\phi\tilde G^{-1}_k\tilde\phi
-\ih S_k^I[\phi+\tilde\phi]},
\ee
up to a constant. This motivates the introduction of the blocked action
$S_{k-\dk}[\phi]$ defined as
\be\label{sblock}
e^{-S_{k-\dk}[\phi]}=\int D[\tilde\phi]
e^{-{1\over2\hbar}\tilde\phi\tilde G^{-1}_k\tilde\phi
-\ih S_k^I[\phi+\tilde\phi]}.
\ee

The higher loop corrections were suppressed in the WH equation
by restricting the functional space of the blocking,
the volume of the loop integration in momentum space. But there is
another way to suppress the radiative corrections, to decrease the
propagator. Such a suppression method is better suited for smooth cut-off
where the U.V. field can not be constrained into a restricted,
'small' functional space. This strategy is implemented by
requiring that in the so far arbitrary split \eq{split}
$\tilde G_k(p)=\ord{\dk}$. The obvious choice is
\be\label{jump}
\tilde G_k(p)=\dk\partial_kG_k(p).
\ee
By assuming the convergence of the perturbation expansion
we expand the action in $\tilde\phi$ and perform the integration
over $\tilde\phi$. Since the new small parameter is the propagator
we retain graphs where the propagator appears in a minimal number.
By assuming that the saddle point is at most $\ord{\dk}$ we can
expand the action in $\tilde\phi$,
\bea\label{lordeff}
e^{-\ih S_{k-\dk}^I[\phi]}
&=&\int D[\tilde\phi]
e^{-{1\over2\hbar}\tilde\phi\tilde G^{-1}_k\tilde\phi
-\ih S_k^I[\phi]-\ih\tilde\phi\cdot{\delta S_k^I[\phi]\over\delta\phi}
-{1\over2\hbar}\tilde\phi\cdot{\delta^2 S_k^I[\phi]\over\delta\phi\delta\phi}
\cdot\tilde\phi+\ord{\tilde\phi^3}}\nonu
&=&e^{-\ih S_k^I[\phi]+{1\over2\hbar}{\delta S_k^I[\phi]\over\delta\phi}\cdot
\left(\tilde G^{-1}_k+
{\delta^2 S_k^I[\phi]\over\delta\phi\delta\phi}\right)^{-1}\cdot
{\delta S_k^I[\phi]\over\delta\phi}-\hf\tr\log\left(\tilde G^{-1}_k+
{\delta^2 S_k^I[\phi]\over\delta\phi\delta\phi}\right)+\ord{\dk^2}}\nonu
&=&e^{-\ih S_k^I[\phi]+{\dk\over\hbar}
\left[\hf{\delta S_k^I[\phi]\over\delta\phi}\cdot
\partial_kG_k\cdot{\delta S_k^I[\phi]\over\delta\phi}-{\hbar\over2}\tr\left(
\partial_kG_k{\delta^2 S_k^I[\phi]\over\delta\phi\delta\phi}\right)\right]
+\ord{\dk^2}}\nonu
&=&\left[1-{\hbar\dk\over2}{\delta\over\delta\phi}\cdot
\partial_kG_k\cdot{\delta\over\delta\phi}\right]
e^{-\ih S_k^I[\phi]}+\ord{\dk^2}.
\eea
The third equation was obtained by expanding the logarithmic function
and gives the differential equation \cite{polch}, \cite{morrp}
\be\label{polch}
\partial_kS_k^I[\phi]=\hf{\delta S_k^I[\phi]\over\delta\phi}\cdot
\partial_kG_k\cdot{\delta S_k^I[\phi]\over\delta\phi}-{\hbar\over2}\tr\left[
\partial_kG_k{\delta^2 S_k^I[\phi]\over\delta\phi\delta\phi}\right].
\ee
The fourth line leads to a linear equation,
\be\label{linpol}
k\partial_ke^{-\ih S^I_k[\phi]}=\hbar{\cal B}e^{-\ih S_k^I[\phi]},\ \
{\cal B}=\hf{\delta\over\delta\phi}\cdot
k\partial_kG_k\cdot{\delta\over\delta\phi}.
\ee

It is instructive to notice the similarities and differences with
the WH equation. Eq. \eq{sblock} looks as the starting point for
the derivation of the WH equation and could have been obtained in
a trivial manner for sharp cut-off. The crux of the argument
leading to this relation is that this simple-looking equation is
actually valid for smooth cut-off. But in a surprising manner it
is just the sharp cut-off limit when the rest of the derivation of
Polchinksi equation is invalid. This is because $\tilde G_k$ as
given in \eq{jump} is $\dk\ord{\dk^{-1}}=\ord{\dk^0}$.

Eq. \eq{polch} is a leading order equation in the perturbation rather
than the loop expansion. It is formally similar to the WH equation, \eq{morgwh}
when only the leading order is retained in the expansion \eq{morgwhe}
which corresponds to keeping graphs with a single vertex only in Fig.
\ref{whgrf}(b) and using the dashed lines to denote the propagation
of $\tilde\phi$. Notice that even if the higher order contributions of
the perturbation expansion are suppressed in the limit $\dk\to0$, the
convergence of the perturbation expansion was assumed.

The difference between WH equation and \eq{polch} is that the
tree-level first term on the right hand side is always non-vanishing
in Eq. \eq{polch}, contrary to the case of the WH equation.
It is easier to understand the reason of the non-trivial
saddle point when the blocking \eq{sourcebl} is considered
in the presence of the source term. The source can be
omitted and the cut-off independence of a single quantity, the partition
function can be considered as the basic equation when the source has
vanishing component in the functional subspace to integrate over in
the blocking.
In case of a smooth cut-off the background field has arbitrary
momentum component and we can not really omit the source. One ought to
follow the more cumbersome blocking which keeps the generator functional
RG invariant. When classical physics (action) is modified by changing the
cut-off, the field induced by the source changes, as well. This is the
origin of the tree-level renormalization. But we made a short-cut and
omitted the source. The price is that the tree-level renormalization
can only be understood by letting the background field, $\phi$ vary freely.

The understanding how the $\phi$-dependence may give rise the
tree-level renormalization in \eq{lordeff} is
an interesting demonstration of the well know theorem stating that the
expansion in loops and in $\hbar$ are equivalent. In fact, the first
equation in \eq{lordeff} introduces an effective theory for
$\tilde\phi$ and the first two terms in the exponent of the second
line give the effective action corresponding to a given background field
$\phi$ in the {\em classical approximation}. In other words,
the action with lowered cut-off which is supposed to reproduce the
action with the original cut-off for arbitrary configuration $\phi$
requires tree-level adjustment. The second term in the exponent
has two factors of $\hbar^{-1}$ coming from the 'source'
$\hbar^{-1}\delta S_k^I[\phi]/\delta\phi$ of the field $\tilde\phi$ and
a factor of $\hbar$ from the $\tilde\phi$ propagator. The overall
$1/\hbar$ disappears in the differential equation \eq{polch}
which displays $\hbar$ times the exponent. This is the way
the graphs of Fig. \ref{whgrf}(b) appear on the $\ord{\hbar^0}$
tree-level in Eq. \eq{polch}. Another reason is that they are
graphs with no freely changing loop-momentum variable.
One can understand in a similar manner that both the tree- and the
loop-level graphs appear with a common factor of $\hbar$ in Eq. \eq{linpol}.

\subsubsection{Gradient expansion}\label{polgre}
We shall use here the results obtained in section \ref{whap} with the ansatz
\be
S^I[\phi]=\int_x\left[\hf(Z(\phi_x)-1)\partial_\mu\phi_xK^{-1}(-\Box/k^2)
\partial_\mu\phi_x+U(\phi_x)\right]
\ee
to derive the gradient expansion for Eq. \eq{polch}. For the left hand side
we find in the given order of the gradient expansion
\bea
&&{\dot Z\over2}\int_p\eta_{-p}\eta_p{p^2\over K_p}
+{Z-1\over2}\int_p\eta_{-p}\eta_pp^2\dot K^{-1}_p+V\dot U
+{\dot U^{(2)}\over2}\int_p\eta_{-p}\eta_p  \nonu
&&\approx{\dot Z\over2}\int_p\eta_{-p}\eta_p{p^2\over K_p}+V\dot U
+{\dot U^{(2)}\over2}\int_p\eta_{-p}\eta_p.
\eea
By means of the relation $k\partial_kG=-2K'_p/k^2$ where $K'_p=dK(z)/dz$,
$z=p^2/k^2$ the right hand side gives
\bea
&&-\int_p\Biggl\{\delta_{p,0}U^{(1)}
+\eta_p\left(U^{(2)}+(Z-1){p^2\over K_p}\right)
+\hf\int_r\eta_r\eta_{p-r}\left[U^{(3)}
-Z^{(1)}\left({r\cdot(p-r)\over K_{p-r}} -{p\cdot r\over
K_r}-{p\cdot(p-r)\over K_p}\right)\right]\Biggr\}\nonu
&&\times{K'_p\over k^2}\Biggl\{\delta_{p,0}U^{(1)}
+\eta_{-p}\left(U^{(2)}+(Z-1){p^2\over K_p}\right)+
\hf\int_s\eta_{-p-s}\eta_s\left[U^{(3)}-Z^{(1)}\left({p\cdot
s\over K_p} -{(p+s)\cdot s\over K_s}-{p\cdot(p+s)\over
K_{-p-s}}\right)\right]\Biggr\}\nonu
&&+\hbar\int_p{K'_p\over k^2}\Biggl\{V\left(U^{(2)}+(Z-1){p^2\over
K_p}\right) +\hf\int_r\eta_{-r}\eta_r\left[U^{(4)}
+Z^{(2)}\left({p^2\over K_p}+{r^2\over
K_r}\right)\right]\Biggr\}\nonu
&&\approx-VU^{(1)2}{K'_0\over k^2}-\int_r\eta_{-r}\eta_r
\biggl[{K'_0+K''_0K_0r^2/(k^2K_r)\over k^2}\left(U^{(2)}
+(Z-1){r^2\over K_r}\right)^2 +{K'_0\over
k^2}U^{(1)}\left(U^{(3)}+Z^{(1)}{r^2\over K_r}\right)\biggr]\nonu
&&+\hbar\int_p{K'_p\over k^2}\Biggl\{V\left(U^{(2)}+(Z-1){p^2\over
K_p}\right) +\hf\int_r\eta_{-r}\eta_r\left[U^{(4)}
+Z^{(2)}\left({p^2\over K_p}+{r^2\over K_r}\right)\right]\Biggr\}
\eea
This equation is consistent since the terms $\ord{\eta^2}$
are comparable thanks to the smoothness of the cut-off. We
compare the coefficients of the same $\eta$-dependent expressions
and find
\bea\label{polgrexp} k^2\dot U&=&-U^{(1)2}K'_0
+\hbar\int_pK'_p\left(U^{(2)}+(Z-1){p^2\over K_p}\right),\nonu
k^2\dot Z&=&-4K'_0U^{(2)}(Z-1)-2K'_0U^{(1)}Z^{(1)}
-2{K''_0K_0\over k^2}U^{(2)2}+\hbar Z^{(2)}\int_pK'_p, \eea and
\be\label{pouket} k^2\dot
U^{(2)}=-2K'_0U^{(2)2}-2K'_0U^{(1)}U^{(3)}
+\hbar\int_pK'_p\left(U^{(4)}+Z^{(2)}{p^2\over K_p}\right) \ee We
have two independent equations for two functions because the Eq.
\eq{pouket} follows from the first equation of \eq{polgrexp}.

It is instructive to check the first equation in the local potential
approximation $Z=1$,
\be\label{polbeta}
\dot U={U^{(1)2}\over2}{k\partial_kK_p\over p^2}_{\vert p=0}
-{\hbar U^{(2)}\over2}\int_p{k\partial_kK_p\over p^2}.
\ee
The first term describes the tree-level adjustment of the bare action
in keeping the physics cut-off independent and some
of the corresponding graphs are depicted in Fig. \ref{whgrf}(a).
The present form, characteristic of the gradient expansion,
is non-vanishing when the change of the cut-off at $p=k$
is felt at the base point of the gradient expansion, $p=0$.
The second term is the leading order perturbative contribution to the
WH equation \eq{wh} when the mass term is treated perturbatively,
as expected and the beta functions,
\be
\beta_n(\phi)=-{\hbar U^{(n+2)}\over2}\int_p{k\partial_kK_p\over p^2},
\ee
for $n>2$ correspond to the leading order first column in Eqs. \eq{betapol}
with $g_2=0$.
The independent mode approximation reproduces the leading order
renormalization of the coupling constants and the numerical
integration resums the perturbation expansion in the approximation
where the effective vertices are replaced by their values at $p=0$.
The leading order perturbative contributions, retained
in the RG equation are linear in the coupling constants and the classical,
non-linear terms drop out in the linearized evolution around the
Gaussian fixed point therefore the usual critical exponents are reproduced.

\subsection{Composite operator renormalization}\label{compops}
The renormalization of composite operators \cite{collins}
seems to be a highly technical and subtle issue, dealing with
the removal of the U.V. divergences from Green functions where
composite operators are inserted. But it becomes more elementary
and general \cite{comp} as soon as we are ready to give up
the perturbative approach and can reformulate the procedure in
a general, non-perturbative manner.
In order to see this let us go back to the remark made at Eq.
\eq{sourcebl}. It is not enough to impose the RG invariance of the partition
function only since it yields a single equation for infinitely many
coupling constants. Instead one should require the RG invariance of the
generator functional for Green functions because any observable can be
reconstructed from this functional. This turns out to be rather
cumbersome due to the source dependence generated for the blocked action
\cite{source}. We took another direction by imposing the RG invariance of another functional,
the blocked action, cf Eqs. \eq{obl} and \eq{sblock}. Instead of this strategy
we shall now use the matching of observables computed at different values
of the cut-off to rederive the RG equation. This is a non-perturbative, blocking
inspired generalization of the multiplicative RG schemes mentioned in the
Introduction in what the scale dependence of the observables is followed
instead of those of non-physical bare coupling constants.

We start with a toy model to demonstrate the natural relation between blocking
and composite operator renormalization and the field theoretical
application follows next by generalizing the source term in the
generator functional. This term which usually involves the elementary
field only is extended here for any local operator with inhomogeneous
source and we consider it as part of the action which now contains inhomogeneous
coupling constants. Our main point is that the beta functions corresponding to
this extended action provide a bridge between the blocking and
the composite operator renormalization.

\subsubsection{Toy model}
Consider the two dimensional integral
\be
Z=\int dxdye^{-S(x,y)}
\ee
over $x$ and $y$ which play the role of low- and high-frequency variables,
respectively, where the bare action is given by the expression
\be
S(x,y)={1\over2}s_xx^2+{1\over2}s_yy^2 +\sum_{n=0}^\infty g_n (x+y)^n
\ee
in terms of bare coupling constants $g_n$. We use $\hbar=1$ in this section.
The blocked action is defined as
\be
e^{-S(x)}=\int dye^{-S(x,y)}.
\ee
The elementary, bare operators are $(x+y)^n$ with
$n=0,1,2,\ldots$ and their expectation values are
\be
\label{expv1}
{\int dxdy(x+y)^ne^{-S(x,y)}\over\int dxdye^{-S(x,y)}}
={\int dx{\partial S(x)\over\partial g_n}e^{-S(x)}\over\int dxe^{-S(x)}}.
\ee
The right hand side of this equation expresses the expectation
value of the bare operator in terms of an operator of the blocked,
'thinner' system.
The action $S(x)$ can be expanded in the terms of the operators $x^n$,
\be
S(x)=\hf s_xx^2+\sum_{m=0}^\infty g_m'x^m,
\ee
giving rise the composite operators
\be
\{ x^n\}={\partial S(x)\over\partial g_n}=\sum_{m=0}^\infty  x^mS_{m,n}
\ee
where
\be\label{mixmat}
S_{m,n}={\partial g_m '\over\partial g_n}.
\ee

We have chosen a basis for operators in the bare theory and searched for
operators in the blocked theory. The opposite question, the construction of
the composite operators $[(x+y)^n]$ of the bare theory which reproduce
the expectation values of the blocked operators gives
\be
{\int dxdy\lbrack(x+y)^n\rbrack e^{-S(x,y)}\over\int dxdye^{-S(x,y)}}
={\int dxx^ne^{-S(x)}\over\int dxe^{-S(x)}},
\ee
with
\be
\lbrack(x+y)^n\rbrack=\sum_{m=0}^\infty (x+y)^m(S^{-1} )_{m,n}.
\ee

It is instructive to generalize the toy model for three variables,
\be
S(x,y,z) ={1\over2}s_xx^2+{1\over2}s_yy^2+{1\over2}s_zz^2
+\sum_{n=0}^\infty g_n (x+y+z)^n,
\ee
with
\be
e^{-S(x,y)}=\int dze^{-S(x,y,z)},\ \ e^{-S(x)}=\int dye^{-S(x,y)},
\ee
where the operator mixing looks like
\bea\label{toymix}
\{x^n\}_z&=&{\partial S(x,y,z)\over\partial g_n}=(x+y+z)^n,\nonu
\{x^n\}_y&=&{\partial S(x,y)\over\partial g_n}=
-{{\partial\over\partial g_n}
\int dze^{-S(x,y,z)}\over\int dze^{-S(x,y,z)}},\nonu
\{x^n\}_x&=&{\partial S(x)\over\partial g_n}=
-{{\partial\over\partial g_n}\int dydze^{-S(x,y,z)}\over
\int dydze^{-S(x,y,z)}}.
\eea
These relations yield
\be\label{toyopm}
\{x^n\}_y={\int dz\{x^n\}_ze^{-S(x,y,z)}\over\int dze^{-S(x,y,z)}},\ \
\{x^n\}_x={\int dy\{x^n\}_ye^{-S(x,y)}\over\int dye^{-S(x,y)}},
\ee
indicating that the evolution of the operators comes from the elimination
of the field variable in their definition. We can compute
the expectation value of $\{x^n\}$ at any level,
\be\label{toyrginv}
{\int dxdydz\{x^n\}_ze^{-S(x,y,z)}\over\int dxdydze^{-S(x,y,z)}}
={\int dxdy\{x^n\}_ye^{-S(x,y)}\over\int dxdye^{-S(x,y)}}
={\int dx\{x^n\}_xe^{-S(x)}\over\int dxe^{-S(x)}}.
\ee

The lesson of this toy model is twofold. First, one sees that the formal definition
of the renormalized operator at two different value of the cut-off
differs because these operators are supposed to reproduce the
same averages by means of different number of degrees of
freedom. Second, it shows that the cut-off dependence of the
renormalized operators can be obtained in a natural and simple
manner by considering the bare coupling constants at one value of
the cut-off as the functions of the coupling constants given at another
value of the cut-off.

\subsubsection{Quantum Field Theory}
We generalize the operator mixing of the toy model for
the scalar field theory given by the bare action
\be\label{bareact}
S_\Lambda[\phi]=\int_x\sum_nG_{n,x}(\Lambda)O_n(\phi_x)
=\sum_{\tilde n}G_{\tilde n}(\Lambda)O_{\tilde n}(\phi_x),
\ee
where $O_n(\phi_x)$ represents a complete set of local operators
(functions of $\phi_x$ and its space-time derivatives) and $G_{n,x}(\Lambda)$
denotes the coupling constant. We simplify the expressions by introducing a
single index $\tilde n$ for the pair $(n,x)$ labeling the basis elements for
the local operators and $\sum_{\tilde n}=\sum_n\int_x$.
The decomposition $\phi_x=\phi_{k,x}+\tilde\phi_{k,x}$ of the scalar field
into a low- and high-frequency parts is carried out as in Eqs. \eq{obl}
and \eq{split}.

The Kadanoff-Wilson blocking
\be\label{blcompop}
e^{-S_k[\phi_k]}=\int D[\tilde\phi_k]e^{-S_\Lambda[\phi_k+\tilde\phi_k]}
\ee
for the action
\be\label{coact}
S_k[\phi_k]=\sum_{\tilde n}G_{\tilde n}(k)O_{\tilde n}(\phi_k)
\ee
generates the RG flow
\be\label{comprgfl}
k\partial_kG_{\tilde n}(k)=\beta_{\tilde n}(k,G).
\ee

The blocked operators are defined by
\be
\{O_{\tilde n}(\phi_k)\}_k=
{\delta S_k[\phi_k]\over\delta G_{\tilde n}(\Lambda)}
={\int D[\tilde\phi_k]O_{\tilde n}(\phi_k+\tilde\phi_k)
e^{-S_\Lambda[\phi_k+\tilde\phi_k]}\over
\int D[\tilde\phi_k]e^{-S_\Lambda[\phi_k+\tilde\phi_k]}}
\ee
in agreement with Eq. \eq{toymix}. They satisfy the equation
\be\label{match}
{\int D[\phi_k]{\delta S_k[\phi_k]\over\delta G_{\tilde n}(\Lambda)}
e^{-S_k[\phi_k]}\over\int D[\phi_k]e^{-S_k[\phi_k]}}
={\int D[\phi_k]D[\tilde\phi_k]O_{\tilde n}(\phi_k+\tilde\phi_k)
e^{-S_\Lambda[\phi_k+\tilde\phi_k]}\over\int D[\phi_k]D[\tilde\phi_k]
e^{-S_\Lambda[\phi_k+\tilde\phi_k]}},
\ee
c.f. Eq. \eq{toyrginv}, showing that they are represented in the effective theory by the functional
derivative of the blocked action with respect to the microscopical
coupling constants. The form \eq{coact} of the action gives the operator mixing
\be\label{omix}
\{O_{\tilde n}(\phi_k)\}_k=\sum_{\tilde m}
{\delta G_{\tilde m}(k)\over\delta G_{\tilde n}(\Lambda)}
{\delta S_k[\phi_k]\over\delta G_{\tilde m}(k)}
=\sum_{\tilde m}O_{\tilde m}(\phi_k)S_{\tilde m\tilde n}(k,\Lambda),
\ee
in a manner similar to Eq. \eq{toyopm}. We introduced here the sensitivity matrix
\be\label{sensm}
S_{\tilde m,\tilde n}(k,\Lambda)
={\delta G_{\tilde m}(k)\over\delta G_{\tilde n}(\Lambda)}.
\ee
It is the measure of the sensitivity of an effective strength of interaction
on the initial condition of the RG trajectory, imposed in the U.V. domain,
c.f. Eq. \eq{sensit}. The composite operator $\{O_{\tilde n}(\phi_k)\}_k$
introduced in Eq. \eq{omix} replaces the bare operator
$\{O_{\tilde n}(\phi)\}_\Lambda$ in the Green functions of the
effective theory for the I.R. modes $\phi_k$.

The differential equation generating the operator mixing \eq{omix} is obtained
in the following manner. The relation
\be
G_{\tilde n}(k-\dk)=G_{\tilde n}(k)-{\dk\over k}\beta_{\tilde n}(k,G)
\ee
is used to arrive at the sensitivity matrix,
\bea
S_{\tilde m,\tilde n}(k-\dk,\Lambda)
={\delta G_{\tilde m}(k-\dk)\over\delta G_{\tilde n}(\Lambda)}
&=&\sum_{\tilde\ell}{\delta G_{\tilde m}(k-\dk)\over\delta G_{\tilde\ell}(k)}
{\delta G_{\tilde\ell}(k)\over\delta G_{\tilde n}(\Lambda)} \nonu
&=&\sum_{\tilde\ell}\left[\delta_{\tilde m\tilde\ell}-
{\dk\over k}{\delta\beta_{\tilde m}(k,G)\over\delta G_{\tilde\ell}(k)}\right]
S_{\tilde\ell,\tilde n}(k)
\eea
and to find
\be
k\partial_kS_{\tilde m,\tilde n}(k,\Lambda)
=\sum_{\tilde\ell}{\delta\beta_{\tilde m}(k,G)\over\delta G_{\tilde\ell}(k)}
S_{\tilde\ell,\tilde n}(k,\Lambda).
\ee
The scale dependence of the operator mixing matrix is governed by
\be\label{mixing}
\gamma_{\tilde n\tilde m}(k)={1\over k}
{\delta\beta_{\tilde n}(k,G)\over\delta G_{\tilde m}(k)}.
\ee

Summary: Eq. \eq{comprgfl} represents the link between Kadanoff-Wilson blocking
and composite operator renormalization. On the one hand, $\beta_{\tilde n}$
describes the evolution of the action in the traditional blocking scheme,
Eq. \eq{blcompop}. On the other hand, interpreting the coupling
constants in the action as sources coupled to composite operators  $\beta_{\tilde n}$
determines the mixing of composite operators in Eq. \eq{mixing}. Therefore
all conepts and results of the blocking, e.g. fixed point, universality, etc.
has a counterpart in composite operator renormalization.

\subsubsection{Parallel transport}
The operator mixing discussed above has a nice geometrical interpretation, a
parallel transport of operators along the RG trajectory \cite{connection}
with the connection $\Gamma_{\tilde n,\tilde m}(k)$.
The independence of the expectation value from the cut-off,
\be
{\int D[\phi_{k'}]D[\tilde\phi_{k'}]O_{\tilde n}(\phi_{k'}+\tilde\phi_{k'})
e^{-S_{k'}[\phi_{k'}+\tilde\phi_{k'}]}\over
\int D[\phi_{k'}]D[\tilde\phi_{k'}]e^{-S_{k'}[\phi_{k'}+\tilde\phi_{k'}]}}
={\int D[\phi_k]D[\tilde\phi_k]O_{\tilde n}(\phi_k+\tilde\phi_k)
e^{-S_k[\phi_k+\tilde\phi_k]}\over\int D[\phi_k]D[\tilde\phi_k]
e^{-S_k[\phi_k+\tilde\phi_k]}}
\ee
defines the parallel transport of composite operators. The linearity
of the mixing \eq{omix} assures that this parallel transport is indeed
linear and can therefore be characterized by a covariant derivative
\be\label{covder}
D_kO_k=(\partial_k-\Gamma)O_k
\ee
where
$\Gamma_{\tilde m,\tilde n}=(S^{-1}\cdot\gamma\cdot S)_{\tilde n,\tilde m}$
in such a manner that $D_kO_k=0$ along the RG flow.

The dynamical origin of the connection comes from the fact that there are
two different sources the scale dependence of $\langle O_k\rangle$
comes from: from the explicit $k$-dependence of the operator and from
the implicit $k$-dependence due to the cut-off in the path
integration. The operator mixing is to balance them. In fact,
the covariant derivative could have been introduced by the relation
\be\label{covcon}
\partial_k\langle O_k\rangle=\langle D_kO_k\rangle,
\ee
requiring that the operator mixing generated by the connection
is to make up the implicit $k$-dependence of the expectation value
coming from the cut-off.

It is obvious that $\Gamma$ is vanishing in the basis $\{O_{\tilde n}\}_k$,
\be
\partial_k\langle\sum_{\tilde n}c_{\tilde n}(k)\{ O_{\tilde n}\}_k\rangle
=\sum_{\tilde n}\partial_k c_{\tilde n}(k)\langle\{ O_{\tilde n}\}_k\rangle
=\langle\partial_k\sum_{\tilde n}c_{\tilde n}(k)\{ O_{\tilde n} \}_k\rangle.
\ee
The connection can in principle be found in any other basis by
simple computation.

\subsubsection{Asymptotical scaling}
We linearize the beta-functions around a fixed point $G^*_{\tilde m}$,
\be
\beta_{\tilde n}\approx\sum_{\tilde m}
\Gamma^*_{\tilde n\tilde m}(G_{\tilde m}-G^*_{\tilde m}),
\ee
and write the scaling coupling constant, the left eigenvectors of $\Gamma$
\be
\sum_{\tilde m}c^\mr{left}_{\tilde n,\tilde m}
\Gamma^*_{\tilde m,\tilde r} =\alpha_{\tilde n}c^\mr{left}_{\tilde n,\tilde r},
\ee
as
\be
G^{sc}_{\tilde n}=\sum_{\tilde m}c^\mr{left}_{\tilde n,\tilde m}
\left(G_{\tilde m}-G^*_{\tilde m}\right).
\ee
They display the scale dependence
\be\label{scl}
G^{sc}_{\tilde n}\sim k^{\alpha_{\tilde n}}.
\ee

Furthermore let us fix the overall scale of the local operators. This can be
done by using the decomposition
\be
O(\phi_x)=\sum_nb_nO_n(\phi_x)
\ee
where $O_n(\phi(x))$ is the product of the terms
$\partial_{\mu_1}\cdots\partial_{\mu_\ell}\phi^m(x)$ with coefficient 1.
The norm $||O||=\sqrt{\sum_n b_n^2}$ is introduced with the notation
\be
\overline{O}={O\over||O||},
\ee
for the operators of unit norm. We shall use the convention that
the coupling constants $G_{\tilde n}(\Lambda)$ always multiply
operators of unit norm in the action.

The scaling operators
\be
O^{sc}_{\tilde n}=\sum_{\tilde m}c^\mr{right}_{\tilde n\tilde m}O_{\tilde m}
\ee
are obtained by means of the right eigenvectors of $\Gamma^*$,
\be
\sum_{\tilde m}\Gamma^*_{\tilde r\tilde m}c^\mr{right}_{\tilde m\tilde n}
=\alpha_{\tilde n}c^\mr{right}_{\tilde r\tilde n},
\ee
and they satisfying the conditions of completeness
$c^\mr{right}\cdot c^\mr{left}=1$ and orthonormality
$c^\mr{left}\cdot c^\mr{right}=1$. The coupling constants of the action
\be
S_k=\sum_{\tilde n}G_{\tilde n}(k)\overline{O^{sc}_{\tilde n}(\phi_k)}
\ee
obviously follow \eq{scl}. The operator
\be
O=\sum_{\tilde n}b_{\tilde n}\overline{\{O^{sc}_{\tilde n}\}_k},
\ee
written at scale $k$ in this basis yields the parallel transport trajectory
\be
\{O\}_{k'}=\sum_{\tilde n}b_{\tilde n}
\{\overline{\{O^{sc}_{\tilde n}\}_k}\}_{k'}
=\sum_{\tilde n}b_{\tilde n}
\left({k'\over k}\right)^{\alpha_{\tilde n}}
\overline{\{O^{sc}_{\tilde n}\}_k}
\ee
in the vicinity of the fixed point.

The beta function introduced by Eqs. \eq{blcompop}-\eq{comprgfl}
agrees with the usual one. One finds that only relevant operators
have non-vanishing parallel transport flow. Universality manifests itself
at a given scaling regime in the suppression of the parallel transported
irrelevant operators.

The blocked action $S_k[\phi_{k,x},G_{\tilde n}(k),G_{\tilde n}(\Lambda)]$
possesses two interpretations:
\begin{itemize}
\item The value of the coupling constant at the running cut-off,
$G_{\tilde n}(k)$, reflects the scale dependence of the physical parameters.
\item The dependence on the bare coupling constants, $G_{\tilde n}(\Lambda)$,
the initial condition of the RG flow provides us the generator functional for
composite operators.
\end{itemize}

Finally, the composite operator renormalization
represents an alternative way to arrive at the RG equation. In fact,
the beta functions arising from the blocking \eq{blcompop} are obtained
in this scheme by  the parallel transport, the matching \eq{match}
of the expectation values.

\subsubsection{Perturbative treatment}
Let us finally compare the matching of the observables, described above
with the traditional method of perturbative composite operator renormalization.
The inversion of Eq. \eq{omix} gives
\be\label{trcompop}
[O_{\tilde n}(\phi_k+\tilde\phi_k)]_k
=\sum_{\tilde m}O_{\tilde m}(\phi_k+\tilde\phi_k)
(S^{-1}(k,\Lambda))_{\tilde m,\tilde n}.
\ee
It is not difficult to see that the operator $[O_{\tilde n}(\phi_k+\tilde\phi_k)]_k$
of the bare theory which corresponds to the operator $\{O_{\tilde n}(\phi_k)\}_k$
of the effective theory agrees with the result of the usual composite operator renormalization.
In fact, the bare action \eq{bareact} can be split into the sum of the renormalized part and
the counterterms the framework of the renormalized perturbation expansion,
giving $G_{\tilde nB}=G_{\tilde nR}+G_{\tilde nCT}$,
$G_{\tilde n}(\Lambda)=G_{\tilde nB}$ and $G_{\tilde n}(k)=G_{\tilde nR}$.
The counterterms are introduced just to render certain Green functions with composite operator
insertions cut-off independent \cite{source}. The composite operator corresponding to the
renormalized (ie blocked) operator $O_{\tilde n}(\phi_k)$ in this perturbative
framework is
\be
\sum_{\tilde m}O_{\tilde m}(\phi_k+\tilde\phi_k)
{\delta G_{\tilde mB}\over\delta G_{\tilde nR}}
\ee
which agrees with \eq{trcompop}.

It is worthwhile recalling that the parameters and operators of a bare theory correspond
to the cut-off scale in a natural manner. Therefore $\{O_{\tilde n}(\phi_k)\}_k$
represents the bare operator $[O_{\tilde n}(\phi_k+\tilde\phi_k)]_k$
in the effective description at the scale $k$ and reproduces the observational scale
dependence by construction in a manner similar to the scale dependence of the
hadronic structure functions obtained in the framework of the composite operator
renormalization convey the same information \cite{structfct}.

\subsection{Continuous evolution}
The RG equations obtained so far deal with the
evolution of the bare action during the gradual lowering of the U.V. cut-off.
There are two different reasons to look for alternative schemes where
the evolution of the Green functions rather than the bare action is followed.

One reason is usefulness. In the traditional RG strategy followed so far
the renormalized trajectories give the bare coupling constants as functions of the running
cut-off. Though one can extract number of useful information from the
trajectories they remain somehow qualitative because the
bare parameters of the theory with the running cut-off are not
physical quantities. Though this remark does not apply to the multiplicative
RG scheme where the effective parameters are constructed by means of
Green functions this scheme is seriously limited as mentioned in the Introduction.
Returning to blocking, one can say at most that the difference between bare parameters
and physical quantities arises due to the fluctuations in the path
integral and this latter decreases with the number of degrees of
freedom as we approach the IR end point of the trajectories.
Therefore it is the blocked action with very low cut-off only which
is supposed to be directly related to physical quantities
(in the absence of IR instability).

Another more formal point of view is to construct a strictly non-perturbative
scheme. The RG equations \eq{morgwh}, \eq{polch} represent
the complete resummation of the perturbation expansion but
they are not really non-perturbative equations. This situation is
reminiscent of the Schwinger-Dyson equations. They were first obtained
by resumming the perturbation expansion and only later by a genuine
non-perturbative method, by the infinitesimal shift of the integral variable
in the path integral formalism.
The fact that the naive result derived by assuming the
convergence of the perturbation expansion is correct is presumably
related to the unique analytic continuation of the Green functions in the
coupling constants. Can we find in a similar manner
the truly non-perturbative RG equations? If possible, it will come by
a different route: by relating two path integral averages instead
of computing them by brute force. The quantities in question are
one-particle irreducible (1PI) amplitudes and we shall give up
to follow the evolution of the bare coupling constants, the
parameters in the path integral. This will allows us to avoid any
reference to the perturbation expansion. It remains to be seen if
\eq{morgwh} and \eq{polch} can be derived in a similar manner or they
remain valid for strongly coupled models.

We begin this program by writing the generator functional for the
connected Green functions in the form
\be\label{crggf}
e^{\ih W_k[j]}=\int D[\phi]e^{-\ih(S_B[\phi]+C_k[\phi]-j\cdot\phi)},
\ee
where the term $C_k[\phi]$ is introduced to suppress fluctuations.
What we require is that (i) for $k=\infty$ the fluctuations be suppressed,
$C_\infty[\phi]=\infty$, (ii) the original generator functional be
recovered for $k=0$, $C_0[\phi]=0$ and (iii) the fluctuations
are suppressed only, ie $C_k[\phi']=0$ for some configurations
close to the vacuum expectation value, $\phi'\approx\la\phi\ra$.
The simplest choice for models without condensate is a quadratic functional,
\be\label{qsup}
C_k[\phi]=\hf\phi\cdot C_k\cdot\phi
\ee
but in certain cases higher order terms in the field variables are
necessary in the suppression.
We distinguish two kinds of suppression. The length scale of the
modes 'released' $\ell(k)$ is well defined when
\be
{\cal M}^2_k(p)=
{\delta^2C_k\over\delta\phi_{-p}\delta\phi_p}_{\vert\phi=\la\phi\ra}
\cases{>>1&$|p|<1/\ell(k)$,\cr\approx0&$|p|>1/\ell(k)$.}
\ee
The evolution generated by such
suppressions shows the scale dependence in the spirit of the
Kadanoff-Wilson blocking, the contribution of modes with a
given length scale to the dynamics. Examples are,
\be\label{scaldep}
C_k(p)=\cases{ap^2{f(p)\over1-f(p)}&\cite{wetteq}\cr
a\left({k^2\over p^2}\right)^b&\cite{morriseq}\cr
a(k^2-p^2)\Theta(k^2-p^2)&\cite{litimo}}
\ee
where $f(p)=e^{-b({p^2/k^2})^c}$ and $a,b,c>0$. Another kind of suppression
for which ${\cal M}^2_k(p)$ shows no clear structure will be considered
in section \ref{ints} below.

The evolution equation for $W[j]$ is easy to obtain,
\bea\label{evw}
\partial_kW_k[j]&=&-e^{-\ih W_k[j]}\int D[\phi]
\partial_kC_k[\phi]e^{-\ih(C_k[\phi]+S_B[\phi]-j\cdot\phi)} \nonu
&=&-e^{-\ih W_k[j]}\partial_kC_k\left[\hbar{\delta\over\delta j}\right]
e^{\ih W_k[j]}.
\eea
This is already a closed functional differential equation but of little use.
The reason is that it is difficult to truncate $W[j]$ being a
highly non-local functional.
In order to arrive at a more local functional we shall make a Legendre
transformation and introduce the effective action $\Gamma_k[\phi]$ as
\be\label{ltr}
\Gamma_k[\phi]+W_k[j]=j\cdot\phi,\ \ \phi={\delta W[j]\over\delta j},
\ee
with
\be
\partial_k\Gamma_k[\phi]=-\partial_kW_k[j]-{\delta W[j]\over\delta j}
\partial_kj+\partial_kj\phi=-\partial_kW_k[j].
\ee
It is advantageous to separate the auxiliary suppression term $C_k[\phi]$
off the effective action by the replacement
$\Gamma[\phi]\to\Gamma[\phi]+C_k[\phi]$ resulting in
\be\label{evltr}
\partial_k\Gamma_k[\phi]=e^{-\ih W_k[j]}\partial_kC_k
\left[\hbar{\delta\over\delta j}\right]e^{\ih W_k[j]}-\partial_kC_k[\phi].
\ee
This relation takes a particularly simple form for the quadratic suppressions
\eq{qsup}. Since the quadratic part of the effective action
is the inverse connected propagator we find \cite{wetteq}, \cite{morrp},
\cite{evoleq}, \cite{ellwev}, \cite{adamsev},
\be\label{cevol}
\partial_k\Gamma_k[\phi]
={\hbar\over2}\tr\left[\partial_kC_k\cdot\la\phi\phi\ra_{conn}\right]
={\hbar\over2}\tr\left[\partial_kC_k\cdot{1\over C_k+{\delta^2\Gamma_k\over
\delta\phi\delta\phi}}\right].
\ee

Two remarks are in order in comparing the evolution equation obtained
here with the pervious RG equations. First, the Kadanoff-Wilson
blocking is constructed to preserve the generator functional $W[j]$ for the
thinner system and this is realized by eliminating modes
at the U.V. side, by changing $S_k[\phi]$. One is not aiming at keeping
anything fixed in the continuous evolution scheme. The 1PI generator
functional does depend on $k$ which appears as
the lowest momentum of modes considered in the theory whose effective
action is $\Gamma_k[\phi]$. The second remark is that
the right hand side is proportional to $\hbar$. But the apparent
absence of the mixing of tree and loop-levels is misleading since the
propagator may be the sum of $\ord{\hbar^0}$ tree-level and $\ord{\hbar}$
fluctuation contributions. The explicit tree-level term is missing
because the modification of the bare action was carried out in the
part $\ord{\phi^2}$ and this correction to the free propagator was
removed by the step $\Gamma[\phi]\to\Gamma[\phi]+C_k[\phi]$,
the substraction of the tree-level suppression term from the effective action.

It is illuminating to compare the evolution and the
Schwinger-Dyson (SD) equations. Though the formal aspects appear similar
the content of the equations differs. A similarity appearing immediately is
that both set of equations determine the Green functions in a hierarchical
manner. The derivation of the equations shows some similarity, as well.
In fact, the crucial step in the derivation of the
evolution equation is the first equation  in \eq{evw}.
This step, the expression of the derivative of the functional integral
formally, without its actual evaluation is the hallmark of the
non-perturbative demonstration of the SD equations. Both the SD
and the evolution equations are genuinely non-perturbative
because we actually do not evaluate the functional integral,
instead we relate to another one by bringing a derivative inside
the path integral. In both schemes on compares two functional
integrals which differ slightly, either in the infinitesimal shift
of the integral variables or in the infinitesimal change in the action.
Let us write the first equation in \eq{evw} for finite $\dk$,
\bea
&&\int D[\phi]\left(e^{-\ih(C_k[\phi]+S_B[\phi]-j\cdot\phi)}
-e^{-\ih(C_{k-\dk}[\phi]+S_B[\phi]-j\cdot\phi)}\right)\nonu
&&=\ih\int D[\phi]\left(\Delta k\partial_kC_k[\phi]e^{-\ih C_k[\phi]}
+\ord{(\dk)^2}\right)e^{-\ih(S_B[\phi]-j\cdot\phi)}.
\eea
The small parameter $\dk$ is used to suppress the
insertion of more field variables in the evolution equation,
to cut off the higher order Green functions from the evolution,
the strategy common with the non-perturbative proof of the SD equation.
The higher order contributions in $\dk$ bring in higher order Green functions
and we suppress them by the smallness of the step in 'turning on' the
fluctuations. This is how the RG idea, the replacement of the higher loops by
running effective coupling constants and the dealing with small number of
modes at each step, is realized in the evolution equation scheme.

The obvious difference between the evolution and the SD equations
is that the latter express the invariance of the functional
integral in a manner  similar to the RG equations but the former
is simply an expression of the derivative of the functional
integral with respect to a parameter. The similarity of the evolution
equation and the RG strategy leads to another difference. The
subsequent elimination of modes and  taking into account their
dynamics by the introduction/modification of effective vertices
generates a 'universal' evolution equations which does not depend
on the theory in question. In fact, the evolution equation
\eq{cevol} and the functional RG equations Eqs. \eq{sourcebl},
\eq{sblock} {\em do not contain any reference to the model
considered}. The effective or the bare action of the model appear
in the initial condition only. The SD equation is based on a
careful application of the equation of motion within the
expectation values and therefore contains the action of the model
in an obvious manner.

The evolution equation method seems to be better suited to
numerical approximations then the SD equations. This is because
the 'dressing', the summing up the interactions is achieved
by integrating out differential equations rather then coupling
the Green functions in the SD hierarchy, a numerical problem we can
control easier.

\subsection{Blocking in the internal space}\label{ints}
We present now a generalization the RG method. The traditional
RG strategy is aiming at the scale dependence of observables
and consequently follows the cut-off dependence in the theories.
The results are obtained by the successive modification of the cut-off
and the accumulation of the resulting effects. One can generalize this
method by replacing the cut-off by $any$ continuous parameters of the theory
on which the observables depend upon in a differentiable manner.
Such a generalization replaces the RG equation by an evolution equation
corresponding to the parameter or coupling constant in question.

One gains and looses in the same time during such a generalization.
One looses intuition of and insight into the dynamics since the trajectories
generated by the evolution equation have no relation to
scale dependence. To make things even more complicated
the trajectories do not correspond anymore to a fixed physical content,
instead they trace the dependence of the dynamics in the parameter considered.
But we gain in flexibility. In fact, the parameter we select to evolve
can be the Plank constant or a coupling constant and the integration
of the resulting evolution resums the semiclassical
or the perturbation expansion. We may apply this method in
models with non-trivial saddle point structure or where the
cut-off would break important symmetries, e.g. in gauge theories.

A more technical aspect of this generalization touches the U.V. divergences in the
theory. The traditional RG procedure is based on the tacit assumption that the
U.V. divergences are properly regulated by the blocking, the moving cut-off. This
seems to be a natural requirement if the RG flow is interpreted as
scale dependence. In fact, an U.V. divergence left-over by the
blocking would indicate the importance of the processes at the
regulator and would lead to the appearance of a second scale.
In the generalization of the RG procedure we introduce below
there may not be a well defined scale related to the evolution and
we have to regulate the U.V. divergences.

There are two different kind of physical spaces in Field Theory.
The space-time or the momentum-energy space where the 'events'
are taking place will be called external space. The 'events' are
characterized by further quantities, the field amplitudes. The space
where the field amplitudes belong will be called internal space.
Therefore the field configuration $\phi_x$ realizes
a map $\phi:external\ space\to internal\ space$. The traditional
Kadanoff-Wilson blocking orders the degrees of freedom to be eliminated
according to their scale in the external space. We may realize a similar
scheme by ordering the modes according to their scale in the internal
space, their amplitude \cite{int}. We present two different implementations
of this idea, one for Wilsonian and another for the effective action.

\subsubsection{Wilsonian action}
We shall use the flexibility of the method of deriving the Polchinksi equation
in section \ref{resperts} to separate off and eliminate the fluctuation
modes with the largest amplitude in the path integral. Since the fluctuation
amplitude is controlled by the mass this leads to the description of
the dynamics by highly massive modes, small fluctuations at the final
point of the evolution.

Let us consider the model \eq{zor} where the role of the parameter
$k$ is played by the mass $M$ and $G^{-1}_{M^2}=p^2+M^2$. We shall follow
the evolution from $M^2=0$ to $M^2>>\Lambda^2$ where $\Lambda$ is
the U.V. cut-off. The decomposition
\be
G_{M^2}=G_{M^2+\Delta M^2}+\tilde G_{M^2},\ \ \Phi=\phi+\tilde\phi,
\ee
will be used with
\be\label{deci}
\hf\phi\left(M^2+\Delta M^2-\Box\right)\phi
+\hf\tilde\phi{(M^2-\Box)^2\over\Delta M^2}\tilde\phi=
\hf\Phi\left(M^2-\Box\right)\Phi+\hf\tilde\Phi\tilde G^{-1}_D\tilde\Phi.
\ee
We replace the degrees of freedom $\tilde\Phi$ by the
field $\tilde\phi$ with infinitesimal fluctuations. After eliminating
$\tilde\phi$ the remaining field $\phi$ has smaller fluctuations than
$\Phi$. We use the ansatz
\be
S^I_M[\phi]=\int_x\left[\hf(Z_M(\phi_x)-1)\partial_\mu\phi_x
\partial_\mu\phi_x+U_M(\phi_x)\right]
\ee
where the U.V. regulator with smooth cut-off is not shown explicitly. We
follow the steps outlined in section \ref{polgre} which gives rise the
evolution equation with left hand side
\be
\partial_{M^2}\left\{\hf\int_p\eta_{-p}\eta_p\left[(Z-1)p^2+U^{(2)}\right]
+VU\right\}.
\ee
The right hand side reads as
\bea
&&-\hf\int_p\Biggl\{\delta_{p,0}U^{(1)}
+\eta_p\left[(Z-1)p^2+U^{(2)}\right]+\hf\int_r\eta_r\eta_{p-r}\left[U^{(3)}
-Z^{(1)}\left(r\cdot p-r^2+p^2\right)\right]\Biggr\}\nonu
&&\times{1\over(p^2+M^2)^2} \nonu
&&\times \Biggl\{\delta_{p,0}U^{(1)}+\eta_{-p}\left[(Z-1)p^2+U^{(2)}\right]
+\hf\int_s\eta_{-p-s}\eta_s\left[U^{(3)}
+Z^{(1)}\left(s^2+p\cdot s+p^2\right)\right]\Biggr\}\nonu
&&+{\hbar\over2}\int_p{1\over(p^2+M^2)^2}\Biggl\{V\left[(Z-1)p^2+U^{(2)}\right]
+\hf\int_r\eta_{-r}\eta_r\left[U^{(4)}+Z^{(2)}(p^2+r^2)\right]\Biggr\}\nonu
&&\approx-VU^{(1)2}{1\over2M^4}-\hf\int_r\eta_{-r}\eta_r\left\{
{\left[(Z-1)r^2+U^{(2)}\right]^2\over(r^2+M^2)^2}
+{1\over M^4}U^{(1)}(U^{(3)}+Z^{(1)}r^2)\right\}\nonu
&&+{\hbar\over2}\int_p{1\over(p^2+M^2)^2}\Biggl\{V\left[(Z-1)p^2+U^{(2)}\right]
+\hf\int_r\eta_{-r}\eta_r\left[U^{(4)}+Z^{(2)}(p^2+r^2)\right]\Biggr\}
\eea
The system of evolution equation projected onto the different
$\eta$-dependent terms is
\bea\label{intwi}
2M^4\partial_{M^2}U&=&-U^{(1)2}
+\hbar M^4\int_p{(Z-1)p^2+U^{(2)}\over(p^2+M^2)^2}\nonu
2M^4\partial_{M^2}Z&=&-4U^{(2)}(Z-1)+4{U^{(2)2}\over M^2}-2U^{(1)}Z^{(1)}
+\hbar M^4Z^{(2)}\int_p{1\over(p^2+M^2)^2}\nonu
2M^4\partial_{M^2}U^{(2)}&=&-U^{(2)2}-U^{(1)}U^{(3)}
+\hbar M^4\int_p{Z^{(2)}p^2+U^{(4)}\over(p^2+M^2)^2}.
\eea
The $\ord{\hbar^0}$ terms on the right hand side are to keep the
tree-level observables $M$-invariant within the gradient expansion ansatz.
As far as the loop corrections are concerned, the beta functions
\be
\beta_n=M\partial_Mg_n=\hbar M^2U^{(n+2)}\int_p{1\over(p^2+M^2)^2},
\ee
obtained from the first equation with $Z=1$ agree with the
leading order contribution to the one-loop renormalized potential,
\be\label{ehurmi}
U^{1-\mr{loop}}_M=U_M+{\hbar\over2}\int_p\ln[p^2+M^2+U^{(2)}_M],
\ee
except their sign. This is because we intend to keep the partition
function unchanged as opposed to Eq. \eq{ehurmi} where the mass is
evolving together with the dynamics.

\subsubsection{Effective action}
Let us consider for the sake of simplicity again the scalar model
given by the bare action
\be
S_B=\int_x\left[\hf(\partial_\mu\phi_x)^2+{m_B^2\over2}\phi^2_x
+U_B(\phi_x)\right]
\ee
and the suppression with no structure in the external space \cite{int},
\be\label{mcf}
C_k(p)={M^2\over2}
\ee
which acts as a 'smooth cut-off' for the amplitude of the fluctuations.
The corresponding evolution equation is
\be
\partial_{M^2}\Gamma_M[\phi]={\hbar\over2}\tr\left[M^2+
{\delta^2\Gamma_M[\phi]\over\delta\phi\delta\phi}\right]^{-1}.
\ee
The effective action of the theory is obtained by integrating this
equation from the initial condition $\Gamma_{M_0}[\phi]=S_B[\phi]$
imposed at $M^2=M_0^2>>\Lambda^2>>m_B^2$ to $M=0$.

Let us consider how this scheme looks like for the ansatz
\be
\Gamma_M[\phi]=\int_x\left[\hf Z_M(\phi_x)(\partial_\mu\phi_x)^2
+U_M(\phi_x)\right].
\ee
One finds
\bea
\partial_{M^2}U&=&{\hbar\over2}\int_p{1\over Zp^2+M^2+U^{(2)}}\nonu
\partial_{M^2}Z&=&{\hbar\over2}\int_p\Biggl[
2Z^{(1)}{{p^2\over d}Z^{(1)}+2\left(Z^{(1)}p^2+U^{(3)}\right)
\over\left(Zp^2+M^2+U^{(2)}\right)^3}
-{Z^{(2)}\over\left(Zp^2+M^2+U^{(2)}\right)^2} \nonu
&&-2Z{\left(Z^{(1)}p^2+U^{(3)}\right)^2\over\left(Zp^2+M^2+U^{(2)}\right)^4}
-{8p^2\over d}ZZ^{(1)}{\left(Z^{(1)}p^2+U^{(3)}\right)\over
\left(Zp^2+M^2+U^{(2)}\right)^4} \nonu
&&+{8p^2\over d}Z^2{\left(Z^{(1)}p^2+U^{(3)}\right)^2\over
\left(Zp^2+M^2+U^{(2)}\right)^5}\Biggr].
\eea
The asymptotic form of the first equation for $M^2>>U^{(2)}$
in the local potential approximation,
\be\label{betacs}
M\partial_MU_M\approx-U^{(2)}_M\int_p{M^2\over(p^2+M^2)^2}
\ee
up to field independent constant and it agrees with asymptotic form of
the first equation in \eq{intwi} except
the sign. The integrand in \eq{betacs} corresponds to transformation
rule \eq{propmd} of the propagator under the infinitesimal change of
the mass for the first, leading order graphs in Figs. \ref{betf}.
The numerical integration resums the higher orders in the
loop expansion. In short, we see the Callan-Symanzik scheme at work in the
functional formalism.

The suppression \eq{mcf}
freezes out modes with momentum $p$ if $p^2<M^2$ and generates a
characteristic scale $p_\mr{cr}$ where
$(Zp^2+M^2+U^{(2)})/(Zp^2+U^{(2)})$ deviates from 1.
So long this scale is far from the intrinsic scales of the model,
$M^2>>U^{(2)}$ in the U.V. scaling regime, such an internal space blocking
generates the external scale $p_\mr{cr}\approx M/\sqrt{Z}$. In other words,
the universal part of the beta functions obtained in the U.V. regime by
means of the Callan-Symanzik scheme should agree with the same part of the
beta functions coming from other schemes. This agreement has been
observed in the framework of the scalar model \cite{cowe}.
We shall check this quickly in the local potential approximation.
By comparing Eq. \eq{betacs} with the asymptotic form of the WH
equation \eq{wh},
\be
k\partial_kU_k=-{\hbar U^{(2)}_k\Omega_dk^{d-2}\over2(2\pi)^d},
\ee
we find
\be
{dk^2\over dM^2}=2\int_0^{\Lambda/k}{y^{d-1}dy\over(y^2+M^2/k^2)^2}.
\ee
Below the upper critical dimension, $d<4$, the right hand side is
finite and the external (WH) and the internal (CS) scales are
proportional. At the critical dimension the $M$
dependence of the right hand side is through the proportionality factor
$\ln(1+\Lambda^2/M^2)$. The two scales become approximately
proportional only for much higher value of the cut-off $\Lambda$,
since the RG flow spends more 'time' close to the non-universal
short distance regime due to the tree-level marginality of $g_4$.
There is no agreement beyond $d=4$ where the cut-off scale is
always important. The agreement
between the two schemes is violated by higher
order terms in the perturbation expansion since they represent
the insertion of irrelevant effective vertices and by leaving
the asymptotical U.V. regimes. This explains that similar argument
does not hold when $Z$ displays important $M$-dependence.

It is worthwhile noting that the non-trivial wave function renormalization $Z$
and the corresponding anomalous dimension $\eta$ of the field variable
can be thought as the reflection of the mismatch between the
scale dependence in the external and internal spaces around the U.V. fixed point.
In fact, the phenomenological form
\be
\la\phi_x\phi_y\ra\approx c|x-y|^{2-d-\eta}
\ee
cf Eq. \eq{etabev} below connects the fundamental dimensional objects of the
internal and external spaces.

The flexibility of choosing the suppression functional $C_k[\phi]$ may be
important for certain models. By choosing
\be
C_k[\phi]={k\over\Lambda}S[\phi]
\ee
where $\Lambda$ is the U.V. cut-off the evolution in $k$ resums the
loop expansion since the effective Planck-constant,
$\hbar^{-1}(k)=\hbar^{-1}+k/\Lambda$
evolves from $0$ to $\hbar$. This scheme is advantageous for models
with inhomogeneous saddle point eg solitons or instantons because
their space-time structure is 'RG invariant', being independent of the
gradual control of the amplitude of the fluctuations. Another advantage
offered by the flexibility in choosing the suppression functional is
the possibility of preserving symmetries, the point considered in section
\ref{gauges} below.

\section{Applications}
We shall briefly review a few applications of the functional
evolution equations. An incomplete list of the developments not
followed due to limitation in time is the following. The exciting
competition for the 'best' critical exponents have led to several
works using this method \cite{critexp}, \cite{hh},
\cite{morriseq}, \cite{bhlm}, \cite{comell}, \cite{aoki},
\cite{criexp}, \cite{senben}. Such kind of application opens up
the issue of understanding the impact of truncation on the
blocking \cite{fperrorm}, \cite{fperrors}, \cite{morplb}
\cite{aoki}, \cite{terwet}, \cite{alford} and looking for
optimization \cite{optimlps}, \cite{optiml}, \cite{litimo}. The
phase structure and the nature of phase transitions are natural
subjects \cite{adamsev}, \cite{phtrans}, \cite{matrix}. The
incorporation of fermions is essential to arrive at realistic
models in High Energy \cite{fermions} and Condensed Matter Physics
\cite{cfermions}. The computation of the quenched average of Green
functions, one of the essential obstacle of progress in Condensed
Matter Physics can be approached in a new fashion by applying the
internal space renormalization group method \cite{cdf}. A
promising application is in general relativity \cite{genrel}.
Finally, the infamous problem of bound states can be reconsidered
in this framework \cite{wettb}, \cite{dft}.

Much more to be found in review articles \cite{rev},
conference proceedings \cite{proc} and PhD thesis \cite{phd}.

\subsection{Fixed points}

\subsubsection{Rescaling}
`The original form of the RG procedure consists of two steps, the
blocking, followed by a rescaling. The latter can be omitted if
the RG strategy is used to solve models only. But it becomes
important when we try to understand the scale dependence of the
theories. Our physical intuition is based on classical physics and
particles. After losing much of the substance during the
quantization procedure there is a chance to recover some clarity
by the introduction of quasiparticles, localized excitations with
weak residual interaction. The rescaling step of the RG scheme is
to check a given quasiparticle assumption by removing the
corresponding scale dependence. The remaining scale dependence is
a measure of the quality of the quasiparticle picture. The
deviation from this non-interactive system is parametrized in
terms of anomalous dimensions.

The particle content of a theory
is usually fixed by the quadratic part of its action. Therefore
the rescaling is defined in such a manner that this part of the action
stays invariant during the blocking. For an $O(d)$ invariant
Euclidean or relativistically invariant real-time system the
resulting rescaling reflects the classical dimension of the
dynamical variables and coupling constants. For non-relativistic
systems the fixed point scaling is an artificial device only to
identify the non-interacting part of the action. When smooth cut-off
is used its details may influence of the scaling properties of
the quadratic action, as well, and may induce deviations from
classical dimensions..

Consider the scale transformation
\be\label{etabev}
p\to p'=(1+\epsilon)p,\ \ \ x\to x'=(1-\epsilon)x,\ \ \
\phi((1+\epsilon)x')\to(1-\epsilon d_\phi)\phi(x)
\ee
in an $O(d)$ invariant scalar model where $d_\phi=(d-2+\eta)/2$.
The parameter $\eta$ reflects the deviation of from classical
dimensional analysis. The effect of this rescaling on the action
\be
S[\phi]=\sum_n\int_{p_1,\cdots,p_n}u_{p_1,\cdots,p_n}\delta_{p_1+\cdots+p_n,0}
\phi_{p_1}\cdots\phi_{p_n},
\ee
can be found in the following manner \cite{resc}:
\begin{itemize}
\item The momentum integral measure changes as $d^dp\to(1-d\epsilon)d^dp'$,
giving $S\to[1-\epsilon d\int_p\phi_p{\delta\over\delta\phi_p}]S$.
\item The coupling constants change as $u_{p_1,\cdots,p_n}\to
u_{(1-\epsilon)p'_1,\cdots,(1-\epsilon)p'_n}$. This can be written as the
transformation $S\to[1-\epsilon\int_p\phi_pp\cdot\partial'_p
{\delta\over\delta\phi_p}]S$ of the action where the prime on the
gradient $\partial _p$ indicates that the derivative acts on the
coupling constants only and not on the Dirac-deltas.
\item The Dirac-deltas change as $\delta_{p_1+\cdots+p_n,0}\to
\delta_{(1-\epsilon)p'_1+\cdots+(1-\epsilon)p'_n,0}$,
amounting to the transformation $S\to[1+d\epsilon]S$ of the action.
\item Finally, the field transforms as
$\phi_p\to[1-\epsilon(d_\phi-d)]\phi_p$, inducing
$S\to[1-\epsilon(d_\phi-d)]\int_p\phi_p{\delta\over\delta\phi_p}S$.
\end{itemize}
Adding up these contributions we find the rescaling generator
\be
{\cal G}=-\int_p\left[\phi_pp\cdot\partial'_p
{\delta\over\delta\phi_p}+d_\phi\phi_p{\delta\over\delta\phi_p}\right].
\ee
The RG equation for the rescaled action can be obtained by adding
${\cal G}$ acting on the (effective) action to the right hand side, eg
the evolution \eq{linpol} turns out to be
\be\label{linpolr}
k\partial_ke^{-\ih S^I_k[\phi]}=\left({\cal G}+\hbar{\cal B}\right)
e^{-\ih S_k^I[\phi]}.
\ee

The rescaling of the field may be viewed as a transformation of the
action without changing the physics at the fixed point. The terms
generated on the fixed point action by the rescaling, ${\cal G}S^*[\phi]$
are called redundant at the fixed point in question.

\subsubsection{Reparametrizing}
We considered linear rescaling of the field variable but one can easily
generalize the rescaling to non-linear reparametrization. The
infinitesimal change $\phi_x\to\phi'_x=\phi_x+\epsilon\Psi[\phi;x]$
performed inside the path integral gives
\be
\int D[\phi]e^{-S[\phi]}\to\int D[\phi']e^{-S[\phi']}
=\int D[\phi]\left[1+\epsilon\int_x{\delta\Psi[\phi;x]\over\delta\phi_x}\right]
e^{-S[\phi]-\epsilon\int_x\Psi[\phi;x]{\delta S[\phi]\over\delta\phi_x}}.
\ee
The reparametrization invariance of the integral assures that the
modification
\be\label{reptr}
S[\phi]\to S[\phi]+\epsilon{\cal G}_\Psi S[\phi]
\ee
of the action where
\be
{\cal G}_\Psi=\int_x\left[\Psi[\phi;x]{\delta S[\phi]\over\delta\phi_x}
-{\delta\Psi[\phi;x]\over\delta\phi_x}\right]
\ee
is an invariance of the partition function. It has furthermore been noted
\cite{repar} that some of the infinitesimal blocking relations for the
(effective) action can be written in the form \eq{reptr}.

Though it is certainly very interesting to find a common structure for
the functional RG equations the connection with reparametrization invariance
is not clear. The point is that non-linear reparametrizations can not
be carried out inside of the path integral as in ordinary integrals.
The problem arises from the fact that the typical
configurations are rather singular in the path integration.
Calculus known from classical analysis is replaced by Ito-calculus in
Quantum Mechanics due to the nowhere-differentiable nature of the
quantum trajectories \cite{schulm}. The quantum field configurations are
even more singular. Consider a free massless field $\phi$ in $d$
dimensions and its partition function
\be
Z=\prod_x\int d\phi_xe^{-{a^{d-2}\over2}\sum_x(\Delta_\mu\phi_x)^2}
=\prod_x\int d\phi_xe^{-{1\over2}\sum_x(\Delta_\mu\tilde\phi_x)^2}
\ee
where $\Delta_\mu\phi_x=\phi_x-\phi_{x-\mu}$ and the dimension of
the field was removed in the second equation, $\tilde\phi=a^{d/2-1}\phi$.
The typical configurations have $\Delta\tilde\phi=\ord{a^0}$,
ie the discontinuity of the original field variable is
$\Delta\phi=\ord{a^{1-d/2}}$.

This simple scaling, the basis of the usual U.V. divergences in quantum
field theory, gives non-differentiability in Quantum Mechanics, for $d=1$,
finite discontinuity in $d=2$
and diverging discontinuities for $d>2$. The non-differentiability
can be represented by effective potential vertices in Quantum
Mechanics, \cite{curve}, a reflection of the unusual
quantization rules in polar coordinates, sensitivity for
operator ordering and quantum anomalies \cite{rgqm}. In two dimensions
the discontinuity is finite and the continuous structure can either
be preserved or destroyed by quantum fluctuations, cf. section \ref{sineg}.
In higher dimensions the singularities remain always present. It is this
singular nature of the trajectories which requires that one goes to
unusually high order in the small parameter $\epsilon$ characterizing
the infinitesimal change of variables in the
path integral. This problem, composite operator renormalization,
renders the non-linear change of variables a poorly controlled
subject in quantum field theories.

\subsubsection{Local potential approximation}
We shall search for the fixed points of the Polchinski equation
in the local potential approximation, \cite{bhlm} \cite{comell}.
We introduce dimensionless quantities $\phi=k^{1-d/2}\Phi$ and
$u_k(\phi)=k^dU_k(\Phi)$. The RG equation with rescaling for $u$ is
\be\label{pocunez}
\dot u=-du+{\eta+d-2\over2}\phi u^{(1)}-u^{(1)2}K'_0+\hbar\bar K'u^{(2)}
\ee
where $\bar K'=k^{-d}\int_pK'_p$ and we set $\eta=0$ in the approximation
$Z=1$. Notice the explicit dependence
on the cut-off function $K$. This reflects the fact that
the RG flow depends on the choice blocking transformation. Only the
qualitative, topological features of the renormalized trajectory and the
critical exponents around a fixed point are invariant under non-singular
redefinitions of the action.

\underline{Gaussian fixed point}:
The fixed point equation, $\dot u^*=0$ has two trivial solutions,
\be\label{uvfpu}
u^*(\phi)=0,
\ee
and
\be\label{irfpu}
u^*(\phi)={\phi^2\over2|K'_0|}+\hbar{\bar K'\over dK'_0},
\ee
where $K'_0,\bar K'<0$. For any other solution $u=\ord{\phi^2}$ as
$\phi\to\infty$.

In order to find the scaling operators we introduce a perturbation
around the fixed point by writing $u=u^*+\epsilon k^{-\lambda}v(\phi)$ where
$\epsilon$ is infinitesimal and solve the linearized eigenvalue equation
\be\label{scopu}
\hbar\bar K'v^{(2)}=(d-\lambda)v+\left({2-d\over2}\phi
+2u^{*(1)}K'_0\right)v^{(1)}.
\ee
Having a second order differential equation one can construct
a one-parameter family of solution after having imposed say $v^{(1)}(0)=1$
in a theory with the symmetry $\phi\to-\phi$. But the polynomial solutions
\be
v_n(\phi)=\sum_{\ell=1}^nv_{n,\ell}\phi^{2\ell},
\ee
which are parametrized by a discrete index $n$ correspond to discrete
spectrum \cite{polspectr}.
The critical exponents are identified by comparing the terms
$\ord{\phi^{2n}}$ in Eq. \eq{scopu}, $\lambda_n=d+(2-d)n$, and
$\lambda_n=d-(d+2)n$, for the fixed points \eq{uvfpu} and \eq{irfpu},
respectively \cite{hh}. The leading order critical exponents are given entirely
by the tree-level contributions. The dimensionful coupling constants
are cut-off independent in this case and we have $\lambda_n=[g_{2n}]$
for the fixed point \eq{uvfpu} where the dimension of the coupling constant
$[g]$ is given by Eq. \eq{gdim}. The dominant, largest exponent
$\lambda_T$ and the corresponding coupling constant play distinguished
role in the scaling laws. The dimensionless coupling constant can be
identified with the 'reduced temperature' $t$. According to the definition
$\xi\approx t^{-\nu}$ of the critical exponent $\nu$ we have the mean-field
exponent, $\nu=1/\lambda_T=1/2$. The mass is relevant
at the point \eq{uvfpu} as expected, as shown by the beta function
$\beta_2=-2K'_0\tilde g^2_2-2\tilde g_2$, according to Eq. \eq{polbeta}.
The scaling potentials with exponential growth
for large $\phi$ \cite{huang} correspond to the continuous spectrum
and their physical interpretation is unclear.

All exponents are negative at \eq{irfpu}, this is an IR fixed
point.

\underline{Wilson-Fischer fixed point}: There are non-Gaussian
fixed point solutions for Eq. \eq{pocunez} when $2\le d<4$. One
can construct a one-parameter family of solutions but the fixed
point potentials which remain non-singular for arbitrary $\phi$
correspond to a discrete set \cite{wffpr}, \cite{polspectr}: the
Wilson-Fischer fixed point for $3<d<4$ and as many fixed points as
relevant  when $2<d<3$. The perturbation around these fixed points
is a one-parameter family of scaling potentials with continuous
spectrum of critical exponents. The restriction for solution which
are finite and non-singular everywhere produces a discrete
spectrum \cite{wfscp} in good agreement with other methods of
determining the critical exponents \cite{hh}, \cite{criexp}.

The truncation of the fixed point solution at a finite order of $\phi$
introduces error and spurious solutions \cite{fperrorm}, \cite{fperrors}
which can partially be eliminated by expanding the potential along the
cut-off dependent minimum \cite{morplb}.

\subsubsection{Anomalous dimension}
The RG equation for the wave function renormalization constant
$z(\phi)=Z(\Phi)$ is
\bea\label{pocznez}
\dot z
&=& -4K'_0u^{(2)}(z-1)-2K'_0u^{(1)}z^{(1)}-2{K''_0K_0\over k^2}u^{(2)2} \nonu
&&+\hbar z^{(2)}\bar K'-\eta(z-1)+{d+\eta-2\over2}\phi z^{(1)}.
\eea
The lesson of the case $Z=1$ is that the requirement of the existence and
finiteness of $u^*(\phi)$ introduces discrete number of fixed point
solutions. Let us try to follow the same strategy again. The dominant terms
of such fixed point solutions $\dot u=\dot z=0$ of Eqs. \eq{pocunez} and
\eq{pocznez} are
\be
u^*(\phi)\approx{2-\eta\over4}\phi^4+A\phi^{d-2+\eta\over d+2-\eta},\ \
z^*(\phi)\approx B.
\ee
Together with the conditions $u^{(1)}(0)=z^{(1)}(0)=0$ imposed for the
$\phi\to-\phi$ symmetrical models the solution are well determined in terms
of $A$ and $B$. The problem is that such kind of argument does not fix the
value of $\eta$.

There is another condition to be fulfilled by the fixed points, the
critical exponents should be invariant under rescaling. This is
sufficient to determine $\eta$. Unfortunately the rescaling
invariance of the fixed point is lost unless sharp or specially chosen
polynomial smooth cut-off is used \cite{wfscp}, \cite{repinvc}. Furthermore
the truncation of the gradient expansion contributes to the violation
of this invariance, too. An approximation to find the 'best' solution
when the rescaling invariance is not respected is the following
\cite{comell}: Introduce a further condition which violates
rescaling invariance, say fix the value of $z(0)$. This allows
the determination of $\eta$ which would be unique if rescaling
could be used to relax our last condition. We are as close as possible
to the invariant situation within our parametrized problem when the
dependence of $\eta$ on $z(0)$ is the slowest. Therefore the condition
$d\eta/dz(0)=0$ selects the 'best' estimate of $\eta$.

The numerical error due to the truncation of the gradient
expansion has been the subject of extensive studies in the framework
of the scalar model, c.f. Ref. \cite{rev}. The general trend is
that the truncation of the gradient expansion at the local
potential approximation (zeroth-order) or at the wave function
renormalization constant level (first order) yields approximately
$10\%$ or $3\%$ difference in the critical exponents compared with
Monte-Carlo simulations, seven-loop computations in fixed dimensions or
the fifth-order results of the expilon-espansion.

\subsection{Global RG}\label{grgs}
The usual application of the RG method can be called local because
as in the determination of the critical exponent it is performed
around a point in the space of coupling constants. Models with
more than one scale may visit several scaling regimes as the
observational scale changes between the U.V. and the IR fixed
points. The determination of the 'important' coupling constants of
such models which parametrize the physics goes beyond the
one-by-one, local analysis of the scaling regimes. It requires
the careful study of crossovers, the overlap between the scaling
laws of different fixed points, a problem considered in this
section.

\subsubsection{RG in Statistical and High Energy Physics}
It is important to realize the similarity the way RG is used in
Statistical and High Energy Physics despite the superficial
differences. The most obvious difference is
that while the running coupling constants are introduced in Particle Physics
by means of Green functions or scattering amplitudes at a certain scale,
the parameters of Solid State Physics models are defined at the cut-off,
the latter being a finite scale parameter, eg lattice spacing. Since the
bare coupling constants characterize the strength of the physical processes
at the scale of the cut-off, the two ways of defining the scale dependence
are qualitatively similar and reproduce the same universal scaling laws.

The U.V. fixed point where the correlation length diverges in units of the
lattice spacing in Statistical Mechanics corresponds to renormalized
theory in High Energy Physics where the U.V. cut-off is removed.

There are two classification schemes of operators, one comes from
Statistical and the other from High Energy Physics: if the
coupling constant of an operator increases, stays constant or
decreases along the renormalized trajectory towards the IR the
operator is called relevant, marginal or irrelevant in Statistical
Physics. The coupling constants which can be renormalized as the
U.V. cut-off is removed in such a manner that observables converge
are called renormalizable in High Energy Physics. The important
point is these classification schemes are equivalent, in
particular the set of irrelevant operators at an U.V. fixed point
agrees with the set of non-renormalized ones.

One can easily give a simple quantitative argument in the leading order
of the perturbation expansion. First let us establish the power
counting argument about renormalizability. Suppose for the sake of simplicity
that there is only one coupling constant, $g$, playing the role
of small parameter and a physical quantity is obtained as
\be\label{pertser}
\la{\cal O}\ra=\sum_ng^nI_n
\ee
where $I_n$ is sum of loop integrals. This series gives rise the expression
\be\label{pertsera}
[I_n]=[{\cal O}]-n[g]
\ee
for the (mass) dimension of the loop integral. Let us recall that the
degree of the overall (U.V.) divergence of a loop integral is given by
its dimension. We distinguish the following cases:
\begin{itemize}
\item $[g]<0:$ The higher powers of $g$ decrease the dimension
in \eq{pertser} which is compensated for by increasing the degree of
the overall divergence of the loop integrals,
cf Eq. \eq{pertsera}. Graphs with arbitrary high degree of
divergence appear in the perturbation expansion and $g$ is called
non-renormalizable.
\item $g[g]>0:$ The higher order loop integrals are less divergent,
there are finite number of divergent graphs in the perturbation
series of any observable. The coupling constant is super-renormalizable.
\item $[g]=0:$ The maximal degree of (U.V.) divergence is finite for any
observable but there are infinitely many U.V. divergent graphs.
$g$ is a renormalizable coupling constants.
\end{itemize}
Since the IR degree of divergence of massless loop
integrals is just $-[I_n]$ the perturbation expansion of a
super-renormalizable or non-renormalizable massless model is
IR unstable (divergent) or stable (finite), respectively. The compromise
between the U.V. and the IR behaviors is attained by dimensionless,
renormalizable coupling constants. The IR stability can be realized
by a partial resummation of the perturbation expansion in massless
super-renormalizable models \cite{super}.

It is rather cumbersome but possible to prove by induction that
the definition of renormalizability outlined above according to the
overall divergence of the loop integrals remains valid in every
order of the perturbation expansion.

In order to separate off the trivial scale dependence one usually
removes the classical dimension of the coupling constants by means
of the cut-off in Statistical Physics. On the tree-level, in the
absence of fluctuations classical dimensional analysis applies
giving the relation $g=k^{[g]}\tilde g$ between the dimensional
and dimensionless coupling constants, $g$ and $\tilde g$,
respectively. Assuming that there is no evolution on the
tree-level we have the scaling law \be\label{clscla} \tilde
g(k)=\left({k\over\Lambda}\right)^{-[g]+\ord{\hbar}}\tilde
g(\Lambda) \ee for the dimensionless coupling constants, showing
that the non-renormalizability of a coupling constant, $[g]<0$ is
indeed equivalent to irrelevance, the decrease of the coupling
constant towards the IR direction. Higher loop contributions do
not change this conclusion so long as the loop expansion is
convergent and the anomalous dimension, the contribution
$\ord{\hbar}$ in the exponent can not overturn the sign of the
classical dimension $[g]$. In case of a marginal coupling
constant, $[g]=0$ on has to carry on with the loop expansion until
the first non-vanishing contribution to the anomalous dimension.

\begin{figure}
\centerline{\psfig{file=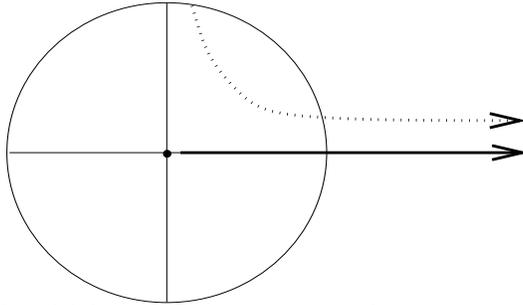,height=4cm,width=7cm,angle=0}}
\caption{An U.V. fixed point and its vicinity. The $x$ and the $y$
axis correspond a relevant and an irrelevant operator, respectively.
The circle denotes the vicinity of the fixed point where the blocking
relation is linearizable.\label{fixp}}
\end{figure}

One can construct a much simpler and powerful argument for the
equivalence of the irrelevant and the non-renormalizable set of
coupling constants in the following manner. Consider the U.V. fixed
point $P$ and the region around it where the blocking relations
can be diagonalized, as depicted in Fig. \ref{fixp}. The solid
line shows a renormalized trajectory of a model which contains
relevant operator only at the cut-off. The Lagrangian of another
model whose trajectory is shown by the dotted line contains an
irrelevant operator, as well. The difference between the two
models becomes small as we move towards the IR direction and the
physics around the end of the U.V. scaling regime is more
independent on the initial value of the irrelevant coupling
constants\footnote{More precisely all the irrelevant coupling
constants can modify is an overall scale.} longer the U.V. regime
is. This is what is called universality of the long range, low
energy phenomena. By looking "backwards" and increasing the
cut-off energy, as done in Particle Physics, the non-vanishing
irrelevant coupling constants explode and the trajectory is
deflected from the fixed point. As a result we cannot maneuver
ourselves into the U.V. fixed point in the presence of irrelevant
operators in the theory. Since the infinite value of the cut-off
corresponds to the renormalized theory, represented by the fixed
point, the irrelevant operators are non-renormalizable. Despite its
simplicity, this argument is valid without invoking any small
parameter to use in the construction of the perturbation
expansion.

Renormalizable models were thought to be essential in High Energy
Physics. These models in which a sufficiently large class of observables can
be made convergent are distinguished by their simplicity,
their interactions exist without any cut-off. Examples are Yang-Mills models in
four dimensions and the sine-Gordon, Thirring and the X-Y models in two
dimensions. The non-renormalizable models are those whose existence
requires a finite value of the U.V. cut-off, like QED, the $\phi^4$ model
in four dimensions, the Standard Model and models in Solid State
Physics. It is believed that asymptotically free models are renormalizable
only.

The traditional reason to discard non-renormalizable models was
their weak predictive power due to the infinitely many
renormalization conditions they require. Since we can never be
sure what kind of Lagrangian to extrapolate up to high energies we
need another point of view. According to the universality scenario
the non-renormalizable models can be excluded because they cannot
produce anything different from renormalizable theories. The
cut-off independent physics of the latter is parametrized by means
of renormalized coupling constants. But the subtle point to study
below is that this reasoning assumes the presence of a single
scaling regime with non-trivial scaling laws in the theory which
is a rather unrealistic feature \cite{glrg}.

Thus renormalizability is the requirement of a simple extrapolation to
the U.V. regime without encountering "new physics". The evolution of
High Energy Physics shows that this is a rather unrealistic assumption,
any model with such a feature can be an approximation at best, to study
a given interaction and to sacrify the rest for the sake of
simplicity. The goals are less ambitious in Statistical Physics
and apart of the case of second order phase transitions the
renormalizability of models is required seldom, for convenience only,
not to carry the regulator through the computation. All models in
Solid State Physics are effective ones given with a physically
motivated cut-off.

\begin{figure}
\centerline{\psfig{file=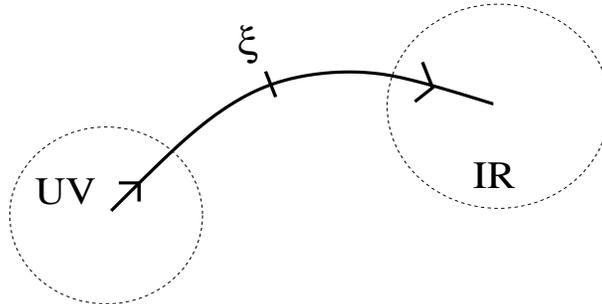,height=4cm,width=8.cm,angle=0}}
\caption{The RG flow of a model with massive particles. The
renormalized trajectory connects the U.V. and IR scaling regimes,
indicated by circles which are separated by a crossover at the intrinsic
scale of the model, $p\approx1/\xi$.\label{scal}}
\end{figure}

Let us start the discussion of the possible effect of the co-existence
of several scaling regimes along the RG flow with the simplest case,
a model with a gap in the
excitation spectrum above the ground state, ie with finite correlation
length, $\xi<\infty$. Suppose that the RG flow of the model starts in the
vicinity of an U.V. fixed point and reaches an IR fixed point region as shown
in Fig. \ref{scal}. Universality, the parametrizability of
the physics beyond the U.V. scaling regime by the relevant coupling constants
of the U.V. scaling laws, tacitly assumes the absence of any new
relevant coupling constants as we move along the renormalized
trajectory towards the IR, as shown in Fig. \ref{fixp}. This
assumption is in fact correct for massive models whose IR scaling
laws are trivial, meaning that that only the Gaussian mass term is
relevant. To see why let us
consider the RG flow in the IR side of the crossover, when the
U.V. cut-off, say the lattice spacing $a>>\xi$ and make a blocking
step, $a\to a'$ which generates the change $g_n(a)\to g_n(a')$
in the coupling constants. The change $\Delta g_n(a)=g_n(a')-g_n(a)$
is due to fluctuation modes whose characteristic length is $a<\ell<a'$.
Owing to the inequality $\xi<<a<\ell$ these fluctuations are suppressed
by $e^{-a/\xi}$ and the flow slows down, $g_n(a)\approx g_n(a')$.
This implies the absence of run-away trajectories, the absence
of relevant non-Gaussian operator. But more realistic models with
massless particles or with dynamical or spontaneous symmetry breaking
occurring at finite or infinite length scales, respectively, or with
condensate in the ground state the IR scaling may generate new relevant
operators and universality, as stated in the introduction, referring
to a single scaling regime is not a useful concept any more.

\subsubsection{Scalar model}\label{singli}
In order to understand better the generic case shown in Fig. \ref{scal}
we return to its simplest realization, the scalar model in the local potential
approximation, Eq. \eq{wh}, \cite{glrg}. One does not expect anything
unusual here since the model possesses a gap and the
IR scaling regime is trivial in either the symmetrical or symmetry broken
phase. But we shall see that the model has soft large amplitude
fluctuations and the true vacuum is just there where this
non-perturbative and the well known perturbative domains join.
The second derivative of the local potential
is discontinuous there, depends on the direction from which
the vacuum is approached. We shall arrive at some understanding
of the soft modes in two step, by separating the loop contributions from
the tree-level structure. We start with the more traditional
loop contributions.

To facilitate the dealing with soft modes we introduce an external
constraint which controls
the expectation value $\int_x\la\phi_x\ra_j=V\Phi$.
The usual realization of this constraint is the introduction of an
external source as in Eq. \eq{cggf}.
We shall consider the model in the symmetry broken phase where
$\int_x\la\phi_x\ra_0=V\Phi_\mr{vac}\not=0$
and choose homogeneous external source $j_x=J$ which can destabilize
the naive symmetry broken vacuum whenever $J\Phi_\mr{vac}<0$ and
can induce $|\Phi|<\Phi_\mr{vac}$. Note that such a control of
the field expectation value by an external source preserves the
locality of the action.

\begin{figure}
\centerline{\psfig{file=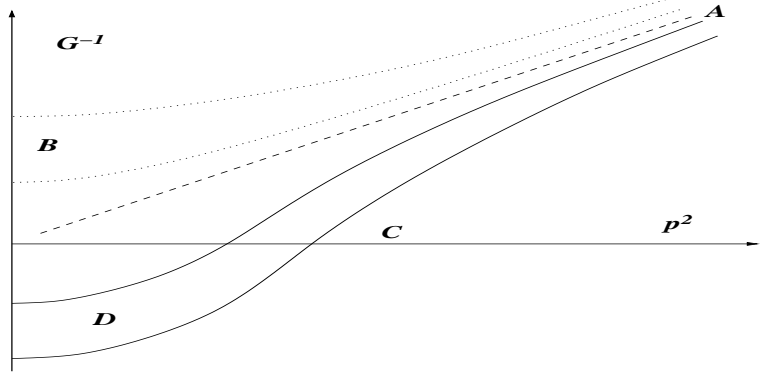,height=5cm,width=10.cm,angle=0}}
\caption{The inverse propagator $G^{-1}(p^2)$ of the scalar model as the
function of the momentum square $p^2$ for $\Phi=0$.
The dotted and solid lines correspond to the symmetrical and the
symmetry broken phase, respectively. The massless case is
shown by dashed line. One can distinguish four
different scaling regimes, namely A: U.V., B: symmetrical massive,
C: precursor of condensation and D: spinodial phase separation regime.
\label{ipro}}
\end{figure}

The beta-functions, as given by Eq. \eq{betapol}, depend strongly
on the propagator $G(k^2)=1/(k^2+g_2(k))$, represented by the the internal
lines in Figs. \ref{betf}. The qualitative features of the function
$G^{-1}(k^2)$ are shown in Fig. \ref{ipro} for $\Phi=0$.
According to the loop expansion
\be\label{loopegm}
G^{-1}(k^2)=k^2+U^{(2)}_k(\Phi)=\cases{
k^2+U^{(2)}_\Lambda(\Phi)+\ord{\hbar}&$U^{(2)}_k<<k^2$,\cr
k^2+U^{(2)}_0(\Phi)+\ord{\hbar}&$k^2<<U^{(2)}_k$,}
\ee
the bending of the lines in the figure is due to radiative corrections,
as far as the perturbative region, $G^{-1}>>1$ is concerned. The dashed line
corresponds to the massless Coleman-Weinberg case \cite{cowe}. The flow
above or below this separatrix is in the symmetrical or symmetry broken
phase, respectively. This qualitative picture is valid so long
$|\Phi|<\Phi_\mr{vac}$.

It is important to recall that the argument of the logarithmic
function in the WH equation in Eq. \eq{wh}, $G^{-1}(k^2)$,  is
the curvature of the action at the cut-off and serves as a
measure of the restoring force driving fluctuations back to their
trivial equilibrium position. Thus perturbative treatment is reliable
and the loop contributions are calculable so long
$\tilde G^{-1}(k^2)=G^{-1}(k^2)/k^2>>\tilde g_4$.

In order to have a qualitative ides if what is happening let us consider
the model where $U_\Lambda(\Phi)=m^2\phi^2/2+g\phi^4/4!$ at the cut-off,
$k=\Lambda$. By simply ignoring the loop corrections we have
$g_2(k)=m^2$, $g_4(k)=g$ and $g_n(k)=0$ for $n>4$ on the tree-level,
ie $m^2$ and $g$ can be identified with renormalized parameters
and $m^2<0$. For an isolated system ($j=0$) where
$|\Phi|=\Phi_\mr{vac}=-6m^2/g$  and the use of the perturbation expansion
appears to be justified for $\tilde G^{-1}(k^2)=1-m^2/k^2>g_4k^{d-4}$.
Spontaneous symmetry breaking, or the appearance of a condensate in
the ground state in general, is characterized by instability of fluctuations
around the trivial, vanishing saddle point. When an external
source is coupled to $\phi_x$ to dial $|\Phi|<\Phi_\mr{vac}$
in the symmetry broken phase then non-perturbative effects set in
as $k$ decreases. We can see this easily when the field expectation value
is squeezed in the concave part of the potential,
$|\Phi|<\Phi_\mr{infl}=-2m^2/g=\Phi_\mr{vac}/3$. In fact,
$G^{-1}$ approaches zero and the amplitude of the fluctuations explodes
at $k=\sqrt{-m^2-g\Phi^2/2}$. Below this non-perturbative regime
the plane wave modes become unstable since $G^{-1}(k^2)<0$
and a coherent state is formed, reflected in the appearance of a
non-homogeneous saddle point with characteristic scale
$1/\sqrt{-m^2-g\Phi^2/2}$

The mechanism responsible of spreading the instability over the whole region
$|\Phi|<\Phi_\mr{vac}$ comes from the tree-level structure and will be
discussed later, in section \ref{tree}.

We identify at this point four different scaling regimes:

\begin{itemize}
\item[A:] In the U.V. scaling regime $k^2\gg g_2(k)$:
$G_n\approx g_n/k^2$, the mass term is negligible.

\item[B:] The explicit scale dependence disappears in the IR regime
$k^2\ll g_2(k)$ of the symmetrical phase, where $G_n\approx g_n/g_2$.
This is a trivial scaling, there are no interactive relevant operators.

\item[C:] The onset of the condensation is at $G^{-1}(k^2)=k^2+g_2(k)\approx0$
and the higher loop effects start to dominate.

\item[D:] The spinodal unstable region where the homogeneous ground state
becomes unstable against infinitesimal fluctuations, $k^2+g_2(k)<0$.
The leading scaling laws are coming from tree-level, classical physics.
\end{itemize}

There are two crossover regimes. One is between the scaling regimes $A$ and
$B$ at $k^2\approx g_2(k)$. Another one is between $A$ and $C$
when $|\Phi|<\Phi_\mr{infl}$. This will be extended to
the region $|\Phi|<\Phi_\mr{vac}$ by tree-level contributions. We shall
look into the crossover of the symmetry broken phase.

\begin{figure}
\centerline{\psfig{file=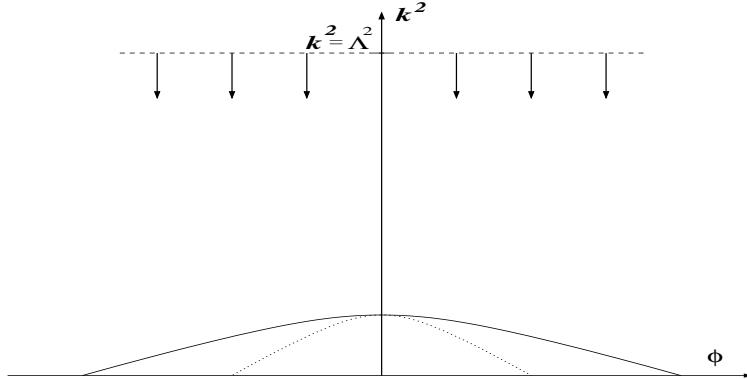,height=5cm,width=10.cm,angle=0}}
\caption{The $(\Phi,k^2)$ plane where the local potential
$U_k(\Phi)$ is defined. The potential should be imagined as
a surface above the this plan. The initial condition gives
$U_\Lambda(\Phi)$ on the horizontal dashed line and the
RG equation is integrated from this line in the direction of the arrows.
The extension of the minimum and the inflection point of the
potential from $k^2=0$ to $k^2>0$ is shown by solid and
dotted lines, respectively. \label{phik}}
\end{figure}

The best is to follow the evolution of the local potential the plane $(\Phi,k^2)$,
as shown in Fig. \ref{phik}. The potential $U_k(\Phi)$ should be imagined as a
surface above the plane  $(\Phi,k^2)$. The RG equations for the
dimensionless coupling constants, $\dot{\tilde g}_n=\tilde\beta_n$
where $\tilde\beta_n$ is given by Eq. \eq{dlbefunct}
are integrated in the direction of the arrows. The initial condition
$\tilde U_\Lambda(\tilde\phi)=m^2\tilde\phi^2/2k^2+gk^{d-4}\tilde\phi^4/4!$,
set at $k=\Lambda$, along the horizontal line on the top. As the running cut-off $k$
is decreased and the RG equations are integrated the potential
becomes known along horizontal line intercepting the
ordinate at $k^2$. The spontaneous breaking of the symmetry in the vacuum
implies that for sufficiently small $|\Phi|$ the propagator explodes
by approaching a singularity at $k=k_\mr{cr}(\Phi)$ as $k$ decreases.
This happens first at $\Phi=0$ when the running horizontal line touches
the two curves shown in the Figure. These curves correspond to the generalization of the
minimum and the inflection point of the potential for $k^2>0$ and are given by the relations
$k^2+m^2_B+g\Phi^2_\mr{min}/6+\ord{\hbar}=0$ and
$k^2+m^2_B+g\Phi^2_\mr{infl}/2+\ord{\hbar}=0$, respectively. We shall argue in
section \ref{tree} that the singularity lies close to the extension of the
minimum,
\be\label{crline}
k_\mr{cr}(\Phi_\mr{min}(k))\approx k.
\ee

The numerical integration of the WH equation requires a truncation of the
summation in Eq. \eq{locpot} for the local potential. By using
$n_\mr{max}\le22$ the singular line can be located with
reasonable accuracy. Close
to the singular line on the plane $(\Phi,k^2)$ the coupling
constants increase rapidly and the truncation of the potential
is not acceptable. In the U.V. scaling regime (denoted by A in
Fig. \ref{ipro}) the propagator is
small, $G=\ord{k^2}$, and the first term on the right hand side of
Eq. \eq{betapol} is dominant. As we approach $k_\mr{cr}$ the
propagator starts to increase, all term contribute equally in the
beta functions and we enter into a new scaling regime (C in Fig. \ref{ipro}).

In order to connect the two scaling regimes we shall consider
$f_{m,n}(k/\Lambda)=\partial\tilde\beta_{m,k}(\phi)/\partial\tilde
g_{\Lambda,n}$,
the dependence of the beta functions at $k$ on the microscopical
initial condition, imposed at $k=\Lambda$. The simplest way to estimate
$f_{m,n}(x)$ is to start with $f_{m,n}(1)\approx1$
which can be inferred from the form \eq{betapol}. The scaling law
\eq{clscla} allows us to write
\be\label{univsc}
f_{m,n}\left({k\over\Lambda}\right)
={\partial\tilde\beta_m(\phi)\over\partial\tilde g_n(\Lambda)}
={\partial\tilde\beta_m(\phi)\over\partial\tilde g_n(k)}
\left({k\over\Lambda}\right)^{-[g_n]}
=\ord{\left({k\over\Lambda}\right)^{-[g_n]}}
\ee
up to corrections $\ord{\hbar}$, confirming the universality
of the U.V. scaling regime. In fact, initial, microscopical
value of the the irrelevant, non-renormalizable coupling constants
($[g]<0$) leave vanishing trace on the dynamics as $\Lambda/k\to\infty$.

The numerical determination of the function $f_{m,n}(x)$ in $d=4$ supports
the prediction \eq{univsc} in the U.V. scaling regime,
namely the impact of a non-renormalizable coupling constants $g_n(\Lambda)$
on the beta functions weakens as $(k/\Lambda)^{-[g_n]}$ for $x\approx1$.
But this trend changes for $x<1$ as the critical value of the cut-off,
$k_\mr{cr}$, found to be in agreement with Eq. \eq{crline},
is approached in what $f$ starts to increase. The potential must not
be truncated close to the singularity and the numerical results
appear inconclusive. But the increase of $f_{m,n}(x)$ at the crossover
actually starts already far enough from $k_\mr{cr}$ when all couplings
are small, $\tilde g_n<<1$ and the truncation is safe. The
lesson of this numerical result is that universality as used around
the U.V. scaling regime alone is not valid anymore. The instability at the
onset of the condensation at $k\approx k_\mr{cr}$ introduces
divergences which are strong enough to overwrite the
U.V. scaling laws and generate new relevant operators. Such an
instability of the RG flow enhances the sensitivity of the
physics at finite length scales on the microscopical parameters.

The need of generalizing universality in a global manner has been
shown for $|\Phi|<\Phi_\mr{vac}$. It remains to see if such a
phenomenon can be observed at the highly singular point
$|\Phi|=\Phi_\mr{vac}$, in the true vacuum.

A similar phenomenon has already been noticed in connection with
the BCS ground state. It has been pointed out that kinematical factors
turn the four electron operator which is irrelevant according to
the power counting into marginal for processes close to the Fermi
level \cite{bcsrgp}, \cite{bcsrgw} \cite{bcsrgs}.
The collinear divergences at the Fermi level
drive the instability of the non-condensed vacuum and generate
new scaling laws. Another similar mechanism is the origin of
the strong long range correlations in the
vacuum state of Yang-Mills models. They appear as a consequence of
non-renormalizable, U.V. irrelevant Haar-measure term of the
path integral \cite{haarons}.

We note finally that the view of the RG flow as a parallel transport,
mentioned in section \ref{compops} is particularly well suited for
the studies of crossovers. This is because the sensitivity matrix
\eq{sensm} is a global quantity displaying clearly the sensitivity
of the IR physics on the U.V. parameters by construction.

\subsubsection{RG microscope}
We now embark on a rather speculative subject, playing with the
possibility of matching different scaling laws. Let us start with
the generic case, a model with U.V. and IR scaling regimes,
separated by a crossover, as sketched in Fig. \ref{scal}. We have
a classification of the very same operator algebra at both scaling
regimes. One can write any local operator $A$ as a linear
superposition of scaling operators, $A=\sum_nc_{A,n}{\cal O}_n$,
the latter being the eigen-operators of the linearized blocking
relations of a scaling regime, ${\cal
O}_n(k)\approx(\Lambda/k)^\nu_n{\cal O}(\Lambda)_n$. Let us denote
by $\nu_A$ the largest scaling dimension of scaling operators
which occur in the a linear decomposition of $A$,
$\nu_A=\max_n\nu_n$. We shall simplify matters by calling a local
$A$ operator relevant, marginal or irrelevant if $\nu_A>0$,
$\nu_A=0$ or $\nu_A<0$, respectively. We ignore marginal case by
assuming that radiative corrections always generate non-vanishing
scaling dimensions.

\begin{figure}
\centerline{
\begin{minipage}{4.4cm}
\psfig{file=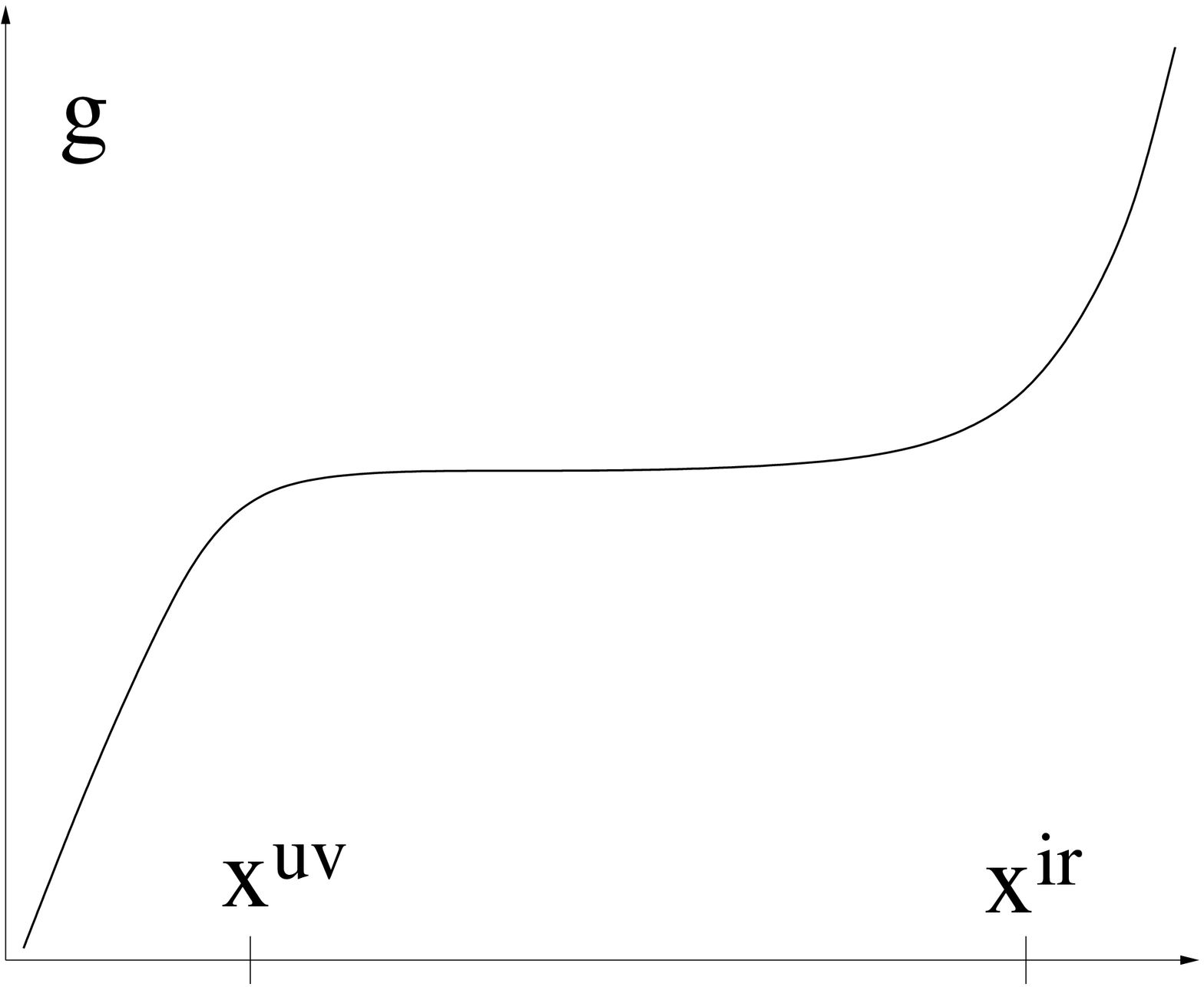,height=3cm,width=3.8cm} \centerline{(a)}
\end{minipage}
\begin{minipage}{4.4cm}
\psfig{file=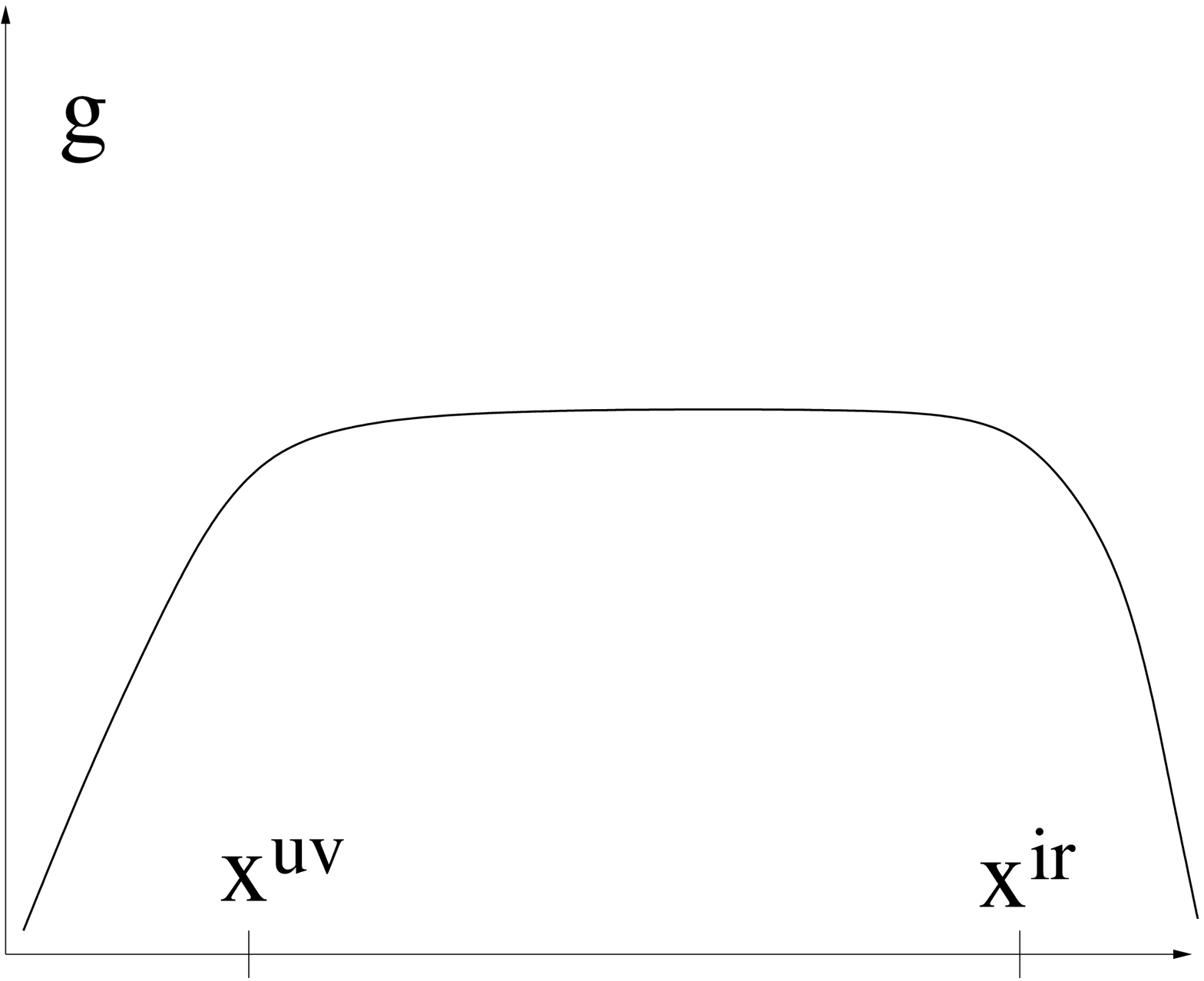,height=3cm,width=3.8cm}
\centerline{(b)}
\end{minipage}
}
%
\centerline{
\begin{minipage}{4.4cm}
\psfig{file=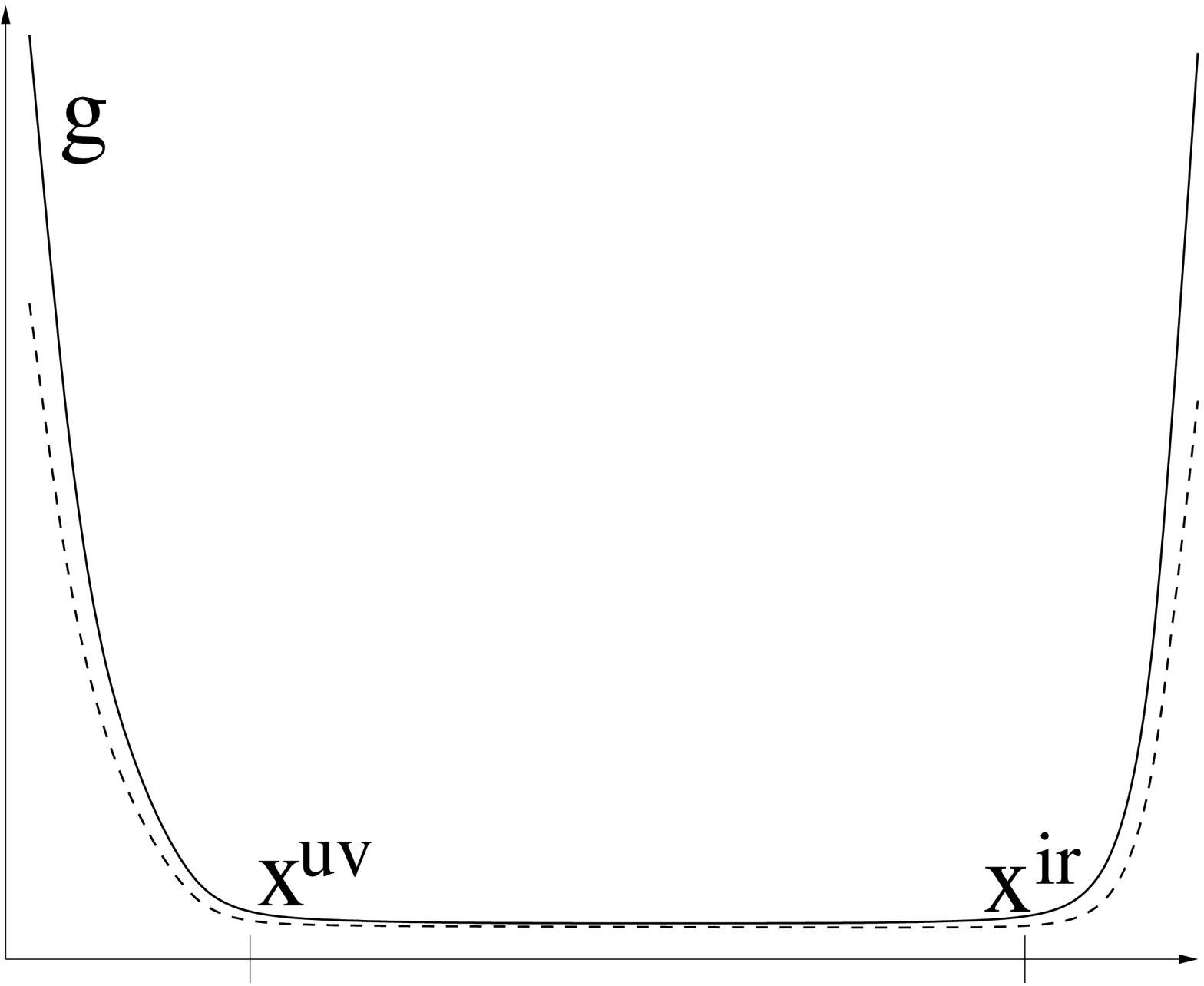,height=3cm,width=3.8cm}
\centerline{(c)}
\end{minipage}
\begin{minipage}{4.4cm}
\psfig{file=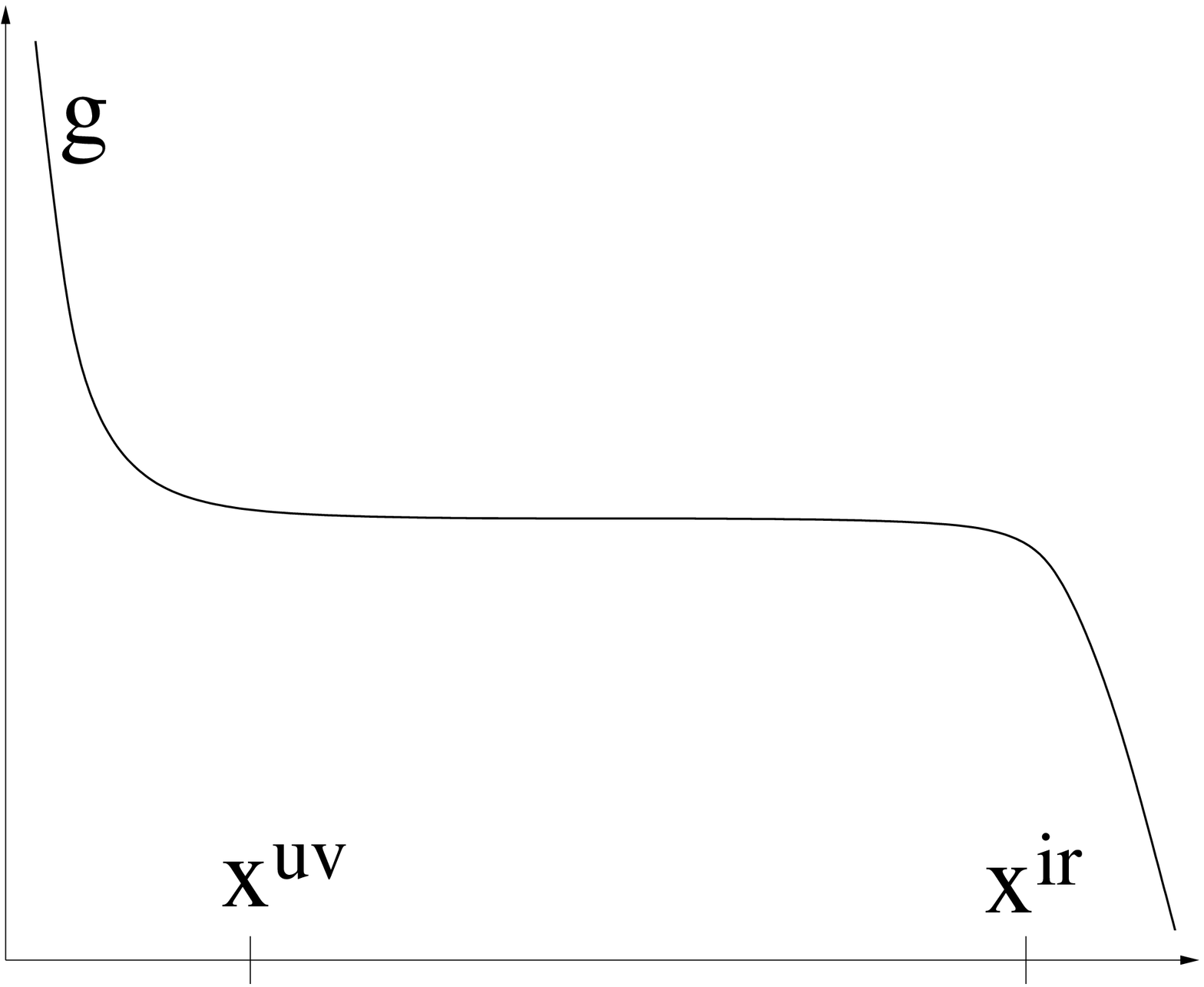,height=3cm,width=3.8cm}
\centerline{(d)}
\end{minipage}}
\caption{The qualitative dependence of the four types of coupling
constants on the observational length scale. The U.V. and the IR
scaling regimes are $0<x<x^{uv}$ and $x^{ir}<x$, respectively. The
dashed line in (c) corresponds to a different initial condition
for the RG flow.\label{gensc}}
\end{figure}

One can distinguish between the following four cases:
\begin{itemize}
\item[(a)] $r_\mr{U.V.}r_\mr{IR}$: relevant in both regimes,
\item[(b)] $r_\mr{U.V.}i_\mr{IR}$: relevant in the U.V., irrelevant in
the IR, \item[(c)] $i_\mr{U.V.}r_\mr{IR}$: irrelevant in the U.V.,
relevant in the IR, \item[(d)] $i_\mr{U.V.}i_\mr{IR}$: irrelevant in
both regimes.
\end{itemize}
The qualitative dependence of the corresponding coupling constants
is shown in Figs. \ref{gensc}.
%

Let us consider for example QED containing electrons, muons and ions
with metallic density. There is an U.V. scaling regime for energies
well above the electron mass.
Suppose that the ground state is a BCS type superconductor where the IR
scaling laws are driven by long range phonon mediated interactions.
The electron mass is an example of the class (a). The muon mass belongs
to class (b) owing to the decoupling of muons at low energy. By assuming
that radiative corrections turn the four fermion vertex relevant
it serves as an example of class (c). Finally the six fermion vertex is
doubly irrelevant and belongs to class (d).

The interesting class is (c). Suppose that we modify the values of
the bare coupling constants at the initial condition and find the
dashed line. Due to the universality of the U.V. scaling regime the two
renormalized trajectories converge towards each other for $0<x<x^{uv}$
and run very close between the two scaling regimes, $x^{uv}<x<x^{ir}$.
But in the IR scaling regime where the coupling constant is
relevant the two trajectories start to diverge from each other.
The question is whether the difference of the trajectories
in the U.V. scaling regime has any impact on the difference found
at the IR. It may happen that the IR phenomena is independent of the U.V. regime
and the crossover smear completely out the small effects of the U.V.
initial condition of the trajectory. But one can imagine that
the extremely small differences
left at the crossover remain important and lead to an initial condition
dependent divergence at the IR. The answer is model dependent
and can be given by the detailed analysis of the set of coupled
differential equations.

In order to make this point clearer let us imagine a model
with two parameters, say a mass $m$ and a coupling constant $g$,
both expressed in a dimensionless manner by means of the U.V. cut-off,
the lattice spacing $a$. Suppose the following rather simple scaling laws:
The beta function of the mass is approximately constant,
\be
\gamma(m,g)={a\over m}{\partial m\over\partial a}\approx\beta_m>0
\ee
and of the coupling constant $g$ can be written as
\be
\beta(m,g)=a{\partial g\over\partial a}=\left[\chi\left({m\over m_{cr}}\right)\nu_{U.V.}
+\chi\left({m_ {cr}\over m}\right)\nu_{I.R.}\right]g
\ee
where $\chi(z)$ is interpolating smoothly between
$\chi(0)=1$ and  $\chi(\infty)=0$. The running mass is given by
\be
m(a)=m(a_0)\left({a\over a_0}\right)^{\beta_m},
\ee
and  the length scale of the crossover between the U.V. and the
I.R. scaling regimes is at $a_{cr}\approx a_{U.V.}(m_{cr}/m_{U.V.})^{1/\beta_m}$.
The asymptotical scaling for the coupling constant is
\bea
g(a_{U.V.})&\approx c_{U.V.}g(a_{cr})\left({a_{U.V.}\over a_{cr}}\right)^{-\nu_{U.V.}},~~~
a_{U.V.}&\ll a_{cr},\nonu
g(a_{I.R.})&\approx c_{I.R.}g(a_{cr})\left({a_{I.R.}\over a_{cr}}\right)^{\nu_{I.R.}},~~~
a_{I.R.}&\gg a_{cr}.
\eea
Let us furthermore assume that $\nu_{U.V.}<0$ and $\nu_{I.R.}>0$, i.e. this coupling
constant belongs to the class (c). The sensitivity of the coupling constant in the
asymptotical regions, $g(a_{U.V.})$  and $g(a_{I.R.})$ on $g(a_{cr})$  is
\bea\label{sensit}
{\partial g(a_{U.V.})\over\partial g(a_{cr})}&\approx&c_{U.V.}\left({a_{U.V.}\over a_{cr}}\right)^{-\nu_{U.V.}},\nonu
{\partial g(a_{I.R.})\over\partial g(a_{cr})}&\approx&c_{I.R.}\left({a_{I.R.}\over a_{cr}}\right)^{\nu_{I.R.}}
\eea
therefore
\be\label{ratio}
{\partial g(a_{I.R.})\over\partial g(a_{U.V.})}\approx{c_{I.R.}\over c_{U.V.}}
{a_{I.R.}^{\nu_{I.R.}}a_{U.V.}^{\nu_{U.V.}}\over a_{cr}^{\nu_{I.R.}+\nu_{U.V.}}}.
\ee
If the IR scaling regime is long enough then the difference between renormalized
trajectories with different initial conditions imposed in the U.V. regime
can be as large or even larger than at the initial condition. When the
U.V. and the I.R. observational scales are related by
\be
a_{U.V.}\approx a_{cr}\left({a_{cr}\over a_{I.R.}}\right)^{-\nu_{I.R.}/\nu_{U.V.}}.
\ee
then \eq{ratio} is $\ord{a^0}c_{I.R.}/c_{U.V.}$ and despite the 'focusing' of
the universality in the U.V. regime
the 'divergence' of the trajectories in the I.R. can amplify the extremely
weak dependence on the initial condition at finite scales.
The increase of $a_{I.R.}$ in this case enhances the sensitivity
on the initial condition and we can 'see' the initial value of the
non-renormalizable coupling constants at smaller distance $a$.

The lesson of the numerical results mentioned above is that the
scalar theory in the unstable regime realizes such a RG
'microscope'. The loss of the U.V. based universality as we approach
the critical line on the plane $(\Phi,k^2)$ suggests that there is
at least one relevant operator at the new scaling laws which is
irrelevant in the U.V.. We do not know the eigen-operators of the
linearized blocking relations in this region but this new relevant
operator must contain the local monomials $\phi_x^n$ with $n>4$.
To make it more difficult, it is not obvious that this is a local
operator as opposed to the scaling operators of an U.V. fixed point.

In order to parametrize the physics of the scalar model
we have to use this new coupling constant
at the U.V. cut-off, as an {\em additional free parameter}.
This is rather un-practical not only because we do not know
the operator in question but mainly due to the smallness
of this coupling constants in the U.V. and the crossover region.
It seems more reasonable to use a mixed parametrization,
consisting of the renormalizable coupling constants at the U.V. cut-off and the
new coupling constant taken close to the singularity where
it has large enough value, ie $k\approx k_\mr{cr}$. In the scalar model
$k_\mr{cr}\to0$ as the external source is turned off.
We shall call the new coupling constants appearing in this manner
hidden parameters.

What operator corresponds to the hidden
parameter of the scalar model? It seems reasonable to assume that
the only observables whose value is determined at the onset of the
condensate is just the magnitude of the condensate.
Therefore the conjecture is that the strength of the condensate,
$\Phi_{vac}$, is dynamically independent of the renormalizable
parameters $g_2$, $g_4$ in $d=4$. We have therefore the following
possibilities in parametrizing the scalar model with spontaneously
broken symmetry:
\begin{itemize}
\item We identify a non-renormalizable coupling constant, $g_{nr}$, which influences
the value of the condensate and use the bare values of
$g_2(\Lambda),g_4(\Lambda),g_{nr}(\Lambda)$ at the
U.V cut-off.
\item One uses the renormalizable bare coupling constants from the U.V.
end and the strength of the condensate from the I.R. regime,
$g_2(\Lambda),g_4(\Lambda),\Phi_{vac}$.
\item One may use I.R. quantities only, say $g_2(0),g_4(0),\Phi_{vac}$.
\end{itemize}

Our point is that the scalar model with spontaneously broken
symmetry has three free parameters instead of two. This rather surprising
conjecture can be neither supported nor excluded by perturbation expansion
because one can not connect the U.V. and the I.R. regimes in a reliable manner.
The numerical studies of the scalar model carried out so far
are inconclusive, as well, because they were constrained to
the quartic potential and the possible importance of higher order
vertices in forming the condensate was not considered.

The modification of the scaling laws of the scalar model,
necessary to generate a hidden parameter comes from the
onset of a condensation. As discussed below in Section
\ref{tree} there are non-trivial tree-level
scaling laws in a condensate which overwrite the loop-generated
beta functions and may provide the dynamical origin of the
hidden parameter. In a similar manner one can speculate about
the role of the Cooper-pair condensate in the BCS vacuum.
If there turns out to be a hidden parameter as in the
one-component scalar model then the non-renormalizable effective
coupling constants of the Standard Model would influence the
supercurrent density in ordinary metals! Another possible example
is the Higgs sector of the Standard Model, where the hidden parameter,
if exists, would be an additional free parameter of the model.

One may go further and inquire if dynamical symmetry breaking
can modify the scaling law in a similar manner. The bound state
formation, responsible to the generation of large anomalous
dimensions in strong coupling QED \cite{scqed}, in strong extended
technicolor scenario \cite{wtechn,mir}, in the Nambu-Jona-Lasinio
model \cite{njl} or in the top-quark condensate mechanism
\cite{tquark,mir} may be a source of hidden parameters, as well.

Hidden parameters represent an unexpected 'coupling' between
phenomena with very different scales and question of our traditional
strategy to understand a complex system by analyzing its
constituents first. The axial \cite{axan} or scale \cite{scan}
anomalies represent a well know problem of this sort, except that
the regulator independence of the anomaly suggests that for each
classical IR fixed point there is a single 'anomalous' one without
continuous fine tuning.

Universality is expressed in the context of High Energy Physics
by the decoupling theorem \cite{decoupl}. Let us start with a renormalizable
model containing a heavy and a light particle and consider the effective
theory for the light particle obtained by eliminating the heavy one.
There are two possible classification schemes for the effective vertices
for the light particle, generated by the elimination process. One
is according to the U.V. scaling laws in the effective theory, i.e.
there are renormalizable and non-renormalizable vertices.
Another scheme is based on the strength of the effective
interactions, i.e. there are vertices which are stay constant or
diverge when the ratio of the light and the heavy particle mass tends to zero
and there are vertices whose coefficients tend to zero in the same limit.
The decoupling theorem asserts that these two classification schemes
are equivalent, i.e. the non-renormalizable effective coupling constants
are vanishing when the heavy mass diverges. The existence of the
hidden parameter is, at the final count is a violation of this theorem.

\subsubsection{The Theory of Everything}
\begin{figure}
\centerline{\psfig{file=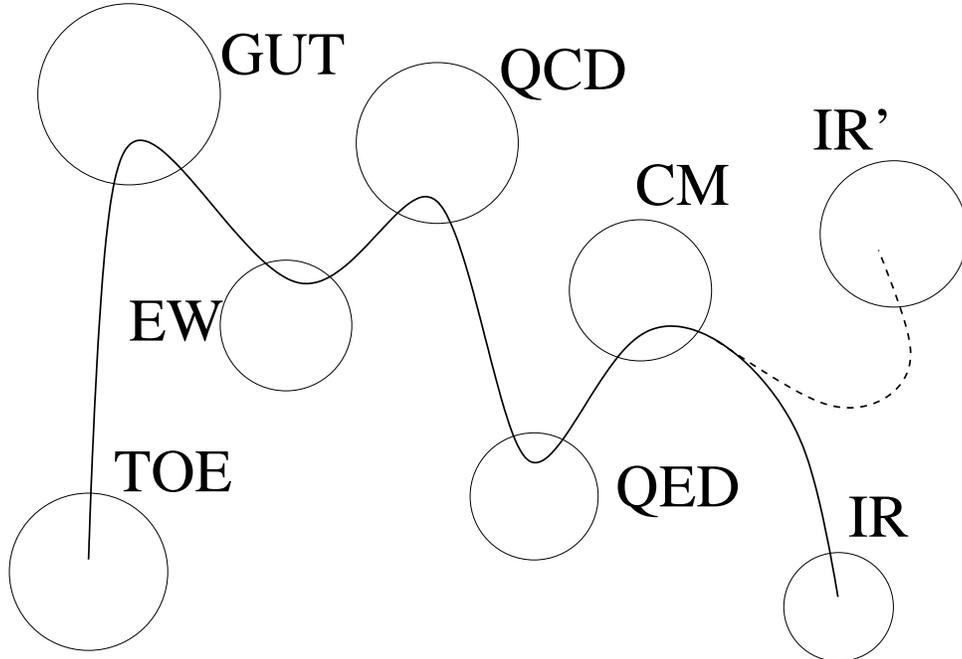,height=8.8cm,width=12.8cm}}
\caption{The renormalization group flow of the Theory of Everything.
The branching drawn at an IR scale is a phase transition driven by
the environment, such as particle or heat reservoirs.\label{toe}}
\end{figure}

The appearance of a hidden coupling constant renders our goal of
understanding the hierarchy of interactions in Nature
extremely complicated. In fact, the RG flow
of the Theory of Everything must be imagined in the space
of all coupling constants in Physics, including the parameters
 of the Grand Unified Models down to different models in Solid
State Physics or hydrodynamics in the classical regime, as indicated
in Fig. \ref{toe}. Different fixed points are approached by the
trajectory when the scale dependence is dominated by a single interaction.
But all fixed point except the U.V. one is avoided due to the
non-renormalizable, irrelevant vertices \cite{decoupl} generated by the
dynamics of the next fixed point in the U.V. direction.

Such a wandering in the space of coupling constants echoes
the age old disagreement between High Energy and Solid State Physics.
It is usually taken for granted in the High Energy Physics
community that the sufficiently precise determination of the
microscopical parameters of the Theory of Everything would
'fix' the physics at lower energies. The obvious difficulties
to extract the high energy parameters from experiments which
render each new experiment expensive, long and large scale
operation indicate that this might not be a practical direction to follow.
In other words, the relevant coupling constants have positive
Lyapunov exponent and render the trajectory extremely sensitive on the
initial conditions. Therefore the characterization of the
RG flow of The Theory of Everything by the initial condition, though
being possible mathematically, is not practical due to the finite resolution
of the measurements. The other side of the coin, the physical
parameters of a fixed point are clearly the relevant coupling
constants of the given scaling regime. They are the object of
Solid State Physics, as far as the scaling regimes
QED, CM and IR are concerned in Fig. \ref{toe}. One looses sight
of the fundamental laws but gains predictive power by restricting
oneself to local studies of the RG flow.

\subsection{Instability induced renormalization}\label{tree}
In the traditional applications of the RG method which are based on
the perturbation expansion the RG trajectory is sought in the
vicinity of the Gaussian fixed point. One can describe in this manner
crossovers which connect scaling regimes which share the same small
parameter. We can enlarge the classes of
accessible crossovers by relying on the saddle point expansion,
the only systematical non-perturbative approximation scheme.
The generic way to induce new crossover in this scheme
is by passing a condensation at a finite length scale \cite{mini}.

One may object that
there is nothing surprising or new in finding that the physics is
fundamentally changed by a condensation mechanism. What
additional knowledge can then be gained by looking into condensation
by the method of the RG? The point is that the classification
of operators around a fixed point which was achieved in the
framework of the perturbation expansion changes essentially by the
condensation, ie the appearance of a non-trivial saddle point.
To see how this happens we return to the strategy followed at
Eq. \eq{trba}. Let us
consider an observable $A$ in a model with a single coupling constant
$g$ for simplicity, computed in the saddle point expansion,
$\la A\ra=F_0(g,\Lambda)+\hbar F_1(g,\Lambda)+\ord{\hbar^2}$,
where $F_\ell(g,\Lambda)$ denotes the $\ell$-loop contribution.
We can obtain the beta function by taking
the derivative of this equation with respect to cut-off $\Lambda$,
\be
\beta=k\partial_kg=-{k\partial_k(F_0+\hbar F_1)\over\partial_g(F_0+\hbar F_1)}
+\ord{\hbar^2}
=-{k\partial_kF_0\over\partial_gF_0}\biggl[1+\hbar
\biggl({\partial_kF_1\over\partial_kF_0}-{\partial_gF_1\over\partial_gF_0}
\biggr)\biggr]+\ord{\hbar^2}.
\ee
The loop corrections $F_\ell(g,\Lambda)$, $\ell\ge1$ are polynomials
of the coupling constants $g$. But the leading order, tree-level
piece usually has stronger dependence on $g$ and may become
singular as $g\to0$. The tree-level contribution may induce
qualitatively new scaling laws with new set of relevant operators.
Due to the singularity at $g=0$ the condensation actually realizes
the dangerous irrelevant variable scenario \cite{dangir}.

At the onset of an instability, region C in Fig. \ref{ipro}
certain modes experience strong nonlinearity and develop large
amplitude fluctuations. But as we enter in the unstable regime
one may hope to recover quasiparticles with weak residual interaction
after having settled the dynamics of the unstable modes.
At least this is what happens
within the framework of the semiclassical, or loop expansion
where the fluctuations remain small after the proper condensate
is found. There are models where the vacuum state is just at
the edge of the instability, like the scalar model discussed
above in the symmetry broken phase.

The tree-level contributions to the blocking transformation, if
exist, are more important then the loop corrections. This is the reason
that we shall consider the tree-level, $\ord{\hbar^0}$ RG flow.
There is no evolution in this order so long the system is stable and the
saddle point is trivial. But as soon as we arrive at the unstable
region the RG flow becomes non-trivial.

\subsubsection{Unstable effective potential}
Before turning to the actual tree-level blocking let us review
briefly the kind of instabilities one expects.
As we enter into the unstable regime $k^2+g_2(k)<0$ the trivial
saddle point, $\phi_k=0$ becomes unstable in the blocking \eq{obl}.
To understand this instability better we introduce the constrained
functional integral,
\be
Z_\Phi=\int D[\phi]e^{-S_B[\phi]}\delta\left({1\over V}\int_x\phi_x-\Phi\right)
\ee
corresponding to a conserved order parameter. By following the RG
trajectory until the IR end point one eliminates all but the homogeneous
mode by keeping the partition function unchanged,
\be
Z_\Phi=e^{-S_{k=0}[\Phi]}=e^{-VU_{k=0}(\Phi)}.
\ee
To find another local representation for this constrained partition function
we consider the generator functional density $w[j]=W[j]/V$ defined by means
of Eq. \eq{cggf} and compute its Legendre transform, $V_\mr{eff}(\Phi)$
in the mean field approximation. For this end we set $j_x=J$ and write
\be\label{meanflt}
-V_\mr{eff}(\Phi)=w[J]-J\Phi=\min_{J,\phi}[-U_{k=0}(\phi)+J(\phi-\Phi)].
\ee
The minimization with respect $\phi$ and $J$ yields $J=dw(J)/dJ$ and
and $\phi=\Phi$, respectively. Finally we have $V_\mr{eff}(\Phi)=U_{k=0}(\Phi)$,
the result announced after Eq. \eq{ukvef} so long the mean field approximation
is reliable, namely in the thermodynamical limit and in the absence of large
amplitude fluctuations. It is further known that $V_{eff}(\Phi)$ is convex.

The instabilities of the kind mentioned above are well known in the
case of first order phase transitions. But similar instabilities may appear
at higher order phase transitions, as well. In fact, the magnetization
of the Ising model as the function of the external magnetic field
shows discontinuous behavior below the critical temperature.
Therefore the free energy constrained into a sector with a given
magnetization displays such kind of instabilities.

Suppose that the $U_{k=0}(\Phi)$ determined perturbatively has degenerate
minima and a concave part. Then it is advantageous to introduce
two curves on the $(\Phi,k)$ plane of Fig. \ref{phik},
$\Phi_\mr{infl}(k)$ and $\Phi_\mr{min}(k)$ defined by
$k^2\Phi_\mr{min}+U_k^{(1)}(\Phi_\mr{min})=0$, and
$k^2+U_k^{(2)}(\Phi_\mr{infl})=0$.
The stable region in the mean-field approximation is
$|\Phi|>\Phi_\mr{min}(0)=\Phi_\mr{vac}$.
For $\Phi_\mr{infl}(0)<|\Phi|<\Phi_\mr{min}(0)$ there are two minima in
$\phi$ for the last equation in \eq{meanflt}. One of them is metastable,
ie is unstable against sufficiently large amplitude modes. The spinodal
phase separation, the instability against infinitesimally small
amplitude fluctuations occurs when some eigenvalues of the second functional
derivative of the action becomes negative. This is region D in Fig. \ref{ipro},
$|\Phi|<\Phi_\mr{infl}(0)$. It is unreachable by the mean-field treatment
because no local minimum is found in $\phi$ which would satisfy
$|\phi|<\Phi_\mr{infl}(0)$. To understand better the
nature of these unstable regions one has to go beyond the mean-field
approximation and to follow the dynamics of the growing inhomogeneous
instabilities. We shall use the tree-level WH equation to deal with
such large amplitude, inhomogeneous fluctuations.

\subsubsection{Tree-level WH equation}
The construction of the condensate within the unstable region
implies the minimization of the action with respect to a large number
of modes. The RG strategy offers an approximation for such a
rather involved problem, it deals with the modes one-by-one
during the minimization \cite{wettinst}, \cite{inrg}.

The negative curvature of the potential makes the saddle point in \eq{pbl}
non-trivial,
\be\label{tpbl}
S_{k-\dk}[\phi]=\min_{\tilde\phi'}S_k[\phi+\tilde\phi']
=S_k[\phi+\tilde\phi_{cl}[\phi]]\not=S_k[\phi].
\ee
Since the saddle point depends in general on the background field,
$\tilde\phi_{cl}=\tilde\phi_{cl}[\phi]$, the action is  modified
during the blocking and we find a non-trivial RG flow \cite{inrg}. The local
potential approximation to the tree-level blocking is
\be\label{ltbl}
U_{k-\dk}(\Phi)=\min_{\tilde\phi_{cl,x}}\int_x\left[
\hf(\partial_\mu\tilde\phi_{cl,x})^2+U_k(\Phi+\tilde\phi_{cl,x,})\right].
\ee

One encounters here a conceptual problem. The $k$-dependence of the
tree-level RG flow might well be singular since there is no
obvious reason that the saddle points which are usually rather
singular functions of the parameters evolve smoothly. This problem
has already been noticed as a possible 'first order phase transition'
in the blocking which induces discontinuous RG flow \cite{srg},
\cite{srgh}, \cite{srgnu}. It was later shown by rigorous methods
that the RG flow is either continuous or the blocked action
is non-local \cite{regrg}. The resolution of this apparent
paradox is that the saddle point actually develops in a continuous manner
as we shall shown below. As the RG flow approaches the onset
of the condensate, $k\to k_{cr}$ then the new scaling laws
generate such an action that the saddle point turns out to be
continuous in $k$.

\subsubsection{Plane wave saddle points}\label{plwase}
The saddle points, the minima in \eq{tpbl} satisfy the highly non-linear
Euler-Lagrange equations whose solutions are difficult to find. The
use of sharp cut-off slightly simplifies the problem since it
reduces the functional space in which the minimum is sought to
$\tilde\phi\in{\cal F}_k\backslash{\cal F}_{k-\dk}$.
We shall retain the the plane wave saddle points only,
\be\label{plkspt}
\tilde\phi_{cl,p}={\rho_k\over2}\left[e^{i\theta_k}\delta_{p,ke_k}
+e^{-i\theta_k}\delta_{p,-ke_k}\right],\ \
\tilde\phi_{cl,x}=\rho_k\cos(ke_k\cdot x+\theta_k).
\ee
The parameter $\theta_k$ and the unit vector $e_k$ correspond
to zero modes, they control the breakdown of translational
and rotational symmetries.

This is a key to what happens at the
unstable line: {\em Despite the discreteness of the internal symmetry
$\phi\to-\phi$ there are soft modes because the inhomogeneous
saddle points break the continuous external symmetries}. The condensation
of particles with non-vanishing momenta automatically generates
Goldstone modes. This phenomenon is well known for solids where
the saddle point is a crystal of solitons which breaks external
symmetries and there is no integration over the zero mode to restore
the symmetrical ground state. Our point is that the soft modes make
their appearance even if the symmetry of the ground state is restored
by the integration over the zero modes.

The amplitude $\rho_k$ is determined by minimization,
\be\label{pwtbl}
U_{k-\dk}(\Phi)=\min_{\rho_k}\left({1\over4}k^2\rho_k^2+{1\over\pi}\int_0^\pi
dyU_k(\Phi+\rho_k\cos y)\right),
\ee
in the local potential approximation. The numerical implementation of
this iterative procedure to find the RG flow with an initial condition
imposed on the potential at $k=\Lambda$ gave the following results \cite{inrg}:
\begin{enumerate}
\item The recursive blocking relation, Eq. \eq{pwtbl} is not a finite
difference equation. Despite of this the flow converges as $\dk\to0$.
\item The amplitude of the saddle point satisfies the equation
$\Phi+\rho(k)=\Phi_\mr{min}(k)$.
\item The potential obtained by the tree-level blocking is
$U_k(\Phi)=-\hf k^2\Phi^2$.
\item The tree-level results above hold independently of the choice of the
potential at the cut-off.
\end{enumerate}

The key is point 3, point 1 follows immediately from it. The
lesson of this result is that the action is degenerate for the
modes at the cut-off, the kinetic and potential energies cancel.
The 'best' effective theory for a given plane wave mode is the one
whose cut-off $k$ is slightly above the wave vector of the mode.
This result suggests the degeneracy of the action within the whole
unstable region. Conversely, if we can show that the action is
degenerate at the cut-off within the unstable regime we
established this potential.

We prove by induction in the number of steps $k\to k-\dk$
that the variation of the action density
within the unstable regime is $\ord{\dk}$. This result protects the
consistency of the saddle point expansion for $d>1$ since $\dk\ge2\pi/L$ where
$L$ is the size of the system and therefore the variation of the action is
$\ord{L^{d-1}}$. Recall that $k_{cr}$ denotes the cut-off where the
kinetic snd the potential energies cancel.

\underline{First step:} Let us denote by $k'$ the value of
the cut-off at the first occurrence of non-trivial saddle point in the
numerical implementation of Eq. \eq{pwtbl}. It obviously satisfies the
inequality $k_\mr{cr}-\dk<k'<k_\mr{cr}$. The trivial saddle point
$\tilde\phi_\mr{cs}=0$ becomes unstable for blockings with
$|\Phi|<\Phi_\mr{infl}(k')$
in which case $|\Phi+\tilde\phi_{\mr{cs},x}|<\Phi_\mr{infl}(k')$, ie
$|\tilde\phi_{\mr{cs},x}|=\ord{\sqrt{\dk}}$. The term $\ord{\tilde\phi}$
is canceled in the action on a homogeneous background field
so the $\ord{\tilde\phi^2_\mr{cs}}$ contribution gives $\ord{\dk}$ variation.

\underline{Induction:} Suppose that the variation of the action is $\ord{\dk}$
in the unstable region and we lower the cut-off, $k\to k-\dk$.
At the new cut-off the balance between the kinetic and the potential energy
is lost by an amount of $\ord{\dk}$ since the potential energy is still
the given by $k$ but the kinetic energy corresponds to the lowered cut-off,
$k-\dk$. Thus the negative curvature potential energy wins and the action
bends downward in the unstable region as the function of the amplitude
of the plane wave. By assuming that the amplitude is stabilized at an
$\ord{\dk^0}$ value (will be checked later) the variation of the action
density $\ord{\dk}$ in the unstable region.

Point 2 can be understood in the following manner: During the
minimization of the action $\rho_k$ slides down on the $\ord{\dk}$
slope until the potential starts to increase again. We can find
where this happens by equating the slope of the original, bare
potential with that stated  in point 3. One would expect that the
deepest point is reached at $\rho_k=\Phi_\mr{infl}(k)$. But the
$\ord{\phi^4}$ term in $U_k(\phi)$ is not yet strong enough to
make the potential increase strong enough here. The two slopes
agree just at $\rho_k=\Phi_\mr{min}(k)$.

Result 4 which follows from 3, as well, can be called super universality
since it reflects scaling laws where all coupling constants are irrelevant.
Note that the $k$-dependence is continuous through the whole RG trajectory.

Point 3 is a generalized Maxwell construction. It reduces to the
traditional Maxwell construction for $k=0$, to the degeneracy of
$U_{k=0}(\Phi)$ for $|\Phi|<\Phi_\mr{min}(0)=\Phi_\mr{vac}$. The
naive Maxwell cut, applied for the concave part of the potential
would give the degeneracy for $|\Phi|<\Phi_\mr{infl}$ only. The
problem with this argument is that it produces an effective
potential which is convex everywhere except at
$|\Phi|=\Phi_\mr{infl}(0)$. The second derivative of the potential
is ill defined and the first derivative is discontinuous at this
point. By placing the system into a finite box the singularity is
rounded off and the second derivative becomes finite, but turns
out to be negative in a vicinity of $|\Phi|=\Phi_\mr{infl}(0)$.
Convexity regained only if the cut is extended between the minima,
$|\Phi|<\Phi_\mr{min}(0)$.

The generalized Maxwell construction in the mixed phase of a first
order phase transition can be understood by the dynamics of the
domain walls. The flatness of certain thermodynamical potentials
in the mixed phase reflects the presence of zero-modes, the
location of domain walls. Such a rather simple kinematical
mechanism which is independent of microscopic details is the
source of the `super universality', point 4 above. The role of the
domain walls is played by saddle points in our computation, the
cosine function in \eq{plkspt} realizes infinitely many equally
spaced, parallel domain walls. The integration over the zero modes
$\theta_k$ and $e_k$ according to the rules of the saddle point
expansion restores translational and rotational symmetries of the
ground state and reproduces the mixed phase.

One might object our independent treatment of the plane wave
saddle points. The RG equation \eq{pwtbl} handles the plane waves
in the consecutive momentum space shells independently which seems
as a vast oversimplification. But one should recall at this point
that according to the general strategy of the RG method the
dynamics of the modes eliminated during the blocking is retained
by the modification of the effective coupling constants. This is
not always possible since there are more modes than coupling
constants, the problem mentioned at the end of section
\ref{blcspts}. More precisely, the general framework of the
blocking what is in principle always applicable is turned into a
powerful scheme when an approximation is made. One truncates the
effective action and assumes that the solution of this
over-determined problem exists. This is the main, and so far
unproven, assumption of the RG method. Accepting this point we can
determine the blocked action by means of a homogeneous background
field and use it for non-homogeneous field configurations at the
next blocking step as written in Eq. \eq{pwtbl}.

\subsubsection{Correlation functions}
One can compute the tree-level contributions to the
correlation functions in a combination of the
mean-field and the saddle point approximation \cite{jean}.
Let us split the complete functional integral into the sum of
contributions which come from the unstable and the stable regions.
The fluctuations from the stable regions are governed by a non-trivial
action and they are taken into account as loop corrections.
The fluctuations in the unstable region experience a flat action and their
contributions must be taken into account on the tree-level.
The complication of integrating up these contributions is the
determination of the region where the action is flat.
Our approximation consists of estimating this region
for each plane wave independently, ie extending the integration
over the amplitude $r_p$ of the degenerate plane wave
\be
\tilde\phi_{cl}(x)=r_p\cos(pe_px+\theta_p)
\ee
over the interval $-\rho_p<r<\rho_p$, where $\rho_p$ is given by \eq{plkspt}.
Let us denote the integration over the resulting domain by $D_\Phi[r]$
and write the correlation function in momentum
space in this single-mode determined flat region as
\bea
G^\Phi_{tree}(p,q)&=&{1\over4}\left[\int D[\theta]D[e]D_\Phi[r]\right]^{-1}
\int D[\theta]D[e]D_\Phi[r]r_pr_q\nonu
&&\times
\left(e^{i\theta_p}\delta_{p,ke_p}+e^{-i\theta_p}\delta_{p,-ke_p}\right)
\left(e^{i\theta_q}\delta_{q,ke_q}+e^{-i\theta_q}\delta_{q,-ke_q}\right).
\eea
Due to the flatness of the action these integrals are purely
kinematical and can easily be carried out. All integration whose
variable does not show up in the integrand drops out. The integration
over the shift of the plane waves, the phase angle $\theta$, restores
the translation invariance,
\bea
G^\Phi_{tree}(p,q)&=&\delta_{p+q,0}\left[\int de\int dr_p\right]^{-1}
\int de\int dr_pr_p^2\nonu
&=&\delta_{p+q,0}{2(2\pi)^dd\over3\Omega_d}
{(\Phi_\mr{min}(p)-\Phi)^2\over\Phi_\mr{min}^{-1}(\Phi)},
\eea
where $\Phi_\mr{min}^{-1}(\Phi)$ is defined as
$\Phi_\mr{min}(\Phi_\mr{min}^{-1}(\Phi))=\Phi$.
The Fourier transform of this propagator describes a diffraction
type oscillation with characteristic length scale $\xi=k_{cr}^{-1}$,
the characteristic feature of domains in a homogeneous state.

\subsubsection{Condensation as crossover}
The scaling laws of the scalar model change already approaching
the condensation in the stable region $C$ of Fig. \ref{ipro}.
Inside the instable region $D$ the action with the potential given
by point 3 of section \ref{plwase} becomes invariant under blocking,
ie the whole region $D$ is an IR fixed point.

The vacuum of the scalar model is a single, homogeneous coherent state
consisting of zero momentum particles when $|\Phi|>\Phi_\mr{vac}$.
When $|\Phi|<\Phi_\mr{vac}$ we encounter the singularities as seen above.
This singularity may not be real, it might be smoothened out by higher
order vertices and a saddle points appearing for $k<k_\mr{cr}$. But the vacuum
is the superposition of inhomogeneous coherent states with characteristic
scales $\ell>1/k_\mr{cr}$. Each of them breaks space-time symmetries but
their sum remains symmetrical.

One may wonder if any of the discussion applies to the true vacuum
$|\Phi|=\Phi_\mr{vac}$. For a weakly coupled system the fluctuations are small
and one may hope that the fluctuations around $\Phi_\mr{vac}$
are stable and un-influenced by what is happening with the unstable modes
$\phi_x<\Phi_\mr{vac}$. But the answer to this question is rather
uncertain and is open because the true vacuum is just at the border of the
instabilities and the typical fluctuations around the true vacuum
$(\phi_x-\Phi_\mr{vac})\approx g^2(0)/g_4(0)$ penetrate into the
unstable region.

The issue of the modification of the scale dependence by a condensate \cite{scqed}
is rather general and points far beyond the simple scalar model.

The vacuum of asymptotically free models supports strong
correlations at long distance. In the case of 4 dimensional
Yang-Mills models the naive, perturbative vacuum is destabilized
by the one-loop level effective action which predicts that the
vacuum is made up by a coherent state of gluons \cite{savid}. But
this state cannot be the true vacuum. The problem is not only that
the field generated by the condensate is strong and spoils the
saddle point expansion but it turns out that there is an unstable
mode. The true vacuum is supposed to be found at even lower energy
densities where the long range fluctuations restore the the
external and color symmetry broken by a homogeneous condensate of
the charged spin one gluons \cite{spagh} and the vacuum is thought
to contain domains of homogeneous field in a stochastic manner
\cite{stoch}. This scenario is close to the view of the mixed
phase of the scalar model developed here with the difference that
the instability comes from the loop or the tree-level renormalized
action in the Yang-Mills or scalar model, respectively.
Furthermore the instability in the Yang-Mills model can be avoided
by extreme environment only, by immersing the system into strong
external field or bringing into contact with heat or particle
reservoir.

Similar instability occurs in vacua containing condensate of bound states.
The BCS vacuum is made homogeneous in a non-trivial manner
when represented in terms of the electrons making up the Cooper pairs.
The spontaneous breakdown of the chiral invariance in QCD
manifests itself in the condensate of quark-anti quark pairs
which is homogeneous after integrating out the instanton
zero modes only \cite{instch}. The crossover from the U.V. scaling regime
to the instability takes place at $k\approx1/\ell$ where $\ell$ is the size
of the bound states. Higher order derivative terms appearing as effective
vertices may generate similar crossover, as well \cite{mini}, \cite{antifer}.

Finally it is worthwhile mentioning the tunneling phenomena,
the dynamical extension of the instabilities considered above.
The interesting feature of the dynamical realization of the
tunneling by means of time-dependent, tree-level instabilities
is that they take place in conservative systems but the
long time distribution agrees with the equilibrium predictions
coming from the canonical ensemble \cite{fcfv}.

\subsection{Sine-Gordon model}\label{sineg}
It is not unusual that one arrives at periodic or variables  in the
construction of effective theories, cf non-linear sigma models
or non-Abelian gauge theories. Such variables represent a
double challenge. One problem which can be considered as local
in the internal space is that the perturbation expansion around a minima
of the potential should keep infinitely many vertices in order
to preserve the periodicity. Another, more difficult, global problem
is a conflict between the two requirements for the effective
potential for periodic variables, namely periodicity and convexity.
One can fulfill both requirements in a trivial manner only, by
constant effective potential. If true, this conclusion has
far reaching consequence for the phenomenological description
of such systems. As a case study we shall consider the simplest
model with periodic variable, the two dimensional sine-Gordon model
\cite{kerson}, \cite{sinegr}.

\subsubsection{Zoology of the sine-Gordon model}
The sine-Gordon model
\be\label{sglagr}
L_\mr{SG}={1\over2}(\partial_\mu\phi)^2+u\cos\beta\phi
\ee
has been shown to be equivalent
with the X-Y model \cite{kt}, \cite{jkkn}, \cite{kerson}, the Thirring
model \cite{col}, \cite{mand} and a Coulomb gas \cite{samu}.
The methods of Refs. \cite{jkkn}, \cite{col} and \cite{mand},
approximate duality transformation, bosonization and semiclassical
approximation are valid in certain regions of the coupling constant
space. The maps used in Refs. \cite{kerson} and \cite{samu} are exact.
The mapping \cite{kerson} $\psi_x=e^{\beta\phi_x}$ transforms the
model \eq{sglagr} into the compactified sine-Gordon model
\be\label{clagr}
L_\mr{CSG}={1\over2\beta^2}\partial_\mu\psi^*\partial_\mu\psi
+{u\over2}(\psi+\psi^*).
\ee
The models \eq{sglagr} and \eq{clagr} are equivalent in any order of
the perturbation expansion in continuous space-time where the
configurations are assumed to be regular.

The X-Y model which is characterized by the action
\be\label{xyaction}
S_\mr{XY}=-{1\over T}\left[\sum_{<x,x'>}\cos(\theta_x-\theta_{x'})
+h\sum_x\cos(\theta_x)\right]-\ln z\sum_xm^2_x,
\ee
where $T=\beta^2$, $h/T=u$, $z$ is the vortex fugacity and
$m_x$ denotes the vortex density. The RG equation of the X-Y model obtained
in the dilute vortex gas limit \cite{kt}, \cite{jkkn}
\be\label{cgasrg}
a{dT\over da}=4\pi^3z^2-\pi T^2h^2,\ \
a{dh\over da}=\left(2-{T\over4\pi}\right)h,\ \
a{dz\over dz}=\left(2-{\pi\over T}\right)z.
\ee

The X-Y model appears as a generalization of the compactified
sine-Gordon model since the lattice regularization transforms
the Lagrangian \eq{clagr} into \eq{xyaction} with $z=1$.
What is surprising here is that the vortex fugacity is a relevant
operator in the high temperature phase of the X-Y model
but it is entirely missed by the lattice regularization!
Does that mean that the sine-Gordon model Lagrangians \eq{sglagr} and
\eq{clagr} are incomplete?

The vortex term could be missed by the regulator because the plane
$z=0$ is RG invariant.
Therefore it is consistent to exclude vortices from the path integral
in the sine-Gordon model given in continuous space-time.
When we recast the model in lattice regularization without
ever thinking about vortices our lattice action neither suppresses,
nor enhances the singular configurations and sets $z=1$. But once the
vortices are not suppressed, $z\not=0$ the renormalized
trajectories moves away from from the plane of fixed $z$.
There is still no problem in the low temperature phase because
all what happens is that an irrelevant coupling constant became fixed.
This is how regulators work. But the problem is more serious
in the high temperature phase. The Lagrangians \eq{sglagr} or
\eq{clagr} can not give account of this phase because they lack a
relevant, renormalizable coupling constant. The perturbative
continuum theory without vortices is consistent
but once vortices are available kinematically their density evolves
in a non-trivial manner.

There is no problem with the lattice regulated
X-Y model because the vortex fugacity gets renormalized due to the
cut-off dependence of the vortex action. In other words, we can
move along the RG flow by adjusting $T$ and $h$ only, the renormalization
of $z$ is carried out 'automatically' in the partition function. The
renormalization of $z$ is a problem in the continuous
formalism only where the vortices are introduced formally as point-like
charges.

This is an unexpected mechanism which brings the singular nature of the
field configurations into play during renormalization and may plague
any quantum field theory. The knowledge of the classical action
in continuous space-time leaves open the possibility that we have of adjust
the fugacity of certain localized singularities or topological
defects not considered in the continuum.

It is interesting to speculate about similar phenomenon in QCD.
Lattice QCD is constructed with
$\ln z=0$ where $z$ is the fugacity of some topological defects, like
instantons, monopoles, merons etc. These topological defects and
singularities are supposed to play an important role in the confining
vacuum, as vortices do in the high temperature phase of the X-Y
model. The renormalization of the fugacities of these objects
should first be studied in lattice QCD in order to construct
a continuum description of the vacuum.

\begin{figure}
\centerline{\psfig{file=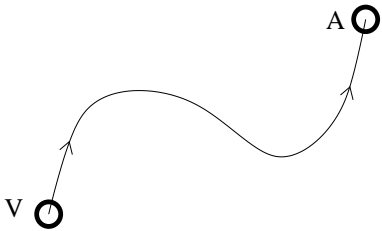,height=3cm,width=5cm}}
\caption{The world line of a soliton starts at a vortex (V)
and ends at an anti-vortex (A) in the two dimensional space-time.\label{vort}}
\end{figure}

The sine-Gordon model possesses a topological current,
$j_{\mu,x}=\beta\epsilon_{\mu \nu}\partial_\nu\phi_x/2\pi$ which is
obviously conserved when the path integral is saturated by
field configurations with analytic space-time dependence. Its flux,
the vorticity, gives the soliton number. The world
lines of the sine-Gordon solitons end at the X-Y model vortices,
cf Fig. \ref{vort}. The distance between the bound vortex-anti vortex
pairs shrinks with the lattice spacing in the low temperature
continuum limit. Any measurement with finite resolution loose sight
of the instability of solitons in the renormalized theory.
The average distance between vortices stays finite, cut-off independent
in the continuum limit of the high temperature phase and the soliton
decay can be observed. One expects the breakdown or at least important
modification of the bosonization transformation in this phase. In fact,
the non-conservation of the topological current requires fermion number
non-conserving terms in the fermionic representation, a fundamental
violation of the rules inferred from the weak coupling expansion.

\subsubsection{Effective potential}
In order to understand better the dynamics of the long wavelength
modes we compute the effective potential in the sine-Gordon model.
We follow the WH method truncated to the local potential approximation
which picks up the tree level evolution, too. It is obvious that period
length of the potential $2\pi/\beta$ remains RG invariant in this
approximation. The tree ands the loop level RG equations are
\be\label{treesg}
U_{k-\dk}(\phi)=\min_{\rho}
\left[{k^2\over4}\rho^2+\int_{-1}^{1}duU_k(\phi+\rho\cos(\pi u))\right],
\ee
and
\be\label{loopsg}
kU_{k-\dk}(\phi)=kU_k(\phi)+\dk{k^2\over4\pi}\ln[k^2+U^{(2)}_k(\phi)]
\ee
in the plane $z=0$.

We use the Fourier expanded form for the potential
\be
U_k(\phi)=\sum_{n=0}^{\infty}u_n(k)\cos\left(n\beta\phi\right)
\ee
and compute the leading order contribution to the WH equation when
expanded in the potential,
\be\label{leadord}
k\partial_k\tilde u_n=\left({\beta^2n^2\over4\pi}-2\right)\tilde u_n,
\ee
in terms of the the dimensionless coupling constants $\tilde u_n=u_n/k^2$.
This agrees with the second equation in \eq{cgasrg}.
The solution of \eq{leadord} is
\be
\tilde u_n(k)=\tilde u_n(\Lambda)
\left({k\over\Lambda}\right)^{{\beta^2n^2\over4\pi}-2}.
\ee

The coupling constants $u_n$ are irrelevant in the disordered
phase $T=\beta^2>8\pi$ and the effective potential obtained for $k=0$ is
flat. The coupling constants $n<8\pi/T$ are relevant in the ordered phase
and the effective potential is non-trivial. At this pointe one suspects
that Maxwell construction interferes with the evolution because a
non-trivial periodic functions necessarily has concave regions. To settle
this question one has to rely on the numerical solution of the
evolution equation \eq{treesg}. We followed the loops-generated evolution
\eq{loopsg} from the initial condition set at $k=\Lambda$ until the stability
is lost at $k=k_\mr{cr}$ for background field values which lie at local
maxima of the periodic potential. For $k<k_\mr{cr}$ equation
\eq{treesg} was used. The result is that the coupling constants approach
zero as $k\to0$ due to the plane wave saddle points. The effective
potential is trivial in both phases.

\subsubsection{Breakdown of the fundamental group symmetry}
The local potential $U_k(\phi)$ flattens without developing singularities
in the high temperature disordered phase. The transformation
\be\label{fund}
\phi_x\to\phi_x+{2\pi\over\beta}
\ee
is a discrete symmetry of the action and is preserved in the vacuum
since the potential barrier between the minima is vanishing
for long range modes. Since we are at the lower critical
dimension, $d=2$, the large amplitude long range modes realize the
'tunneling' between the minima of the periodic potential.
On the contrary to this situation, the potential
develops discontinuous second derivatives in the ordered, low temperature phase
and the instability driven flattening of the potential $U_k(\phi)$
reflects the survival of barriers between the minima of the potential.
All this looks like a spontaneous symmetry breaking, therefore our
conclusion is that the transformation \eq{fund} is not a symmetry
of the vacuum in the ordered, low temperature phase.

The usual circumstance under which such a phenomenon arises is the
multiple connectedness of the internal space. In the present
context the non-linear $U(1)$ $\sigma$-model parametrization, Eq.
\eq{clagr}, is based on the internal space $U(1)$, with the
fundamental group $Z$ generated by the transformation \eq{fund}.
The dynamical breakdown of the fundamental group symmetry is a
genuine quantum effect. In fact, the path integral formally
extends over all homotopy classes, this is the symmetrical phase.
When the dynamics develops sufficiently high barriers between the
homotopy classes the path integral becomes restricted to a single
homotopy class.

It is important to recall that the time evolution described by the
Schrodinger equation can be derived from the path integral by
performing infinitesimal variations on the end point of the
trajectory. Therefore the consistency of the dynamics can be
maintained when the path integration is restricted into any
functional subspace which is closed under continuous deformation
of the trajectories. The loss of the interference between homotopy
classes is the characteristic feature of the symmetry broken
phase.

Such symmetry breaking is the key to understanding the way quarks
become deconfined at high temperature or the droplet phase is
formed for quantum liquids \cite{mech}. The configuration space
for global gauge transformations is $SU(3)/Z_3$ for gluons, with
the fundamental group $Z_3$. The quark propagator vanishes in the
symmetrical phase due to the destructive interference between the
three homotopy classes. In the $Z_3$ symmetry broken deconfined
phase there is no interference and quarks can be observed. Quantum
liquids in the first quantized formalism display similar symmetry
breaking pattern. The coordinate space for $N$ particles is
$R^{3N}/S_N$ where $S_N$ consists of permutations of the
particles. The absence of the overlap among states belonging to
different droplets suppresses the (anti)symmetrization of the
states and reduces the exchange symmetry of the ground state.

How can we recognize the dynamical breakdown of the fundamental group
symmetry? The simplest strategy, to look for the minima of the
effective potential fails because Maxwell construction hides
any structure in a periodic potential. The answer to this question lies
in the topological structure of excitations and the order parameter
will be a topological susceptibility.

This is rather natural since the phase transition is the restriction
of the path integral into a single homotopy class,
the imposition of a topological constraint.
Let us assume that the analysis sketched above for the plane $z=0$
remains qualitatively valid for $z\not=0$, ie the effective potential
is always flat, due to the loop-generated evolution or the Maxwell construction
in the disordered or ordered phase, respectively. The topological
invariant characterizing the homotopy classes is the soliton number.
In the weak coupling expansion (ordered phase) the path integral is
constrained into a single homotopy class hence the soliton number is
conserved, its susceptibility is vanishing in the continuum limit.
The stability of the soliton, based on the continuity of the time
evolution is lost in the disorder phase because the short distance,
large amplitude fluctuations of the typical field configuration
extends the path integral all soliton number sector. Therefore the
susceptibility of the soliton number is non-vanishing in this phase.

\subsubsection{Lower critical dimension}
It is worthwhile noting the manner the Mermin-Wagner-Coleman theorem
\cite{mwc} appears in the framework of the local potential approximation.
By setting $g_2=0$ for $d=2$ the linear part
of ${\cal P}_n$ in Eqs. \eq{betapol}
provides a finite evolution but the higher order
terms make the renormalized trajectory to diverge as $k\to0$
unless the coupling constants approach zero. We can construct
interactive, massless models in two dimensions so long the
running coupling constants approach zero sufficiently fast
in the IR regime as in the case of the sine-Gordon model, presented
above. The retaining of the terms $\ord{\partial^2}$ in the gradient
expansion provides another mechanism to suppress the IR divergences by
generating singular wave function renormalization constant,
$Z_k$ as $k\to0$ \cite{ketdimz}.

In both cases the way to avoid the non-interactive system predicted
by the Mermin-Wagner-Coleman theorem is to use the RG improved
perturbation expansion. In the ordinary perturbation series we have the
contributions like
\be
g^n\int_{p_1,\cdots,p_k}f_{p_1,\cdots,p_k}
\ee
and the integral is IR divergent in $d=2$. One has running coupling
constants, $g\to g_P$, in the RG improved perturbation expansion where the
analogous contribution is
\be
\int_{p_1,\cdots,p_k}g^n_{P(p_1,\cdots,p_k)}f_{p_1,\cdots,p_k}.
\ee
When the IR fixed point is Gaussian, $\lim_{P\to0}g_P=0$, then
the theory remains interactive and the  running coupling
constants suppress the IR divergences.

\subsection{Gauge models}\label{gauges}
The blocking step of the RG scheme, being scale dependent, violates
local symmetries which mix modes with different length scales.

The most natural way to deal with this breakdown of gauge
invariance is to cancel the gauge non-invariant terms by fine
tuning non-invariant counterterms \cite{finet}. But this is
possible in perturbation expansion only. Another, better suited
strategy to the RG method is to argue that the way Ward identities
are violated by blocking shows that the BRST invariance is broken
by the running cut-off only and the BRST symmetry is recovered at
the IR fixed point when the cut-off is removed \cite{ellwev},
\cite{mwi}. This were certainly a good procedure if it could be
implemented without further truncation. In a weakly coupled model
such as QED one could construct approximations with controllable
errors. In asymptotically free models like QCD non-perturbative
long range correlations make any scheme which imposes gauge
invariance in an approximate manner unreliable \cite{haarons}. The
demonstrate this point consider the static force law between two
test charges in a Yang-Mills theory in a scheme where the gauge
invariance, Gauss' law is implemented approximately. The gauge
non-invariant components of the vacuum state appear charged, by
definition. Thus we have an uncontrollable color charge
distribution around the test charges. When the distance between
the test charges are large enough then it will be energetically
favorable for the uncontrollable charges to break the flux tube.
We loose the string tension in a manner similar how it happens in
the true QCD vacuum due to the virtual quark-anti quark
polarizations. In more formal words, an arbitrarily weak
gauge-dependent perturbation can change the long range features of
the vacuum as in the ferromagnetic phase of the Ising model.

The most natural way to guarantee gauge invariance is
to achieve independence on the choice of gauge by using gauge
invariant quantities only. The proposal of Ref. \cite{loops} goes
along this line but the use of loop variables renders the computation
rather involved. We discuss now a version of the RG scheme in the
internal space which at least for Abelian models produces the
effective action without any gauge fixing.
The electron mass combined with chiral transformation
has already been used in generating evolution equation in Ref.
\cite{simin}. We shall follow a simpler and more general by avoiding gauge
dependence in an explicit manner for the photon Green functions
\cite{alposa}. The gauge invariant electron composite operators
are not difficult to include \cite{dft}.

\subsubsection{Evolution equation}
Let us consider the generator functional
\be\label{genfc}
e^{\ih W[j,\jb,J]}=\int D[\psib]D[\psi]D[A]e^{\ih
\int_x[-{1\over4e^2}F^{\mu\nu}F_{\mu\nu}-{\alpha\over2}(\partial_\mu A^\mu)^2
+\psib\left(iD\br-m\right)\psi+\jb\cdot\psi+\psib\cdot j+J^\mu\cdot A_\mu]},
\ee
where $D_\mu=\partial_\mu+iA_\mu$ and the dimensional regularization is used to
render $W$ finite. We control the quantum fluctuations by
modifying the action, $S\to S+S_\lambda$ with
\be
S_\lambda={\lambda\over4e^2}\int_xA_{\mu,x}\Box T^{\mu\nu}A_\nu(x),
\ee
where $T_{\mu\nu}=g_{\mu\nu}-\partial_\mu\partial_\nu/\Box$.
The infinitesimal change of the control parameter
$\lambda\to\lambda+\Delta\lambda$ modifies the free photon propagator
\be\label{propmv}
D^{\mu\nu}(x-y)\to D^{\mu\nu}(x-y)+\Delta\lambda
\int dzD^{\mu\rho}(x-z)\Box_zT_{\rho\kappa}D^{\kappa\nu}(z-y)
+\ord{(\Delta\lambda)^2},
\ee
in a manner reminiscent of the Callan-Symanzik scheme.

The evolution equation \eq{cevol} with the present suppression takes the form
\be\label{egyszeq}
\partial_\lambda\Gamma_\lambda[A]
={\hbar\over2e^2}\tr\left[\Box T^{\mu\nu}\left({\delta^2\tGa[A]\over
\delta A_\mu\delta A_\nu}\right)^{-1}\right]
\ee
in terms of the photon effective action
$\tGa_\lambda[A]=\Gamma_\lambda[A]+S_\lambda[A]$.
We project the evolution equation on the functional space given by the ansatz
\be\label{effa}
\tGa[A]=\hf\int_xA_{\mu,x}D^{-1}(i\partial)T^{\mu\nu}A_{\nu,x}+{\cal C}[A]
\ee
where
\be
D^{-1\mu\nu}=-{1+\lambda\over e^2}\Box T^{\mu\nu}-\alpha\Box L^{\mu\nu},
\ee
$L^{\mu\nu}=\delta^{\mu\nu}-T^{\mu\nu}$ and ${\cal C}[A]$ is a gauge
invariant functional.

The control parameter $\lambda$ 'turns on' the fluctuations of the
photon field. Therefore the electrons loop contributions to the effective
action must already be present at the initial condition which is chosen to be
${\cal C}[A]=-i\tr\ln(iD\br-m)$ at $\lambda=\lambda_0$.

The second functional derivative matrix is written as
\be
{\delta^2\tGa[A]\over\delta A_\mu\delta A_\nu}=
D^{-1\mu\nu}+{\delta^2{\cal C}[A]\over\delta A_\mu\delta A_\nu}
\ee
and the inversion is carried out by expanding in the non-diagonal pieces
to write the evolution equation as
\be\label{evc}
\partial_\lambda{\cal C}[A]={\hbar\over2e^2}\tr\left[\Box T^{\mu\nu}
D\sum_{n=0}^\infty(-1)^n\left(
{\delta^2{\cal C}[A]\over\delta A_\mu\delta A_\nu}D\right)^n\right].
\ee

\subsubsection{Gauge invariance}
We show now that the limit $\alpha\to0$ can be taken in the
evolution equation without hitting any singularity. The gauge
fixing parameter $\alpha$ enters through the photon propagator
$D$ in the Neuman-expansion of the right hand side of Eq. \eq{evc}.
The $\alpha$-dependent longitudinal contributions of the first and the
last $D$ factor are suppressed by the gauge invariance of the suppression
term, represented by the transverse projection $T^{\mu\nu}$ in \eq{evc}.
The longitudinal photon contributions of the internal propagators
are suppressed by the gauge invariance of the effective action,
\be
\partial_\mu{\delta{\cal C}[A]\over\delta A_\mu}=0.
\ee
According to this equation $\delta^2{\cal C}[A]/\delta A_\mu\delta A_\nu$
does not mix the longitudinal and the transverse modes, ie non-longitudinal
contributions appear in the evolution equation.

We choose a gauge in the argument above what was relaxed at the end of
the computation. But the steps followed remain well defined even if
we start with $\alpha=0$, without any gauge fixing. Our argument
about the decoupling of the longitudinal and transverse contributions
to the evolution equation still applies but it is not clear if the
longitudinal part was well defined. This subtle issue is settled
by Feynman's $\epsilon$ parameter, devised to lift infinitesimally the
degeneracies of the action. It is easy to see that it gives
a weak variation of the action along gauge orbits by breaking
gauge invariance. As long as our truncation of the evolution equation,
the functional ${\cal C}[A]$ is explicitly gauge invariant $\epsilon$ plays
the role of an infinitesimally external magnetic field in the Ising
model, ie helps the breakdown of the gauge symmetry only if it is
really broken in the true vacuum.

\section{What has been achieved}
The traditional implementation of the RG idea has proven to be essential
in Statistical and High Energy Physics, starting with the understanding
of critical phenomena \cite{critph}, finite size scaling \cite{fssc},
$\epsilon$-expansion \cite{epsiexp}, dynamical processes with different time
scales \cite{critdyn}, continuing with partial resummation of the perturbation
expansion \cite{resum}, parametrizing the scale dependence at high energies
\cite{asfree} and ending with the construction of effective theories in Particle
Physics \cite{cheffth,soleffth,hmeffth}. This is an
inexcusable short and incomplete list, its role is to demonstrate
variability and the importance of the method only.
Our main concern here was the functional form of the RG method
and its generalizations. From this point of view one may distinguish
conceptual and more technical achievements as the power of the
functional formalism is more exploited.

We should consider the RG method as 'meta-theory', or in
more practical terms as a language, as Feynman graphs are used
in particle physics. But this language is not bound by small parameters
and can provide us a general, non-perturbative approximation method
beyond the semiclassical expansion and numerical simulations.

The path integration was first viewed as a powerful book-keeping
device for perturbation expansion and the truly non-perturbative
application came later, after having gained some experience with
the formalism. This is similar to the development of the functional
RG method where the RG equation refers to the generator function of
the effective coupling constants and provides us with a simple
procedure to keep track of the Feynman graphs.
The steps in functional calculus are more
cumbersome because they deal with generator function(al)s, with
infinitely many coupling constants. The ultimate goal is to
go beyond this level and to use this formalism in a genuinely non-perturbative
manner.

The semiclassical expansion is our first step in this direction.
It will be important to check how
the RG continues to be applicable where the saddle point expansion ceases to be
reliable.

Another field the RG scheme might be compared with is lattice regulated
field theory. Both are general purpose tools to deal
with non-perturbative systems. The bottle-neck of the numerical
simulations on the lattice is the need to send the U.V. and the IR
cut-offs sufficiently far from each others and the restriction to
Euclidean space-time. The RG strategy is set up in continuous,
Minkowski space time and there is no particular problem with keeping
the U.V. and IR cut-offs far from each other.
But the drawback of the RG strategy is that it is rather
lengthy to extend the space of (effective) action functionals
used in the computation. Since the limitations of the two methods are
quite different they might be used in a complementary manner.

The functional formalism is promising because of the possibility of
following the mixing of a much larger number of operators as
in the traditional strategy. This feature gives the hope of extending
the applicability of the method from a single scaling regime
to the whole range of scales covered by the theory. Such an
extension may provide us valuable information about the
competition of interactions in realistic theories.

\section*{Acknowledgments}
I thank Jean Alexandre and Kornel Sailer for several useful
discussions during our collaboration.

\end{document}